\documentclass[]{aa}  

\usepackage{natbib}
\bibpunct{(}{)}{;}{a}{}{,}

\usepackage{graphicx}
\usepackage{revsymb}	
\usepackage{amsmath}
\usepackage{txfonts}
\usepackage{amssymb}
\usepackage{geometry} 
\usepackage[parfill]{parskip} 
\usepackage{slantsc}
\usepackage{array}
\usepackage{amsfonts}
\usepackage{siunitx}
	
\usepackage{epstopdf}	
\usepackage{epsfig}	

\usepackage{url}

\usepackage[colorlinks=true,linkcolor=black, urlcolor=black, citecolor=black]{hyperref}

\usepackage{color}
\usepackage{xcolor}
\colorlet{rouge}{red!70!darkgray}

\usepackage{xfrac}

\usepackage{sidecap}
\usepackage[font=small, labelfont=bf]{caption}
\usepackage{subcaption}

\usepackage{threeparttable}

\begin{document}
   \title{Asteroseismic modelling strategies in the PLATO era}
      \subtitle{I. Mean density inversions and direct treatment of the seismic information}
\author{J. B\'{e}trisey\inst{1} \and G. Buldgen\inst{1} \and D. R. Reese\inst{2} \and M. Farnir\inst{3} \and M.-A. Dupret\inst{4} \and S. Khan\inst{5} \and M.-J. Goupil\inst{2} \and P. Eggenberger\inst{1} \and G.~Meynet\inst{1}}
\institute{Observatoire de Genève, Université de Genève, Chemin Pegasi 51, 1290 Versoix, Suisse\\email: 	\texttt{Jerome.Betrisey@unige.ch}
\and  LESIA, Observatoire de Paris, Université PSL, CNRS, Sorbonne Université, Université Paris Cité, 5 place Jules Janssen, 92195 Meudon, France
\and Centre for Fusion, Space, and Astrophysics, Department of Physics, University of Warwick, Coventry, CV4 7AL, United Kingdom
\and Institut d’Astrophysique et G\'{e}ophysique de l’Universit\'{e} de Liège, Allée du 6 août 17, 4000 Liège, Belgium
\and Institute of Physics, Laboratory of Astrophysics, École Polytechnique Fédérale de Lausanne (EPFL), Observatoire de Genève, Chemin Pegasi 51, 1290 Versoix, Switzerland}
\date{\today}

\abstract{Asteroseismology experienced a breakthrough in the last two decades thanks to the so-called photometry revolution with space-based missions such as CoRoT, \emph{Kepler}, and TESS. Because asteroseismic modelling will be part of the pipeline of the future PLATO mission, it is relevant to compare some of the current modelling strategies and discuss the limitations and remaining challenges for PLATO. In this first paper, we focused on modelling techniques treating directly the seismic information.}
{We compared two modelling strategies, fitting directly the individual frequencies, or coupling a mean density inversion with a fit of the frequency separation ratios.}
{We applied these two modelling approaches to six synthetic targets with a patched atmosphere from \citet{Sonoi2015}, for which the `observed'  frequencies were obtained with a non-adiabatic oscillation code. We then studied ten actual targets from the \emph{Kepler LEGACY} sample.}
{As it is well known, the fit of the individual frequencies is very sensitive to the surface effects and to the choice of the underlying semi-empirical surface effects prescription. This limits significantly the accuracy and precision achievable for the stellar parameters. The mass and radius tend to be overestimated, and the age therefore tends to be underestimated. In contrast, the second strategy based on mean density inversions and on the ratios efficiently damps the surface effects, and allows us to get precise and accurate stellar parameters. The average statistical precision of our selection of targets from the \emph{LEGACY} sample with this second strategy is 1.9\% for the mass, 0.7\% for the radius, and 4.1\% for the age, well within the PLATO mission requirements. The addition of the inverted mean density in the constraints significantly improves the precision of the stellar parameters, on average by 20\%, 33\%, and 16\%, respectively for the stellar mass, radius, and age.}
{The modelling strategy based on mean density inversions and frequencies separation ratios showed promising results for PLATO as it achieved a precision and accuracy on the stellar parameters meeting the PLATO mission requirements with ten \emph{Kepler LEGACY} targets, and leaving some margin for other unaccounted systematics such as the choice of the physical ingredients of the stellar models or the stellar activity.}


\maketitle

\section{Introduction}
With the launch of the space-based photometry missions CoRoT \citep{Baglin2009}, \emph{Kepler} \citep{Borucki2010}, and TESS \citep{Ricker2015} in the last two decades, asteroseismology experienced a rapid development. The field will further expand with the next generation instrument of the future PLATO mission \citep{Rauer2014}. The data quality from these missions enables the use of so-called seismic inversion techniques \citep[see][for a recent review]{Buldgen2022c}, that were until then restricted to helioseismology, where they were applied with tremendous success \citep[see e.g.][for reviews]{Basu&Antia2008,Kosovichev2011,Buldgen2019e,JCD2021}. One of the key challenges of PLATO is the precision requirements on the stellar parameters (1-2\% in radius, 15\% in mass, and 10\% in age for a Sun-like star). In this context and considering the fact that asteroseismic modelling will be part of the PLATO pipeline, it is relevant to combine state-of-the-art modelling strategies exploiting seismic data, classical constraints (e.g. interferometric radius, luminosity, metallicity, effective temperature, etc.), and inversion techniques, and discuss the remaining challenges that could limit the precision and accuracy of the stellar parameters estimated with PLATO data. Among them, we can highlight the so-called surface effects \citep[see e.g.][]{Ball&Gizon2017,Nsamba2018,Jorgensen2020,Jorgensen2021,Cunha2021}, the choice of the physical ingredients \citep[see e.g.][]{Buldgen2019f,Betrisey2022}, and the stellar activity that will be the subject of a future article in this series\citep[see e.g.][]{Broomhall2011,Santos2018,Santos2019_sig,Santos2019_rot,Howe2020,Thomas2021,Santos2021}.

Due to the variety of modelling strategies that were developed over the years, we will divide our discussion into a series of papers. In this first article, we will consider techniques that are treating the seismic information directly, by fitting the individual frequencies or frequency separation ratios. In a future paper of this series, we will consider techniques that treat the seismic information indirectly, by studying indicators that are orthogonalised using a Gram-Schmidt procedure \citep{Farnir2019,Farnir2020}. As a side note, we remark that it is also possible to treat directly the seismic information with the $\varepsilon_{nl}$ matching technique \citep{Roxburgh&Vorontsov2003,Roxburgh2015a,Roxburgh2016a}, and that a comparison between that technique and the ones presented in this study would be relevant for a future study. In addition, there also exists other modelling techniques that can circumvent some of the PLATO challenges, but they are more difficult to implement in a pipeline. To only quote a few examples, it is possible to constrain the stellar structure by applying the differential response technique \citep{Vorontsov1998,Vorontsov2001,Roxburgh&Vorontsov2002a,Roxburgh&Vorontsov2002d,Appourchaux2015}, by using inversions based on so-called seismic indicators \citep{Reese2012,Buldgen2015a,Buldgen2015b,Buldgen2018} and applied to a variety of targets \citep{Buldgen2016b,Buldgen2016c,Buldgen2017c,Buldgen2019b,Buldgen2019f,Betrisey2022,Buldgen2022b,Betrisey2023_rot}, or by constraining the properties of the convective core with an inversion of frequency separation ratios \citep{Betrisey&Buldgen2022}.

In Sec. \ref{sec_spelaion_grid}, we introduce a new high-resolution grid of standard non-rotating stellar models, the \emph{Spelaion} grid. In Sec. \ref{sec_modelling_strategies}, we present the state-of-the art of modelling techniques using directly the asteroseismic data, the classical constraints, and inversion techniques. We first apply them to six synthetic targets with a patched atmosphere from \citet{Sonoi2015}, and then in Sec. \ref{sec_applications_LEGACY} to a selection of ten actual targets from the \emph{Kepler LEGACY} sample. Finally, we draw the conclusions of this work in Sec.~\ref{sec_conclusions}.

\section{The \emph{Spelaion} grid}
\label{sec_spelaion_grid}
The \textit{Spelaion} grid is a large high-resolution grid of standard non-rotating models ($\sim 5.1$ million models) designed to cover main-sequence (MS) stars between 0.8 and 1.6 solar masses. The grid can deal with a large variety of chemical compositions and mixing, with up to three dedicated free parameters (initial hydrogen mass fraction $X_0$, initial metallicity $Z_0$, and overshooting $\rm\alpha_{ov}$). It has a high mesh resolution that brings two advantages. First, the coupling with a minimisation algorithm that can interpolate within the grid allows for a very thorough exploration of the parameter space. Second, the high resolution reduces the issues with the interpolation of higher mass stars. Indeed, these stars can have convective cores or mixed modes at low frequency that are difficult to capture with a lower grid resolution. Presently, the low order mixed modes are unlikely to be observed in main-sequence stars with the actual instruments because they are in a noisy region of the frequency spectrum. For each model of the grid, we computed the theoretical adiabatic frequencies between fixed boundaries in adimensional angular frequency, corresponding approximately to the modes with $n \sim 4-33$ for a solar model, and a few more high radial orders for higher mass stars. This is a broad mode range that goes slightly beyond the actual observational capabilities at low and high radial order, and for reference, the radial order of the frequency of maximal power $\nu_{max}$ of the targets considered in this work is around $n=21$. We considered the $l=0,1,2$ degrees as the grid is ultimately designed to fit the $r_{01}$ and $r_{02}$ ratios.

The grid is composed of three sub-grids that cover specific types of physics (standard, high metallicity, and overshooting). Their statistics and properties are summarised in Tables \ref{tab_Spelaion_statistics} and \ref{tab_Spelaion_properties}. The evolutionary sequences were computed with the Liège Evolution Code \citep[CLES,][]{Scuflaire2008b}, and for each time-step, the frequencies were computed with the adiabatic Liège Oscillation Code \citep[LOSC,][]{Scuflaire2008a}. Regarding the physical ingredients of the models, we used the AGSS09 abundances \citep{Asplund2009}, the FreeEOS equation of state \citep{Irwin2012}, and the OPAL opacities \citep{Iglesias1996}, supplemented by the \citet{Ferguson2005} opacities at low temperature and the electron conductivity by \citet{Potekhin1999}. The microscopic diffusion was described using the formalism of \citet{Thoul1994}, with the screening coefficients of \citet{Paquette1986}, and the nuclear reaction rates are from \citet{Adelberger2011}. The mixing-length parameter $\alpha_{\mathrm{MLT}}$ is fixed at a solar calibrated value of 2.05, following the implementation of \citet{Cox&Giuli1968}. For the atmosphere modelling, we used the $T(\tau)$ relation of model C of \citet{Vernazza1981}.

\begin{figure}[t!]
\centering
\includegraphics[scale=0.58]{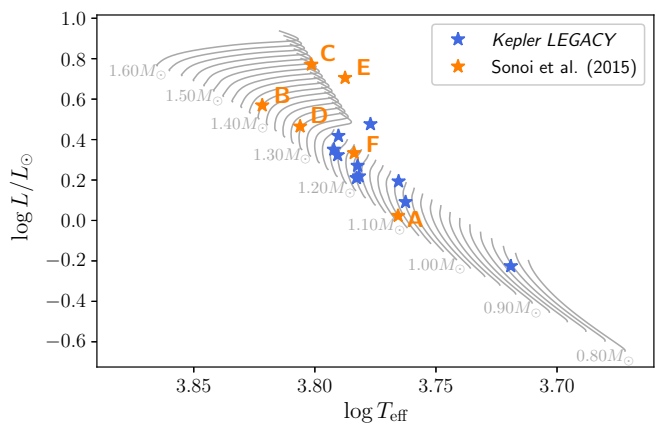} 
\caption{HR diagram of the targets considered in this work. The \citet{Sonoi2015} targets are denoted by the orange stars and the \emph{Kepler LEGACY} targets by the blue stars. The grey lines correspond to the evolutionary tracks from a slice of the \emph{Spelaion} grid with $X_0=0.72$, $Z_0 = 0.018$, and $\rm\alpha_{ov}=0.00$.}
\label{fig_HR}
\end{figure}

In Fig. \ref{fig_HR}, we illustrate the Hertzsprung-Russell (HR) diagram of the two sets of targets considered in this work, in orange for the synthetic targets and in blue for the actual targets from the \emph{Kepler LEGACY} sample \citep[hereafter abbreviated to \emph{LEGACY} sample,][]{Lund2017}. The grey lines correspond to the evolutionary tracks from a slice of the \emph{Spelaion} grid with $X_0=0.72$, $Z_0 = 0.018$, and $\rm\alpha_{ov}=0.00$.


\begin{table}[t!]
\centering
\caption{Statistics of \emph{Spelaion} and its subgrids.}
\begin{tabular}{lccc}
\hline 
 & Tracks & Models & Modes \\
\hline \hline
\emph{Spelaion} & \num{25079} & \num{5121459} & 485 million \\ 
Metallic Sun & 1176 &  \num{149398} & 13.5 million \\  
Standard MS & 7544 & \num{1341813} & 126 million \\  
Overshooting MS & \num{20930} & \num{4600562} & 439\ million \\
\hline 
\end{tabular}
\label{tab_Spelaion_statistics}
\end{table}

\begin{table}[t!]
\centering
\caption{Mesh properties of the \textit{Spelaion} subgrids.}
\begin{tabular}{lccc}
\hline 
 &  Minimum & Maximum & Step \\ 
\hline \hline 
\textit{Metallic Sun} &  &  &  \\ 
Mass $(M_\odot)$ & 0.94 & 1.06 & 0.02 \\ 
$X_0$ & 0.64 & 0.71 & 0.01 \\ 
$Z_0$ & 0.020 & 0.040 & 0.001 \\  
Overshooting & \multicolumn{3}{c}{$\rm\alpha_{ov}=0.00$} \\ 
\hline 
\textit{Standard MS} &  &  &  \\ 
Mass $(M_\odot)$ & 0.80 & 1.60 & 0.02 \\  
$X_0$ & 0.67 & 0.74 & 0.01 \\  
$Z_0$ & 0.008 & 0.030 & 0.001 \\ 
Overshooting & \multicolumn{3}{c}{$\rm\alpha_{ov}=0.00$} \\ 
\hline 
\textit{Overshooting MS} &  &  &  \\  
Mass $(M_\odot)$ & 1.10 & 1.60 & 0.02 \\  
$X_0$ & 0.68 & 0.74 & 0.01 \\  
$Z_0$ & 0.008 & 0.030 & 0.001 \\ 
Overshooting & 0.00 & 0.20 & 0.05 \\ 
\hline 
\end{tabular}
\label{tab_Spelaion_properties}
\end{table}

\section{Modelling strategies}
\label{sec_modelling_strategies}

In this first paper, we focus on modelling strategies treating directly the seismic information, either in the form of individual frequencies, or in the form of frequency separation ratios. Over the years, a variety of methods were developed, such as Levenberg-Marquardt algorithms \citep[see e.g.][]{Frandsen2002,Teixeira2003,Miglio&Montalban2005}, genetic algorithms \citep{Charpinet2005,Metcalfe2009,Metcalfe2014}, Bayesian inference \citep{Silva-Aguirre2015,Silva-Aguirre2017,AguirreBorsen-Koch2022}, machine learning \citep{Bellinger2016,Bellinger2019d}, or Markov Chain Monte Carlo \citep[MCMC,][]{Bazot2008,Gruberbauer2013,Rendle2019}, to only quote a few examples of algorithms.

In this study, we first investigated the fit of the individual frequencies, with a focus on the impact of the surface effects. Then, we tested a more elaborate technique, that uses frequency separation ratios coupled with a mean density inversion, that has already shown its strengths \citep{Buldgen2019f,Betrisey2022}. We also investigated the impact of the correlations between the inverted mean density and the frequency separation ratios, that were neglected in the past studies. For all the minimisations, we used the AIMS software \citep{Rendle2019} and we applied both modelling strategies on synthetic targets from \citet{Sonoi2015} (models A to F). The frequencies of these simulated targets were computed with the MAD oscillation code. This code includes a non-adiabatic non-local time-dependent convection modelling as detailed in \citet{Grigahcene2005}, adapted to the stratification of patched models following the prescriptions of \citet{Dupret2006}. For each target and for the sake of realism, we adopted the observational uncertainty of the frequencies of \emph{LEGACY} targets with similar mode ranges, namely KIC9206432 (model B), KIC10162436 (models C and E), and KIC11081729 (models D and F). For model A, which is a proxy of the Sun, we adopted the uncertainties of \citet{Basu2009} and partially revised by \citet{Davies2014}, degraded by a constant factor to mimic a data quality similar to that of the \emph{Kepler} mission. The classical constraints are the effective temperature, the metallicity, and the absolute luminosity. If the inverted mean density is added to the constraints, it is treated either as another classical constraint or as a seismic constraint to account for the correlation with the ratios. For the effective temperature, we adopted an uncertainty of $90K$ if $T_{\mathrm{eff}} < 6000K$ and $100K$ otherwise, and 0.1 dex for the metallicity. For the luminosity, we adopted 19\% of uncertainty if $L/L_\odot<=3$, 15\% if $3<L/L_\odot<4.5$, and 11\% otherwise, in line with the results from \citet{Silva-Aguirre2017} for the \emph{LEGACY} sample, assuming conservative uncertainties considering the impact of extinction, bolometric correction and uncertainties in the spectral parameters when using Gaia parallaxes. The uncertainties of the effective temperature and of the metallicity are the typical uncertainties recommended for surveys \citep[see e.g.][]{Furlan2018}, and we assumed that if $T_{\mathrm{eff}} < 6000K$, one could expect a slightly better uncertainty. We point out that assuming smaller uncertainties on these quantities would not change the results of our study since the fits are mainly driven by the seismic constraints\footnote{For instance $\sigma_{\nu_{n,l}}/\nu_{n,l}\sim0.01\%$, while $\sigma_{T_{\mathrm{eff}}}/T_{\mathrm{eff}}\sim 1.6\%$.}. Finally, a conservative uncertainty of 0.6\% is assumed for the inverted mean density if it is considered as a classical constraint (see Sec. \ref{sec_mean_density_inversions}).

\subsection{AIMS and convergence assessment}
The AIMS software \citep{Rendle2019} is a MCMC-based algorithm, relying on the \textsc{emcee} package \citep{Foreman-Mackey2013}, an interpolation scheme to sample between the grid points, and a Bayesian approach to provide the posterior probability distributions of the optimised stellar parameters. The coupling of a high-resolution grid with the interpolation scheme allows a very thorough exploration of the parameter space. For the minimisations of this work, we used the \textit{Standard MS} subgrid of \textit{Spelaion} and AIMS therefore included four free variables to optimise (mass, age, and chemical composition with $X_0$ and $Z_0$). We considered uniform uninformative priors for all the free variables, except for the age, for which we used a uniform distribution with the range $[0,13.8]$ Gyr. Regarding the observational constraints, we assumed that the true value of the observations were perturbed by some Gaussian-distributed random noise for the computation of the likelihoods. AIMS accepts two types of constraints, the seismic constraints (individual frequencies, frequency separation ratios, radial frequency of lower order, inverted mean density, etc.) for which all correlations are accounted for to first order, and the classical constraints (stellar radius, absolute luminosity, effective temperature, metallicity, frequency of maximal power $\nu_{max}$, inverted mean density, etc.) for which the correlations with the seismic constraints are neglected. The inverted mean density is an ambivalent constraint as it can be treated either as a classical constraints or as a seismic constraint if the inversion coefficients are provided (see Sec. \ref{sec_mean_density_inversions}). 

By design, a run in AIMS is done in two steps. First, a burn-in phase is computed to identify the relevant part of the parameter space, and then the solution run is performed. By default, AIMS uses 250 walkers, 200 burn-in steps, and 200 steps for the solution. Hence, the stellar parameters are based on \num{50000} samples from the production run, which follows the \num{50000} probability calculations from the burn-in phase. This choice is a compromise between the required computational power and the control on the autocorrelation time. For individual modelling, we would however recommend to modify these default values to 800 walkers, 2000 burn-in steps, and 2000 steps for the solution. In such a configuration, the solution is based on 1.6~million samples from the production run, which follows the 1.6~million probability calculations from the burn-in phase. This ensures that the autocorrelation time is much smaller than the number of steps, at the expense of requiring a higher computational power. In this work, we opted for this new configuration to have a higher degree of confidence in our results, but some tests would be required to find a good compromise in a pipeline. Along with the solution, AIMS provides several diagnostic plots to ensure that the MCMC converged successfully. These plots notably include a triangle plot of the optimized parameters to check that the solution is unique and that the interpolation was smooth, the evolution of the walkers to make sure that they are not drifting, and the échelle diagram \citep{Grec1983}. They allow us to have a good control on the reliability of the MCMC result, but these checks are manual and not pipeline-friendly. In Appendix \ref{appendix_AIMS_convergence}, we provided illustrations of the diagnostic plots for a successful convergence (see Fig. \ref{fig_appendix_successful_convergence}) and for the most common issues that may occur (see Figs. \ref{fig_appendix_drift_walkers} to \ref{fig_appendix_issue_BG1}). We separated the convergence issues into five categories for illustration purposes, but a run can be affected by more than one issue. These categories are described in detail in Appendix \ref{appendix_AIMS_convergence} and we point out here that the most frequent issues occurred when the walkers drifted during the sampling or hit the grid boundaries.

\subsection{Individual frequencies as constraints}
\label{sec_individual_frequencies_as_constraints_takafumi}

\subsubsection{Surface effects}
\label{sec_surface_effects}
Surface effects are related to the poor treatment of near surface layers. Indeed, in these regions, the mixing-length theory (MLT) is an inaccurate description of convection, as it does not account for compressible turbulence for example. The simplistic treatment of convection is especially an issue in asteroseismology, because the perturbation of turbulent pressure can significantly affect the oscillation frequencies. In addition, the thermal timescales in the near surface layers are similar to the oscillation periods and the oscillations are thus highly non-adiabatic there \citep{Houdek&Dupret2015}. Semi-empirical prescriptions were proposed to account for the structural contribution of the surface effects \citep{Kjeldsen2008,Ball&Gizon2014,Sonoi2015}. These prescriptions are described by one or two free parameters that can be added to the optimized variables during the minimisation.

In the following section, we define $\nu_{obs}$ as the observed frequency, and $\nu_{mod}$ the theoretical adiabatic frequency that does not include surface effects.

\citet{Kjeldsen2008} treats the surface effects with a power-law in frequency:
\begin{equation}
\frac{\delta\nu}{\nu_{max}} = a\left(\frac{\nu_{obs}}{\nu_{max}}\right)^b,
\end{equation}
where $a$ and $b$ are parameters to be determined, $\delta\nu = \nu_{obs}-\nu_{mod}$, and $\nu_{max}$ is the frequency of maximal power computed following the scaling relation \citep{KjeldsenBedding&1995}:
\begin{equation}
\frac{\nu_{max}}{\nu_{max,\odot}} = \frac{g}{g_\odot}\left(\frac{T_{\mathrm{eff}}}{T_{\mathrm{eff},\odot}}\right)^{-\frac{1}{2}},
\label{eq_scaling_vmax}
\end{equation}
where $g_\odot\simeq 27420$ cm/s$^2$ \citep{IAU2015-ResolutionB3,CODATA2018}, $T_{\mathrm{eff},\odot}=5777$ K \citep{Allen1976}, and $\nu_{max,\odot}=3090$ $\mu$Hz \citep{Huber2011}. Originally, the parameter $b=4.9$ was determined for the Sun, and the parameter $a$ can then be found with a least square minimisation. \citet{Sonoi2015} showed that $b$ varies significantly with the surface gravity and the effective temperature, and should therefore be determined using the scaling relation:
\begin{equation}
b = -3.16\log T_{\mathrm{eff}} + 0.184\log g + 11.7,
\end{equation}
or treated as an additional free parameter if the prescription is applied to other stars \citep[see e.g. the case of HD52265,][]{Lebreton&Goupil2014}.

\citet{Ball&Gizon2014} proposed two corrections, a one-term and a two-terms, based on the mode inertia. The one-term prescription is:
\begin{align}
\delta\nu = a_{3}\left(\frac{\nu}{\nu_{\mathrm{ac}}}\right)^{3}/\mathcal{I},
\end{align}
and the two-terms prescription is
\begin{align}
\label{eq_BG2}
\delta\nu = \left(a_{-1}\left(\frac{\nu}{\nu_{\mathrm{ac}}}\right)^{-1} + a_{3}\left(\frac{\nu}{\nu_{\mathrm{ac}}}\right)^{3}\right)/\mathcal{I},
\end{align}
where $\mathcal{I}$ is the normalised mode inertia, and $a_{-1}$ and $a_{3}$ are two coefficients to be added in the optimisation procedure. The acoustic cut-off $\nu_{\mathrm{ac}}$ is computed using the scaling relation \eqref{eq_scaling_vmax}, because $\nu_{max}\propto \nu_{\mathrm{ac}}$ as first suggested by \citet{Brown1991}, and we used $\nu_{\mathrm{ac},\odot}=5100$ $\mu$Hz \citep{Jimenez2006}. \citet{Ball&Gizon2014} found that both corrections produced a good fit of the BiSON frequencies \citep{Broomhall2009}, but \citet{Sonoi2015} pointed out that they worked well only in limited frequency ranges of their models, but not in the whole range.

\citet{Sonoi2015} proposed a correction based on patched models, including averaged 3D hydrodynamical models of the upper layer, allowing it to reproduce realistically the frequencies, and based on the frequencies of the corresponding unpatched models. They proposed a correction based on a  Lorentzian function:
\begin{align}
\label{eq_S2}
\frac{\delta\nu}{\nu_{max}} = \alpha\left(1-\frac{1}{1+\left(\frac{\nu_{obs}}{\nu_{max}}\right)^\beta}\right),
\end{align}
where $\alpha$ and $\beta$ can be determined from the surface gravity and effective temperature using scaling relations,
\begin{align}
\log|\alpha| &=\ \ \,  7.69\log T_{\mathrm{eff}} - 0.629\log g - 28.5, \\
\log\beta &= -3.86\log T_{\mathrm{eff}} + 0.235\log g + 14.2,
\end{align}
or be treated as free variables.

These prescriptions were investigated by several works for main-sequence stars \citep{Ball2016,Nsamba2018,Jorgensen2019,Cunha2021} and for more evolved stars \citep{Ball&Gizon2017,Jorgensen2020,Jorgensen2021} using either observational data or synthetic data based on 3D simulations of the surface layers patching 1D models. These works pointed out that the two-terms correction of \citet{Ball&Gizon2014} is in general the most robust prescription, followed by the \citet{Sonoi2015} correction. The \citet{Kjeldsen2008} prescription is less robust and is not recommended in some cases. They also showed that fitting the individual frequencies tends to bias the estimated stellar parameters, especially by overestimating the mass. For post main-sequence stars, these biases are significant as they are comparable to the PLATO precision requirements \citep{Jorgensen2021}. In addition, these works showed that the systematic uncertainty due to the choice of the functional form of the surface effects can be up to twice the statistical uncertainty.

In this work, we considered the prescriptions summarised in Table \ref{tab_surface_prescriptions} when fitting the individual frequencies. We tested them, first with synthetic data whose frequencies were computed using an oscillation code that accounts for non-adiabatic effects\footnote{Models A to F are patched models, non-adiabatic frequencies computed for such models therefore account for the main expected surface effects}, and then with observational data. For the fit of the frequency separation ratios, the surface effects are neglected because the ratios damp them so efficiently that is not possible to estimate them with the MCMC in this configuration. In that case, we account for them in the mean density inversion.

\begin{table}
\centering
\caption{Surface effects prescriptions considered in this work.}
\begin{tabular}{lcc}
\hline 
Prescription & Reference & Parameters \\
\hline 
K1 & Kjeldsen et al. 2008 & $a$ free, $b$ scaled \\ 
K2 & Kjeldsen et al. 2008 & $a$ free, $b$ free \\ 
BG1 & Ball \& Gizon 2014 & $a_{3}$ free \\ 
BG2 & Ball \& Gizon 2014 & $a_{-1}$ free, $a_{3}$ free \\ 
S1 & Sonoi et al. 2015 & $\alpha$ free, $\beta$ scaled \\ 
S2 & Sonoi et al. 2015 & $\alpha$ free, $\beta$ free \\ 
\hline 
\end{tabular} 
\label{tab_surface_prescriptions}
\end{table}

\subsubsection{Application to \citet{Sonoi2015} targets}
\label{sec_takafumi_frequencies}
The direct modelling strategy consists in fitting the individual frequencies and the classical constraints (surface metallicity, effective temperature, and absolute luminosity). Except for model A, the coefficients of the surface effects were poorly estimated, even though the sampling was high (800 walkers, 2000 steps of burn-in, and 2000 steps for the solution). We therefore extended the burn-in to 8000 steps.	This solved the issue for models C, D and F. The runs for model B were still not converging successfully and the stellar parameters of model E were significantly biased. As discussed in more details in Appendix \ref{appendix_additional_data_takafumi}, the impact of the non-adiabatic effects is much stronger for models B to F than for the solar model (model A). If the non-adiabaticity of oscillations is not taken into account in the targets, which corresponds dealing with adiabatic frequencies from the patched 3D simulations, we obtained similar stellar parameters for models A, E and F. The results of models B and D are less accurate and the minimisation failed for model C because the grid boundary was reached. Although it is difficult to draw robust conclusions with a statistics of only six targets, there is a hint with the inaccuracies of models B and D that the non-adiabatic correction may be incompatible with the actual description of surface effects. However, the convergences issues could also be the sign that there is a problem with the structure of the targets, for example with the determination of the position of the connection between the 1D structure and the 3D model of the upper layers. To help the convergence of the MCMC, we discarded the modes above 2400 $\mu$Hz for model B and 1500 $\mu$Hz for model E. This worked for model B, but the stellar parameters of model E were not improved. As we will argue later with additional tests, this disagreement likely originates from the structure of model E and not from the surface effects prescription or from the non-adiabatic correction.

\begin{figure*}[htp!]
\centering
\begin{subfigure}[b]{.43\textwidth}
  \includegraphics[width=.99\textwidth]{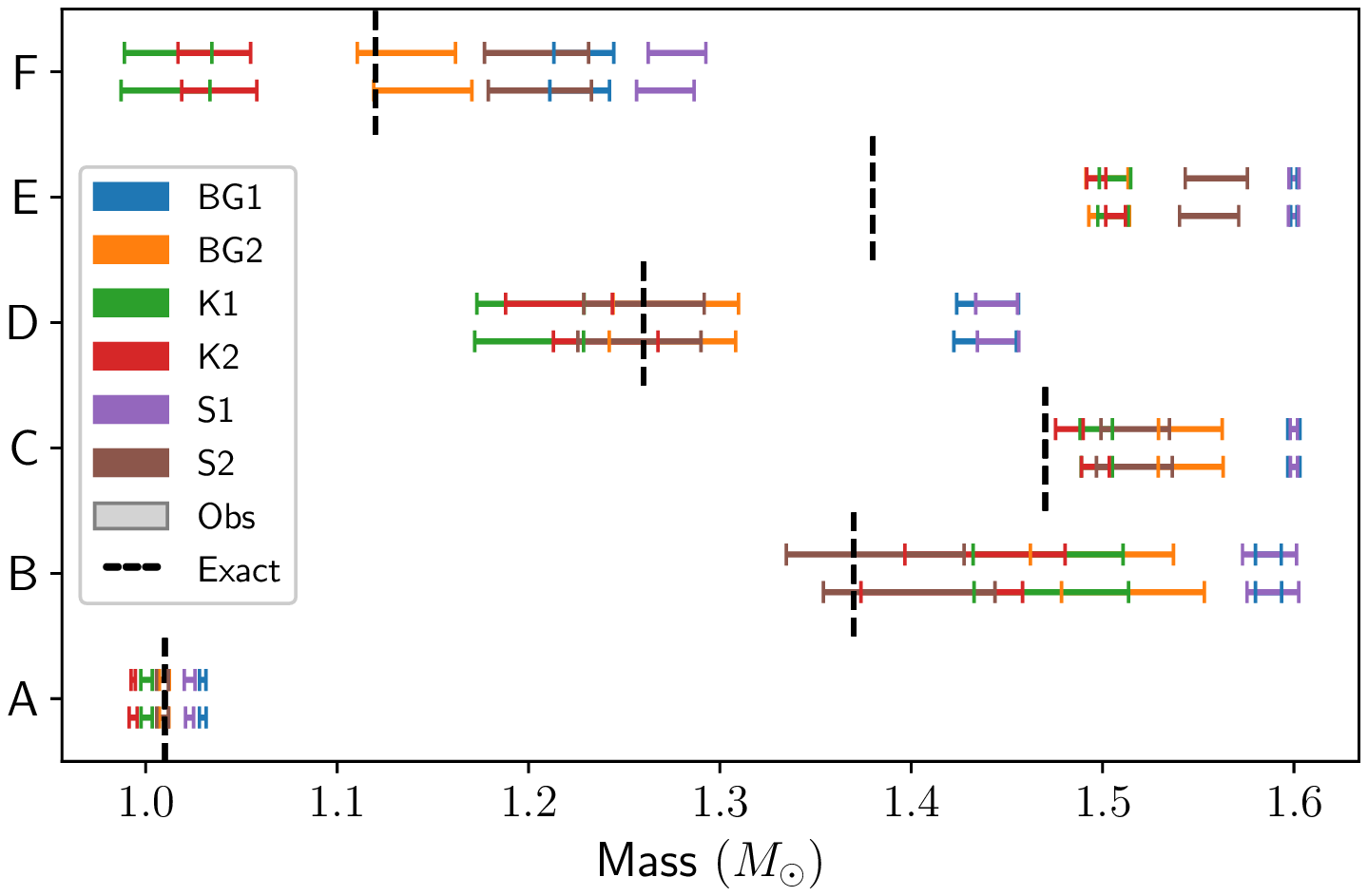}  
  \label{fig_Takafumi_frequencies_naive_mass}
\end{subfigure}
\begin{subfigure}[b]{.43\textwidth}
  \includegraphics[width=.99\textwidth]{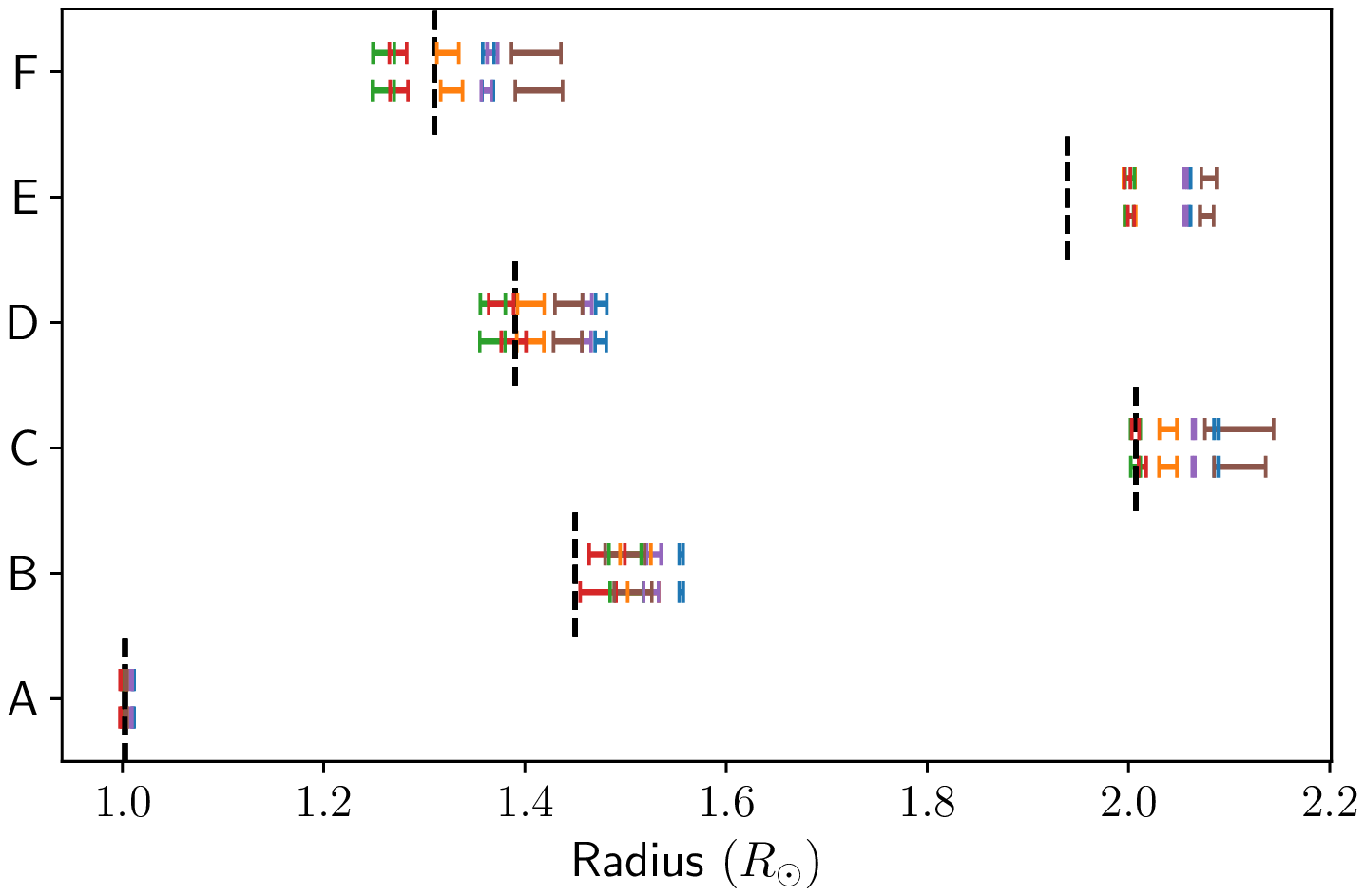}  
  \label{fig_Takafumi_frequencies_naive_radius}
\end{subfigure}
\begin{subfigure}[b]{.43\textwidth}
  \includegraphics[width=.99\textwidth]{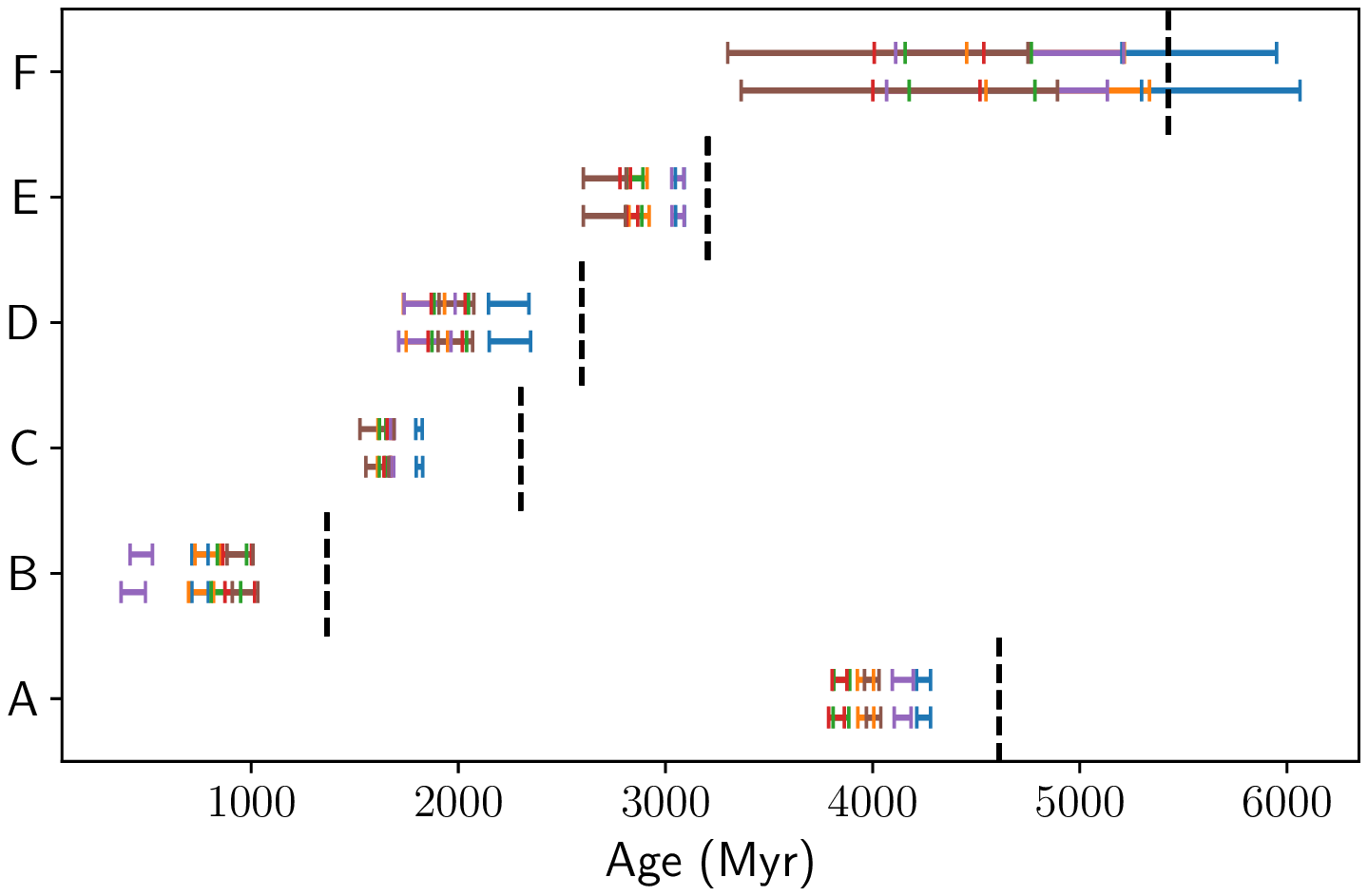}  
  \label{fig_Takafumi_frequencies_naive_age}
\end{subfigure}
\begin{subfigure}[b]{.43\textwidth}
  \includegraphics[width=.99\textwidth]{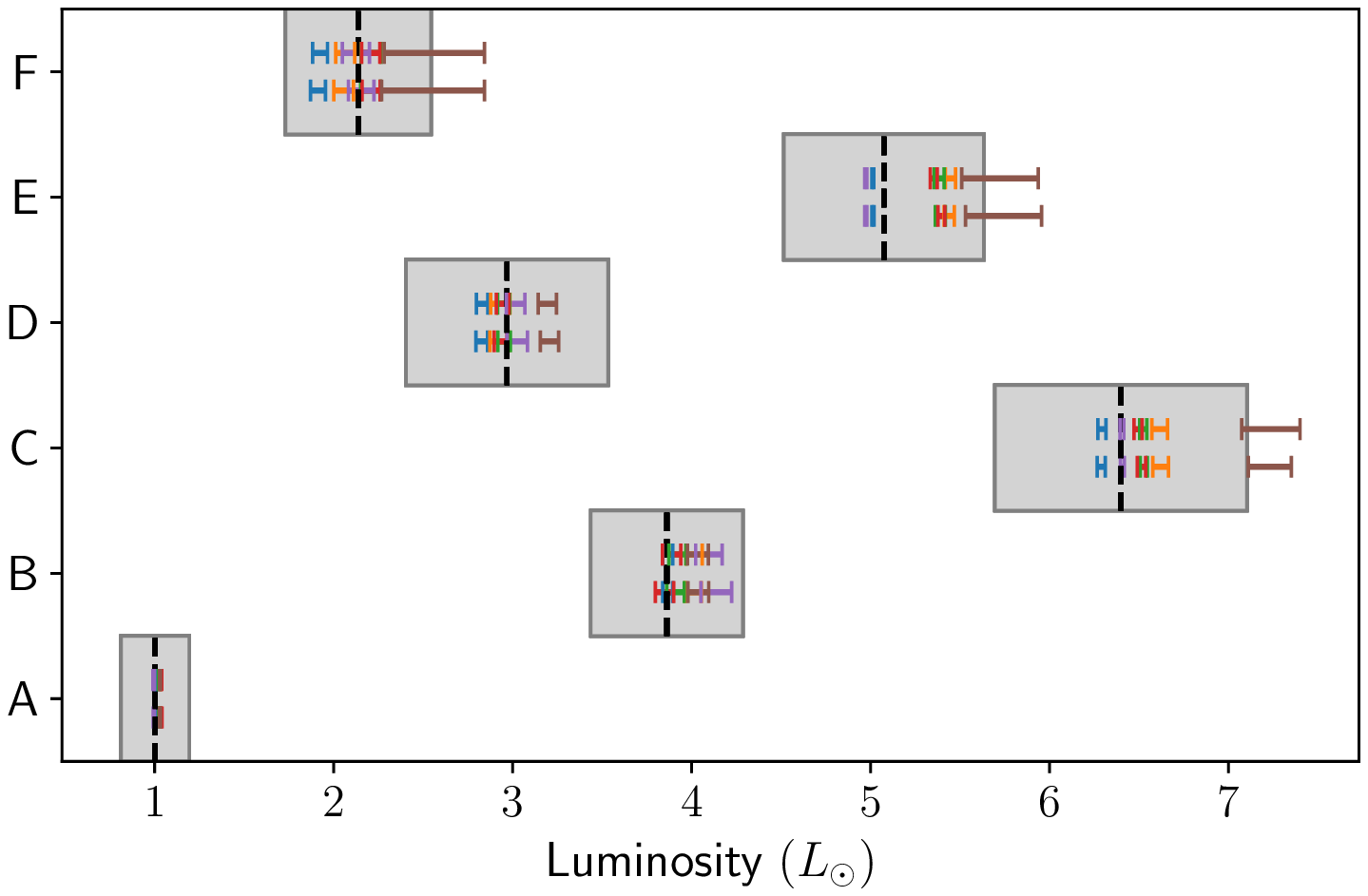}  
  \label{fig_Takafumi_frequencies_naive_luminosity}
\end{subfigure}
\caption{MCMC results for the \citet{Sonoi2015} targets, using individual frequencies and different prescriptions of the surface effects described in Table \ref{tab_surface_prescriptions}. Runs with convergence issues are included. For each target, two sets of classical constraints were considered, including the absolute luminosity (upper line) or excluding it (bottom line). The dashed black lines represent the exact value and the grey boxes represent the observational constraints.}
\label{fig_Takafumi_frequencies_naive_full}
\end{figure*}

\begin{figure*}[htp!]
\centering
\begin{subfigure}[b]{.43\textwidth}
  \includegraphics[width=.99\textwidth]{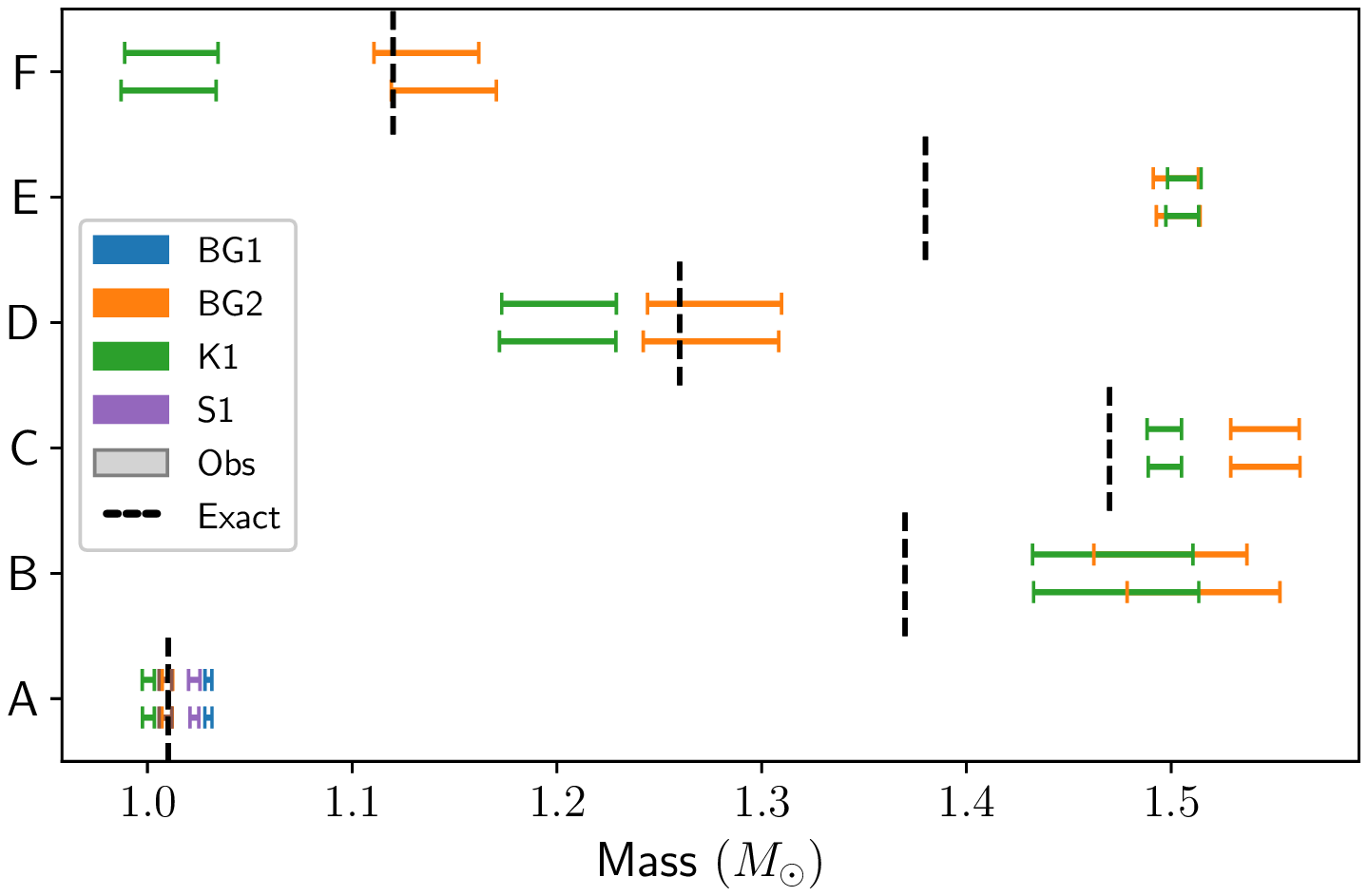}  
  \label{fig_Takafumi_frequencies_clever_mass}
\end{subfigure}
\begin{subfigure}[b]{.43\textwidth}
  \includegraphics[width=.99\textwidth]{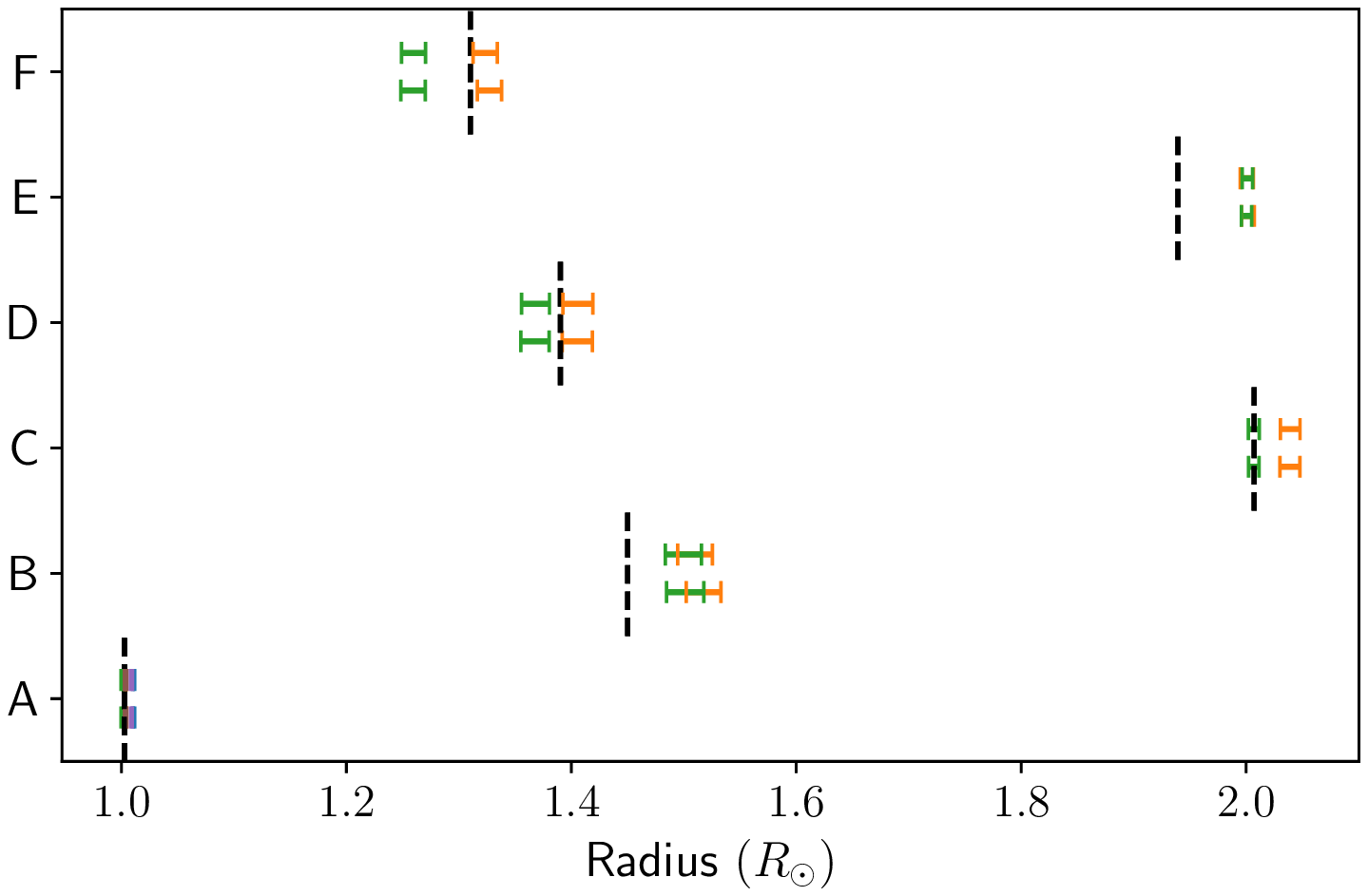}  
  \label{fig_Takafumi_frequencies_clever_radius}
\end{subfigure}
\begin{subfigure}[b]{.43\textwidth}
  \includegraphics[width=.99\textwidth]{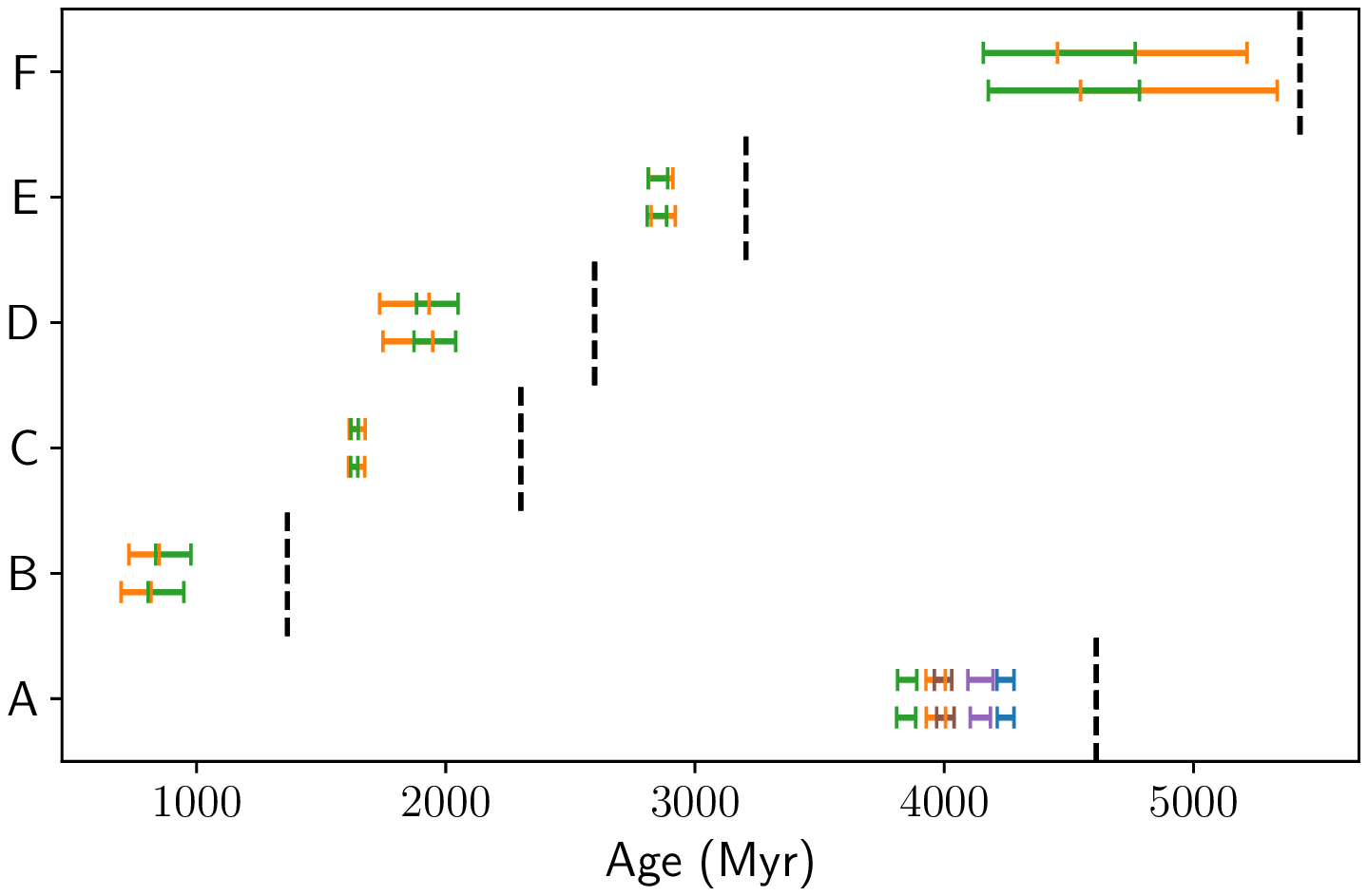}  
  \label{fig_Takafumi_frequencies_clever_age}
\end{subfigure}
\begin{subfigure}[b]{.43\textwidth}
  \includegraphics[width=.99\textwidth]{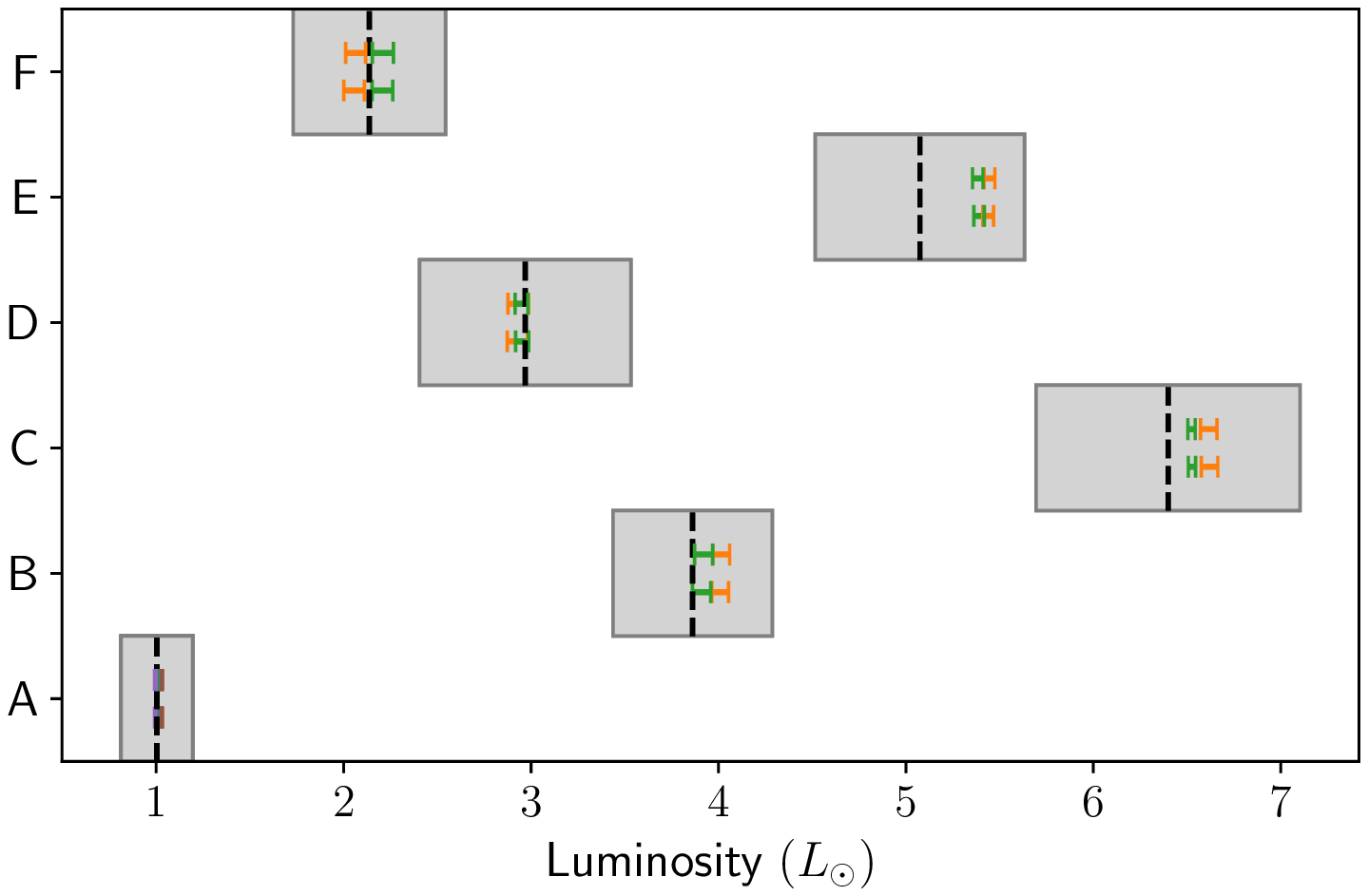}  
  \label{fig_Takafumi_frequencies_clever_luminosity}
\end{subfigure}
\caption{Same as Fig. \ref{fig_Takafumi_frequencies_naive_full} except that runs with convergence issues were discarded.\\}
\label{fig_Takafumi_frequencies_clever_full}
\end{figure*}

In Figs. \ref{fig_Takafumi_frequencies_naive_full} and \ref{fig_Takafumi_frequencies_clever_full}, we show the results of the fit of the individual frequencies before and after manually discarding the runs with issues. Except for the solar model (model A), only the \textit{BG2} and \textit{K1} prescriptions produced runs without issues. Surprisingly, for the unsuccessful runs, the optimized stellar parameters were not significantly biased as we expected. Although it sounds like an advantage, the spread due to surface effects is much larger than the individual uncertainties, as illustrated in Fig. \ref{fig_Takafumi_frequencies_naive_full}, and is therefore incompatible with the PLATO precision requirements. From a pipeline perspective, some of the issues, such as histograms truncated at the grid boundaries, which was the main issue, can be automatically identified with a high level of confidence. However, other issues, such as excessively peaked distribution or walker drifts are harder to identify automatically. This kind of problem is however well suited for machine learning even though it would be difficult to build a robust and comprehensive training set.

In Figs. \ref{fig_Takafumi_frequencies_naive_full} and \ref{fig_Takafumi_frequencies_clever_full}, we tested the impact of the luminosity by including or excluding it in the classical constraints, and verifying that both results are consistent. This test is not mandatory for synthetic models whose absolute luminosity is known exactly (and therefore reliably), but can point out issues with the bolometric corrections or the extinction maps when computing the luminosity of an observational target. As expected, all the models in this section reproduce consistently the luminosity.


\subsection{Frequency separation ratios as constraints}
\label{sec_frequency_ratios_as_constraints_takafumi}

\subsubsection{Mean density inversions}
\label{sec_mean_density_inversions}
In this section, we present a three-step procedure, coupling fits of frequency separations ratios and a mean density inversion, to circumvent the issues due to surface effects. We recall that the ratios are constructed by dividing the small separation by the large separation, therefore suppressing the information about the mean density. Our methodology uses a mean density inversion to recover the lost information in a quasi-model-independent way. We point out that this approach can provide stellar parameters of a PLATO benchmark target whose precision meets the PLATO requirements \citep{Betrisey2022}. The procedure starts by fitting the individual frequencies and the classical constraints as in Sec. \ref{sec_individual_frequencies_as_constraints_takafumi}. Then, a mean density inversion is conducted on the resulting model of this first minimisation. This allows us to constrain the mean density in a quasi-model-independent way and add it to the constraints. If the inverted mean density is treated as a classical constraint in AIMS, because no detailed analysis is conducted at this stage, a conservative uncertainty of 0.6\% is adopted on that quantity. Then, a second minimisation is conducted, fitting this time frequency separation ratios ($r_{01}$ and $r_{02}$), the classical constraints, and the inverted mean density. The $r_{10}$ ratios can be used instead of the $r_{01}$ but they should not be used simultaneously as it will bias the results \citep{Roxburgh2018}. We recall that the surface effects are accounted for in the mean density inversion and are neglected in the fit of the ratios with AIMS. 

\begin{figure}[t!]
\centering
\includegraphics[scale=0.52]{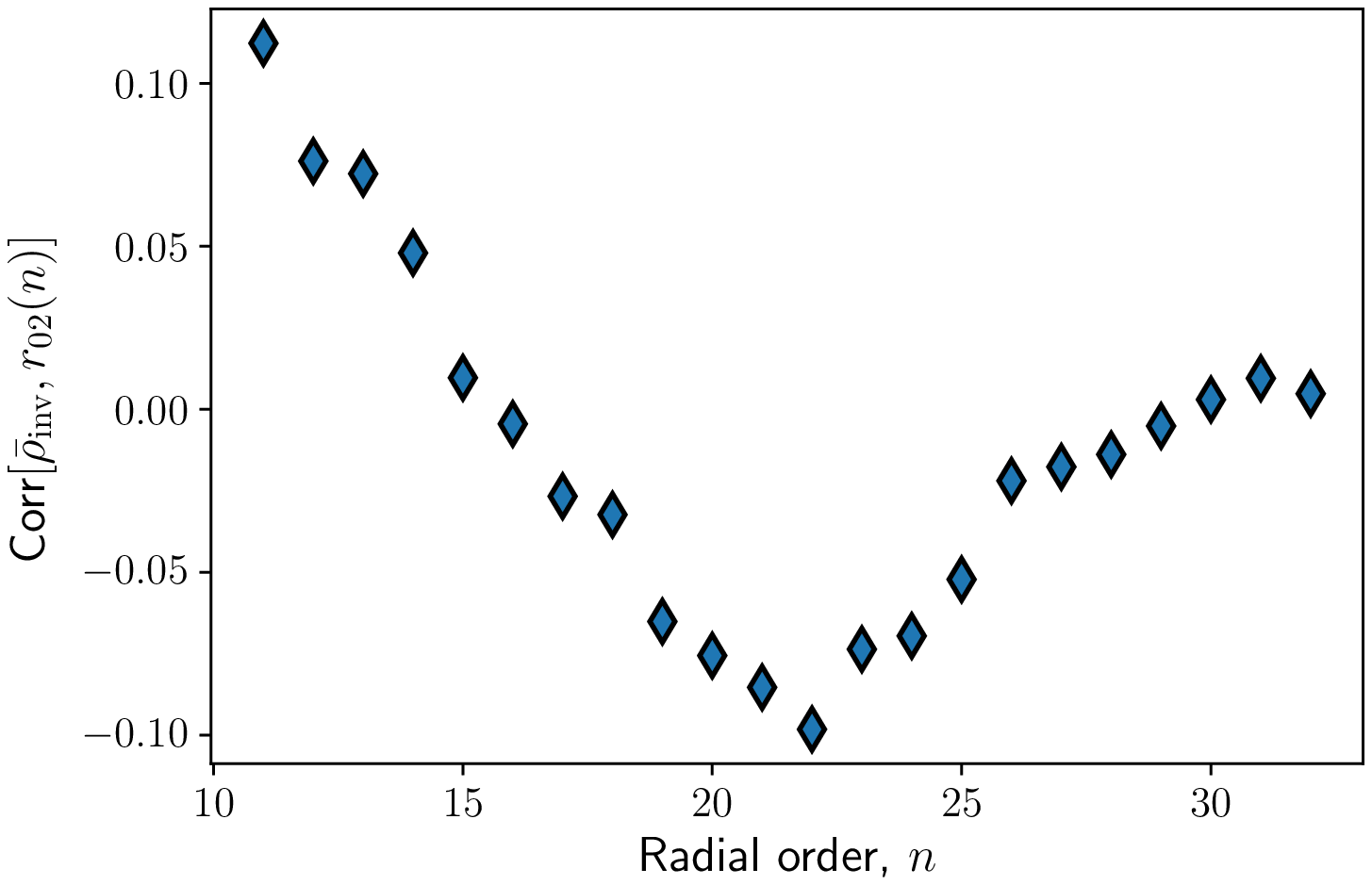}
\caption{Correlations between the inverted mean density and the $r_{02}$ ratios for our toy model, using Model S from \citet{JCD1996} and observational data from \citet{Lazrek1997}.}
\label{fig_correlations_toy_model}
\end{figure}

\begin{figure}[t!]
\centering
\includegraphics[scale=0.57]{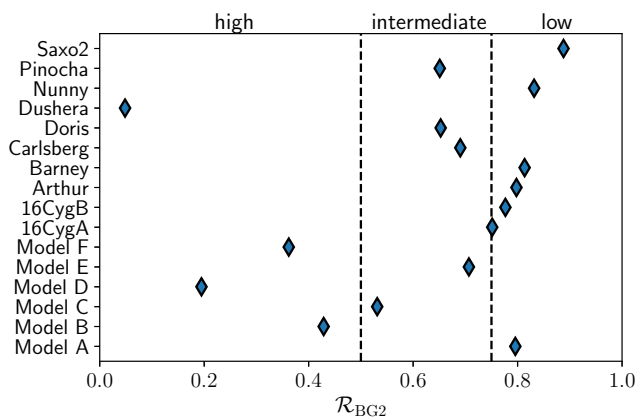} 
\caption{Estimates of the degree of instability in the mean density inversion of the targets considered in this study. The coefficients of the surface correction were estimated by the inversion using the \textit{BG2} prescription. Targets in the high instability regime would require further investigation, while inversion results in the low and intermediate regimes can be used without further investigation.}
\label{fig_instability_assessment}
\end{figure}

The inverted mean density is a combination of frequencies and it is therefore possible to treat it as a seismic constraint to account for the correlations with the other seismic constraints. We computed the inverted mean density using the generalised definition of \citet{Reese2012}:
\begin{align}
\label{eq_s_exact}
\bar{\rho}_{\mathrm{inv}} = \bar{\rho}_{\mathrm{ref}}s^2 \quad \mathrm{with}\quad s=\frac{1}{2}\sum_i c_i \frac{\nu_i^{\mathrm{obs}}}{\nu_i^{\mathrm{ref}}},
\end{align}
where $c_i$ are the inversion coefficients that are optimized by the inversion based on the frequency differences between the reference model (`ref') and the observations (`obs'). The index $i$ denotes the identification pair $(n,l)$ of the corresponding frequency. The inverted mean density is therefore correlated with the frequency separation ratios, as shown in Fig. \ref{fig_correlations_toy_model}, using Model S from \citet{JCD1996} and observational data from \citet{Lazrek1997}. We implemented these correlations in AIMS with two subtleties. First, the inversion coefficients should be updated at each iteration of the MCMC. This would however require an interpolation of the model structure at each step, which is numerically expensive and beyond the actual capabilities of AIMS. Because the variation of the inversion coefficients between close models is small (see Appendix \ref{appendix_inversion_coefficients}), we neglected this effect. We assumed constant coefficients, determined by the original mean density inversion. Second, with the actual definition of $s$, the covariance matrix needs to be updated each time the likelihood is updated as well, which is also numerically inefficient. We therefore modified the definition of $s$ by switching the reference frequency with the observed frequency:
\begin{align}
\label{eq_s_approx}
s' = \frac{1}{2}\sum_i c_i \frac{\nu_i^{\mathrm{ref}}}{\nu_i^{\mathrm{obs}}}.
\end{align}
This switch is only valid if $\nu_i^{\mathrm{obs}}/\nu_i^{\mathrm{ref}} \simeq 1$, which happens in the limit $s^2\rightarrow 1$. This amounts to swapping the roles of the observed star and reference model by looking for what correction is needed for the former to reproduce the mean density of the latter. If it is close to 1, then the two mean densities are close to each other. This approximation allows us to compute the covariance matrix only once at the beginning of the minimisation, which is numerically much more efficient. In addition, the domain of validity of the approximation is well verified because the minimisation converges toward the region where $s^2\rightarrow 1$. For completeness, we computed in Appendix \ref{appendix_inversion_coefficients} the correlations by implementing both definitions in AIMS for the toy model. The actual differences are very small, and are negligible compared to other sources of uncertainty. We note that this implementation has one drawback. It imposes that $s^2\rightarrow 1$ but not that $\bar{\rho}_{\mathrm{ref}}\rightarrow \bar{\rho}_{\mathrm{inv}}$. Depending on the treatment of the surface effects by the inversion, the first condition may imply the second or not. If not, the optimal model may converge toward a wrong mean density while still fulfilling the first condition, or simply not converge. To understand further when issues may occur, it is worth recalling what the inversion is doing. It minimises the following cost function:
\begin{align}
\mathcal{J}_{\bar{\rho}}(c_i) &= \mathcal{F}_{\mathrm{Struc}} + \mathcal{F}_{\mathrm{Uncert}} + \mathcal{F}_{\mathrm{Surf}},
\end{align}
where $\mathcal{F}_{\mathrm{Struc}}$ accounts for the structural differences, $\mathcal{F}_{\mathrm{Uncert}}$ accounts for the observational uncertainties, and $\mathcal{F}_{\mathrm{Surf}}$ accounts for the surface effects. The inversion therefore does a balance between the extraction of structural differences, in our case to provide a correction for the mean density of the reference model, while minimising the observational uncertainties and accounting for surface effects. While the first two terms are well understood \citep[see][]{Reese2012}, $\mathcal{F}_{\mathrm{Surf}}$ is semi-empirical and in practice, it introduces instability in the inversion because it adds two free variables in the minimisation. The degree of instability depends on the strength of the surface effects and in the worst scenario, all the information from the relative frequency differences can be used in the estimation of the surface effects and no information is left for extraction of the structural differences. In that case, the inversion coefficients are poorly estimated, resulting in coefficients with high amplitudes and large variations between two consecutive coefficients. Under such conditions, we say that the inversion is unstable, and this instability is then propagated in AIMS, causing the convergence issue that we mentioned earlier. Although there exists some techniques to check the quality of an inversion \citep[see][or Appendix \ref{appendix_inversion_coefficients}]{Reese2012,Buldgen2015b}, they either require manual checks or a knowledge of the structure of the observed target, and are thus difficult or impossible to adapt in a pipeline. We therefore developed a new test to quantify the quality of an inversion. This test consists in evaluating the Pearson correlation coefficient of the lag plot (with $\mathrm{lag}=1$) of the inversion coefficients. The Pearson correlation coefficient is defined as the covariance of two random variables divided by the product of their standard deviations. For a sample pair $(\vec{x},\vec{y})$,
\begin{align}
\mathcal{R} = \frac{\sum_{i=1}^N (x_i-\bar{x})(y_i-\bar{y})}{\sqrt{\sum_{i=1}^N (x_i-\bar{x})^2}\sqrt{\sum_{i=1}^N (y_i-\bar{y})^2}},
\end{align}
where $\vec{x}=[x_1,...,x_N]$,  $\vec{y}=[y_1,...,y_N]$, and $\bar{x}$ and $\bar{y}$ are the mean of the vectors $\vec{x}$ and $\vec{y}$ respectively. For the sake of conciseness, we do not describe lag in detail here but invite the reader to look at the book of \citet{NIST_lag} and at Appendix \ref{appendix_inversion_coefficients}, where we provided illustrations of lag plots of targets in different instability regimes, additional tests, and a more complete discussion about the regime boundaries. To summarise, we identified three instability regimes: high ($\mathcal{R}<0.5$), intermediate ($0.5<\mathcal{R}<0.75$), and low ($0.75<\mathcal{R}$). If a target is in the intermediate or low instability regimes, we consider that the mean density inversion can be trusted without further investigations. If a target is in the high instability regime, the result of the inversion should be treated with caution. We remark that we identified three regimes, but in a pipeline it would be better to define a unique threshold below which we reject the inversion. Based on the statistics of this work, we would estimate this threshold to be around $\mathcal{R} \sim 0.6$, but it would benefit from further investigations with a larger statistics. In Fig. \ref{fig_instability_assessment}, we show the $\mathcal{R}$ coefficient of the targets considered in this work. Half of the \citet{Sonoi2015} targets lie in the high instablity regime, due to the issues mentioned in Sec. \ref{sec_takafumi_frequencies} (see also Appendix \ref{appendix_additional_data_takafumi}), while only one of the ten \emph{LEGACY} targets is in the high instability regime.

As a side note, we remark that by adding the inverted mean density in the constraints, we re-introduce some uncertainty due to the surface effects, but they are only affecting one constraint with this approach.

\subsubsection{Application to \citet{Sonoi2015} sample}
\label{sec_takafumi_ratios}
By fitting the frequency separation ratios and the classical constraints (metallicity, effective temperature, and luminosity), the relative separations between the frequency ridges can be reproduced but in general not their position, resulting in a horizontal shift in the échelle diagram. The addition of the inverted mean density mitigates this issue but may be insufficient, as was the case for models B and E (see Fig. \ref{fig_ModelD_anchoring}). In that case, we considered an additional seismic constraint, the lowest order radial frequency, because it is the frequency that is the least affected by the surface effects. This addition fixes the position of the ridges but can bring (or emphasise) other minimisation issues, notably a drift of walkers that biases the results (see end of Sec. \ref{sec_frequency_ratios_as_constraints_LEGACY} for further details). Such issues did not occur with the models of this section.

We tested three prescriptions for the inclusion of the inverted mean density in the constraints: not including it, including it as a classical constraint, or including it as a seismic constraint. By classical constraints, we imply that the likelihoods were computed assuming the true value of the observations were perturbed by some Gaussian-distributed random noise (and note that other distributions are also supported by the software), while AIMS accounts for all the correlations for a seismic constraint. A comparison of the three prescriptions is shown in Table \ref{tab_mean_density_ratios_Takafumi}. If the inverted mean density is added to the constraints, the precision of the stellar mass and radius is significantly improved. The precision of the stellar age is mostly dominated by the seismic information contained in the ratios, and a gain in precision is likely to be an indirect consequence of a gain in the precision of the stellar mass and radius. The precision of the stellar parameters by considering the inverted mean density as a classical or seismic constraint is roughly equivalent, but the sources of uncertainties are different. In the former case, we assumed an arbitrary uncertainty of 0.6\% for the inverted mean density, that accounts for the statistical uncertainties ($\sim 0.1-0.2\%$), as well as for the systematic uncertainties due to the choice of the physical ingredients or the prescription for surface effects. Although these effects are difficult to estimate without an individual and detailed analysis of each target, they are unlikely to exceed 0.6\%, which is considered as a very large uncertainty for an inversion. As a reference, for Kepler-93 which is a well-behaved target with moderate surface effects, the total uncertainty of the mean density was 0.2\% \citep{Betrisey2022}. Since this arbitrary choice affects the maximal precision achievable for the stellar parameters, a detailed analysis of several benchmark targets could be relevant to refine this choice in certain mass ranges or types of chemical composition for example. Conversely, if the inverted mean density is treated as a seismic constraint, we can account for the correlations with the ratios, but not for the systematics. As shown in Fig. \ref{fig_Takafumi_VS_full} in orange and green, both prescriptions have an equivalent accuracy. Since both prescriptions lead to a similar precision and accuracy, we would recommend treating the mean density as a classical constraint because it is more stable.

\begin{table*}[htp!]
\centering
\caption{Precision of the stellar parameters obtained by fitting the frequency separation ratios for the \citet{Sonoi2015} models.}
\resizebox{\linewidth}{!}{
\begin{tabular}{lccc|ccc|ccc}
\hline
List of & \multicolumn{3}{c}{Stellar mass} & \multicolumn{3}{c}{Stellar radius} & \multicolumn{3}{c}{Stellar age} \\ 
Constraints & no $\bar{\rho}_{\mathrm{inv}}$ & no corr. & incl. corr. & no $\bar{\rho}_{\mathrm{inv}}$ & no corr. & incl. corr. & no $\bar{\rho}_{\mathrm{inv}}$ & no corr. & incl. corr. \\ 
\hline \hline
\textit{Model A} \\
$[Fe/H],T_{\mathrm{eff}},L,r_{01},r_{02},\nu_{n_{min},l=0}$ & 0.8\% & 0.8\% & 0.7\% & 0.3\% & 0.3\% & 0.3\% & 1.9\% & 1.9\% & 1.8\% \\ 
$[Fe/H],T_{\mathrm{eff}},L,r_{01},r_{02}$ & 3.7\% & 0.9\% & 0.8\% & 2.6\% & 0.4\% & 0.3\% & 2.4\% & 2.0\% & 1.9\% \\ 
$[Fe/H],T_{\mathrm{eff}},r_{01},r_{02}$ & 3.7\% & 0.9\% & 0.8\% & 2.7\% & 0.4\% & 0.3\% & 2.5\% & 2.0\% & 1.9\% \\ 
$[Fe/H],L,r_{01},r_{02}$ & 4.1\% & 0.9\% & 0.8\% & 2.9\% & 0.4\% & 0.3\% & 2.5\% & 2.0\% & 2.0\% \\ 
$T_{\mathrm{eff}},L,r_{01},r_{02}$ & 3.6\% & 0.9\% & 0.8\% & 2.6\% & 0.4\% & 0.3\% & 2.5\% & 1.9\% & 1.8\% \\ 
$[Fe/H],T_{\mathrm{eff}},L,R_{01},R_{02}$ & 4.2\% & 1.4\% & 1.4\% & 3.2\% & 0.6\% & 0.5\% & 4.2\% & 3.3\% & 3.3\% \\ 
$[Fe/H],T_{\mathrm{eff}},L,r_{02}$ & 3.9\% & 2.0\% & 2.2\% & 2.8\% & 0.7\% & 0.8\% & 3.7\% & 2.7\% & 2.2\% \\ 
$[Fe/H],T_{\mathrm{eff}},L,r_{01}$ & 4.7\% & 1.7\% & 1.7\% & 4.1\% & 0.6\% & 0.6\% & 4.0\% & 4.1\% & 4.0\% \\ 
\hline 
\textit{Model B} \\
$[Fe/H],T_{\mathrm{eff}},L,r_{01},r_{02},\nu_{n_{min},l=0}$ & 3.6\% & 3.8\% & 3.6\% & 1.3\% & 1.2\% & 1.3\% & 21.4\% & 18.0\% & 18.8\% \\ 
\hline 
\textit{Model C} \\
$[Fe/H],T_{\mathrm{eff}},L,r_{01},r_{02}$ & 4.3\% & 3.1\% & 2.8\% & 4.3\% & 1.1\% & 1.1\% & 10.5\% & 5.3\% & 4.9\% \\
\hline 
\textit{Model D} \\
$[Fe/H],T_{\mathrm{eff}},L,r_{01},r_{02}$ & 6.7\% & 4.1\% & 4.0\% & 6.5\% & 1.4\% & 1.4\% & 17.4\% & 16.8\% & 16.3\% \\ 
\hline 
\textit{Model E} \\
$[Fe/H],T_{\mathrm{eff}},L,r_{01},r_{02},\nu_{n_{min},l=0}$ & 2.1\% & 2.1\% & 2.2\% & 0.9\% & 0.8\% & 0.8\% & 4.8\% & 5.0\% & 4.0\% \\ 
\hline 
\textit{Model F} \\
$[Fe/H],T_{\mathrm{eff}},L,r_{01},r_{02}$ & 6.5\% & 3.1\% & 3.2\% & 5.1\% & 1.1\% & 1.1\% & 14.8\% & 12.2\% & 11.9\% \\
\hline 
\end{tabular} 
}
{\par\small\justify\textbf{Notes.} We considered three prescriptions for the inclusion of the inverted mean density in the constraints: not including it (no $\bar{\rho}_{\mathrm{inv}}$), including it as a classical constraint (no corr.), or including it as a seismic constraint to account for the correlations with the ratios (incl. corr.). The lowest order radial frequency is denoted by $\nu_{n_{min},l=0}$, and $R_{01}$ and $R_{02}$ are the frequency separation ratios derived by assuming observational uncertainties twice as large on the individual frequencies. \par}
\label{tab_mean_density_ratios_Takafumi}
\end{table*}

\subsection{Comparison and discussion}
\label{sec_comparison_takafumi}
\begin{figure*}[htp!]
\centering
\begin{subfigure}[b]{.45\textwidth}
  \includegraphics[width=.99\textwidth]{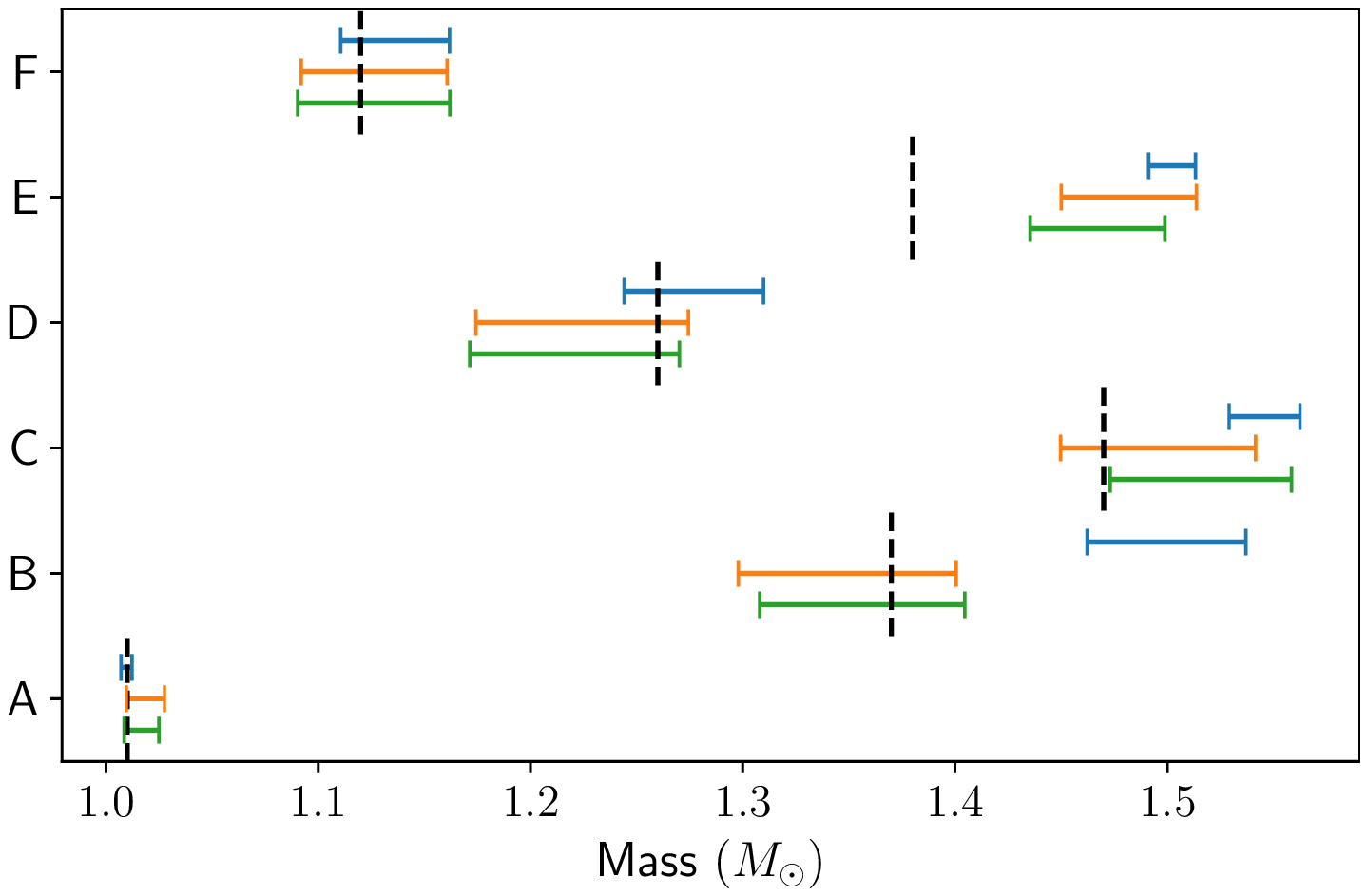}  
  \label{fig_Takafumi_VS_mass}
\end{subfigure}
\begin{subfigure}[b]{.45\textwidth}
  \includegraphics[width=.99\textwidth]{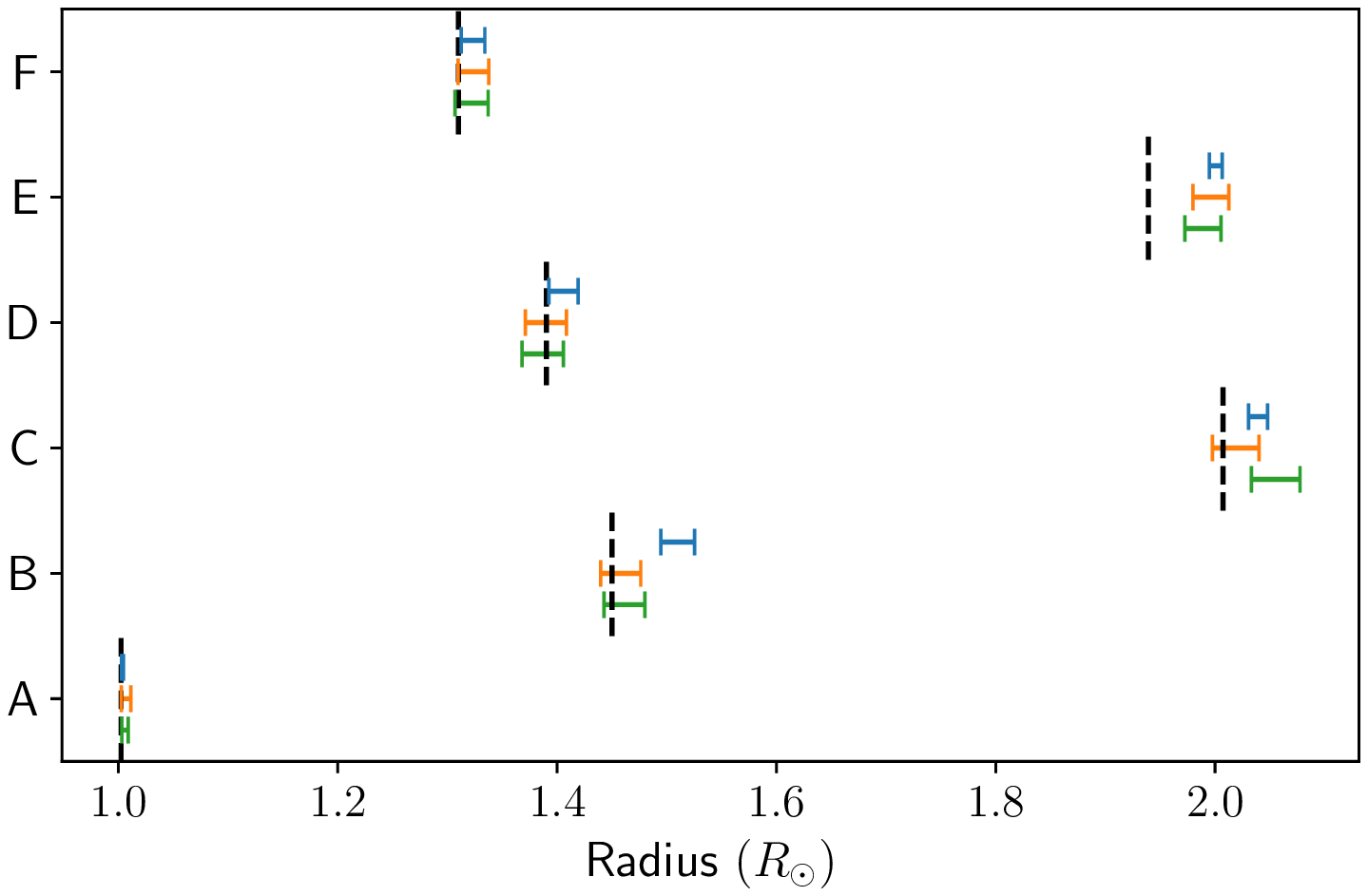}  
  \label{fig_Takafumi_VS_radius}
\end{subfigure}
\begin{subfigure}[b]{.45\textwidth}
  \includegraphics[width=.99\textwidth]{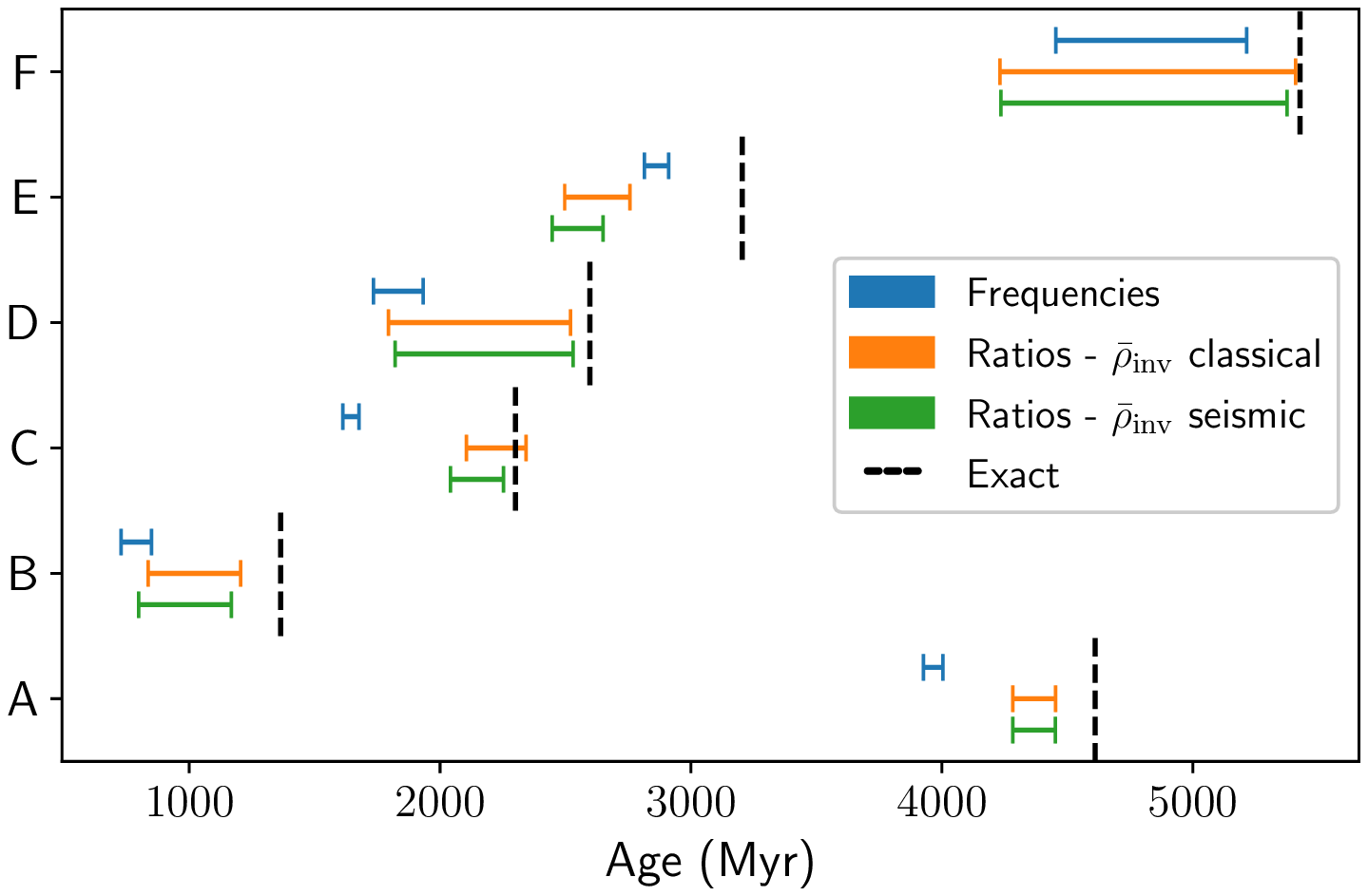}  
  \label{fig_Takafumi_VS_age}
\end{subfigure}
\begin{subfigure}[b]{.45\textwidth}
  \includegraphics[width=.99\textwidth]{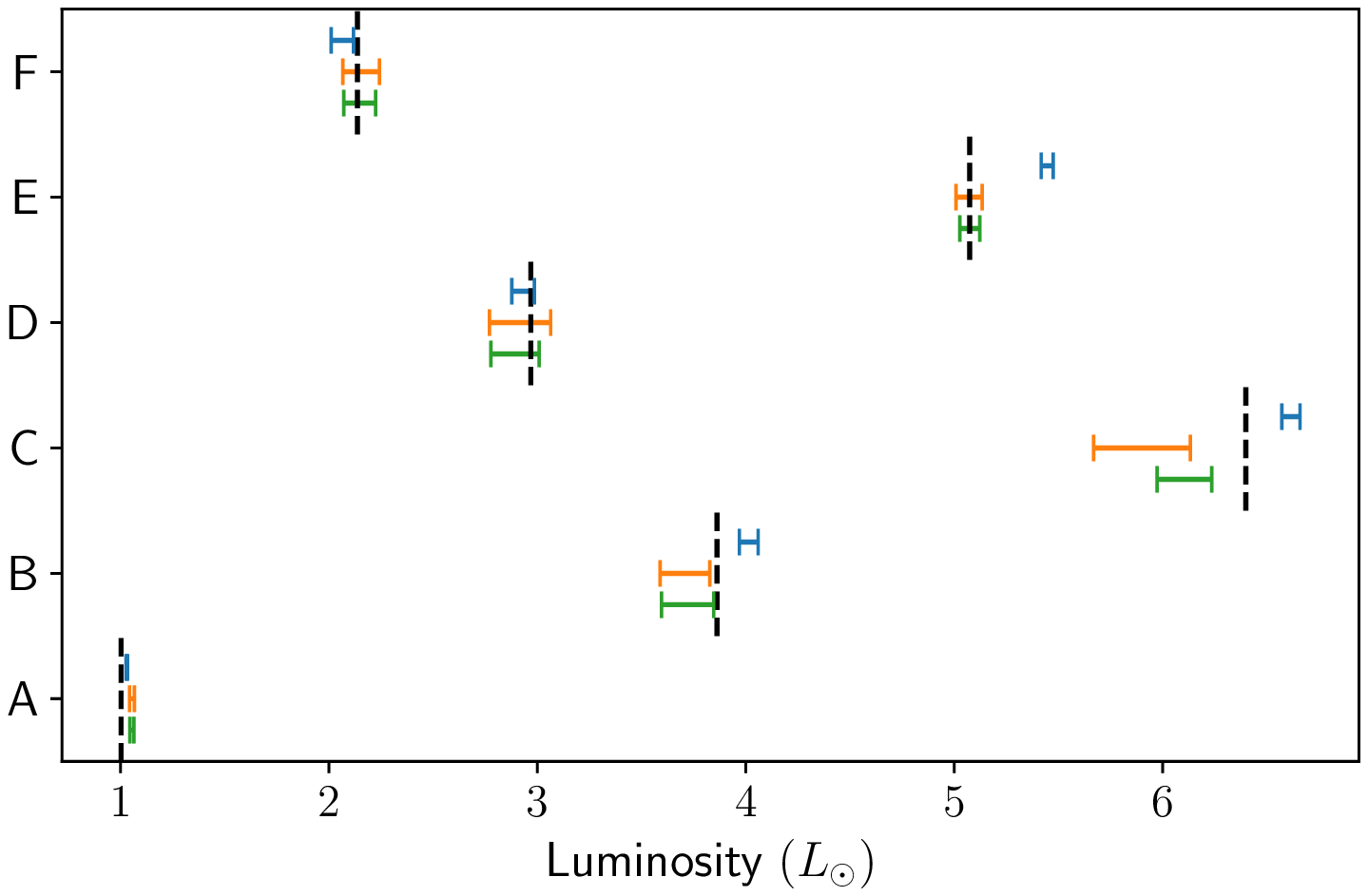}  
  \label{fig_Takafumi_VS_luminosity}
\end{subfigure}
\caption{Accuracy comparison between the results of the modelling strategies fitting the individual frequencies (blue), or the frequency separations ratios by treating the inverted mean density as a classical (orange) or seismic (green) constraint, for the \citet{Sonoi2015} targets. For model A, we show the results of the model using the following constraints: $[Fe/H],T_{\mathrm{eff}},L,r_{01},r_{02}$, and for models B-F, we used the constraints listed in Table \ref{tab_mean_density_ratios_Takafumi}.}
\label{fig_Takafumi_VS_full}
\end{figure*}

In Fig. \ref{fig_Takafumi_VS_full}, we compare the results between the modelling strategy fitting the individual frequencies, and the one fitting the frequency separation ratios and the inverted mean density. For the fit of the individual frequencies (in blue), we selected the models with the \textit{BG2} prescription for the surface effects and the absolute luminosity in the classical constraints, because it provided the most robust models, and for the ratios, we selected the results based on the inverted mean density treated as a classical constraint (in orange) or as a seismic constraint (in green). The stellar parameters are systematically biased with the fit of the individual frequencies. The mass and radius are overestimated and as a consequence, the age is underestimated. These biases are related to the treatment of the surface effects that is too simplistic to model accurately the complex processes taking place in upper stellar layers. Such biases were expected as they are already documented in the literature for other types of stars \citep{Ball&Gizon2017,Nsamba2018,Jorgensen2020,Jorgensen2021,Cunha2021}. In addition, the fit of the frequencies has another issue, also observed in  \citet{Rendle2019}, \citet{Buldgen2019f}, and \citet{Betrisey2022}. The uncertainty is underestimated because the frequencies constitutes a set of constraints that contains too many precise elements, which results in peaked distributions. The fit of the individual frequencies therefore tended to estimate precise but inaccurate stellar parameters. In contrast, the fit of the frequency separation ratios, which damp the surface effects, provided more accurate results. Except for model E, the stellar mass and radius are indeed consistently reproduced. We note some slight inaccuracies in the stellar ages that are likely related to the differences in the physical ingredients between the \citet{Sonoi2015} targets and our grid of models. Especially, the abundances are different, as well as the value assumed for the mixing-length parameter that is fixed at a solar calibrated value of 2.05 in our grid. Because the MCMC cannot modify this parameter, it compensates by modifying the helium mass fraction and the metallicity, resulting in a bias in the stellar age and absolute luminosity. Although it is tempting to let $\alpha_{\mathrm{MLT}}$ be an additional free parameter to avoid this kind of issue, it would be numerically extremely expensive, especially if the overshooting is also free. Regarding model E, none of the models of this work were able to reproduce its stellar parameters. No improvements were observed by removing the non-adiabatic correction of the frequencies. This raises the question whether there is a structural issue with model E, either in the 1D structure of the model itself, in the 3D simulation of the upper layers, or in the connection between the two, or whether the semi-empirical formalism of the surface effects is not suited for this target.

\section{Application to \emph{LEGACY} targets}
\label{sec_applications_LEGACY}
In this section, we applied the two modelling strategies of Sec. \ref{sec_modelling_strategies} on ten targets from the \emph{Kepler LEGACY} sample \citep{Lund2017}, with the best data quality. We divided these targets in two categories differing by the set of classical constraints considered. The 16 Cyg binary system is one of the best studied systems from an asteroseismic point of view \citep[see][and references therein]{Buldgen2022b}. The constraints on 16 Cyg A \&~B are therefore at another level than the other targets in the \emph{LEGACY} sample. An interferometric radius was available for these targets \citep{White2013} and we considered the following classical constraints: effective temperature, metallicity, and interferometric radius, summarised in Table \ref{tab_obs_16Cyg}. We preferred the interferometric radius because it is a more constraining and more accurately determined than the absolute luminosity, which depends on the bolometric correction and extinction map considered. For the eight other targets, we considered three sets of constraints summarised in Table \ref{tab_classical_constraints_LEGACY8}. As discussed in Sec. \ref{sec_takafumi_frequencies}, this is to ensure that the luminosity is estimated consistently with the following formula:
\begin{equation}
\log\left(\frac{L}{L_\odot}\right) = -0.4\left(m_\lambda + BC_\lambda -5\log d + 5 - A_\lambda -M_{\mathrm{bol},\odot}\right) \, ,
\label{eq_luminosity}
\end{equation} 
where $m_\lambda$ is the magnitude, $BC_\lambda$ is the bolometric correction, and $A_\lambda$ is the extinction, given a band $\lambda$, in our case the 2MASS $K_s$-band. We inferred the extinction with the dust map from \citet{Green2018}, and computed the bolometric correction following \citet{Casagrande2014,Casagrande2018}. We adopted a solar bolometric correction of $M_{\mathrm{bol},\odot} = 4.75$. The distance $d$ in pc from \textit{Gaia EDR3} \citep{Gaia2021} was computed by testing two approaches; by inverting the parallax corrected according to \citet{Lindegren2021} or by using the distance from \citet{Bailer-Jones2021}. Both methods lead to consistent results and we adopted the luminosity based on the latter distances as our observational constraint. The precision of the $K_s$ magnitude of Pinocha is very low, which results in a poorly constrained luminosity. In addition, this target and Arthur are flagged as unreliable by GAIA. Indeed, the \textit{RUWE} indicator, for renormalised unit weight error is expected to be around one for single-star sources. If this indicator is much larger than one, as it is the case for Arthur and Pinocha, it may indicate that the source is not single or that another issue affected the astrometric solution. We summarised the constraints of the second category of targets in Table \ref{tab_obs_LEGACY}. 

\begin{table}[t!]
\centering
\caption{Classical constraints and observed luminosity of the 16 Cyg binary system.}
\resizebox{\linewidth}{!}{
\begin{tabular}{lcccc}
\hline 
 & $T_{\mathrm{eff}}$ (K) & [Fe/H] (dex) & $R$ ($R_\odot$) & $L$ ($L_\odot$) \\ 
\hline \hline
16 Cyg A & $5839\pm 42$ & $0.096\pm 0.026$ & $1.22\pm 0.02$ & $1.56\pm 0.05$ \\  
16 Cyg B & $5809\pm 39$ & $0.052\pm 0.021$ & $1.12\pm 0.02$ & $1.27\pm 0.04$ \\ 
\hline 
\end{tabular} 
}
{\par\small\justify\textbf{Notes.} $T_{\mathrm{eff}}$ and $R$ from \citet{White2013}, [Fe/H] from \citet{Ramirez2009}, and $L$ from \citet{Metcalfe2012}. \par}
\label{tab_obs_16Cyg}
\end{table}

\begin{table}[t!]
\centering
\caption{Classical constraints for the second category of targets.}
\begin{tabular}{cc}
\hline 
 & classical constraints \\ 
\hline \hline 
set 1 & $T_{\rm eff}$, [Fe/H], $L$ \\ 
set 2 & $T_{\rm eff}$, [Fe/H], $\nu_{max}$ \\ 
set 3 & $T_{\rm eff}$, [Fe/H] \\ 
\hline 
\end{tabular} 
\label{tab_classical_constraints_LEGACY8}
\end{table}

\begin{table*}[t!]
\centering
\caption{Observational constraints for the second category of \emph{LEGACY} targets.}
\begin{tabular}{lcccccc}
\hline 
KIC & Nickname & $T_{\mathrm{eff}}$ & [Fe/H] & $L$  & $\nu_{max}$ & References \\ 
 &  & (K) & (dex) & $(L_\odot)$  & ($\mu$Hz) & \\ 
\hline \hline 
6116048 & Nunny & $6033 \pm 100$ & $-0.23 \pm 0.10$ & $1.85 \pm 0.07$ & $2126.9 \pm 5.3$ & 1 \\ 
6225718 & Saxo2 & $6203 \pm 100$ & $-0.17 \pm 0.10$ & $2.13 \pm 0.08$ & $2364.2 \pm 4.8$ & 2 \\ 
8006161 & Doris & $5488 \pm 100$ & $0.34 \pm 0.10$ & $0.69 \pm 0.03$ & $3574.7 \pm 11.0$ & 1 \\ 
8379927 & Arthur & $6067 \pm 150$ & $-0.10 \pm 0.15$ & - & $2795.3 \pm 6.0$ & 3 \\ 
9139151 & Carlsberg & $6043 \pm 100$ & $0.05 \pm 0.10$ & $1.60 \pm 0.06$ & $2690.4 \pm 11.8$ & 2 \\ 
10454113 & Pinocha & $6177 \pm 100$ & $-0.07 \pm 0.10$ & - & $2357.2 \pm 8.7$ & 1 \\ 
12009504 & Dushera & $6179 \pm 100$ & $-0.08 \pm 0.10$ & $2.70 \pm 0.11$ & $1865.6 \pm 7.0$ & 1 \\ 
12258514 & Barney & $5964 \pm 60$ & $0.00 \pm 0.10$ & $2.95 \pm 0.11$ & $1512.7 \pm 3.1$ & 1 \\ 
\hline 
\end{tabular}
{\par\small\justify\textbf{Notes.} (1) \citet{Lund2017}; (2) frequencies and $\nu_{max}$ from \citet{Lund2017}, $T_{\mathrm{eff}}$ and [Fe/H] from \citet{Furlan2018}; (3) frequencies from \citet{Roxburgh2017}, $T_{\mathrm{eff}}$, [Fe/H], and $\nu_{max}$ from \citet{Lund2017}. The luminosity $L$ is estimated using the spectroscopic parameters and Eq. \eqref{eq_luminosity}. For Arthur, the RUWE Gaia indicator flags the parallax measurement as unreliable, and for Pinocha, the $K_s$-magnitude measurement is unreliable. Therefore, we could not estimate the absolute luminosity of these two targets reliably. \par}
\label{tab_obs_LEGACY}
\end{table*}

\subsection{Individual frequencies as constraints}
\label{sec_individual_frequencies_as_constraints_LEGACY}
\begin{figure*}[htp!]
  \centering
\begin{subfigure}[b]{.43\textwidth}
  \includegraphics[width=.99\linewidth]{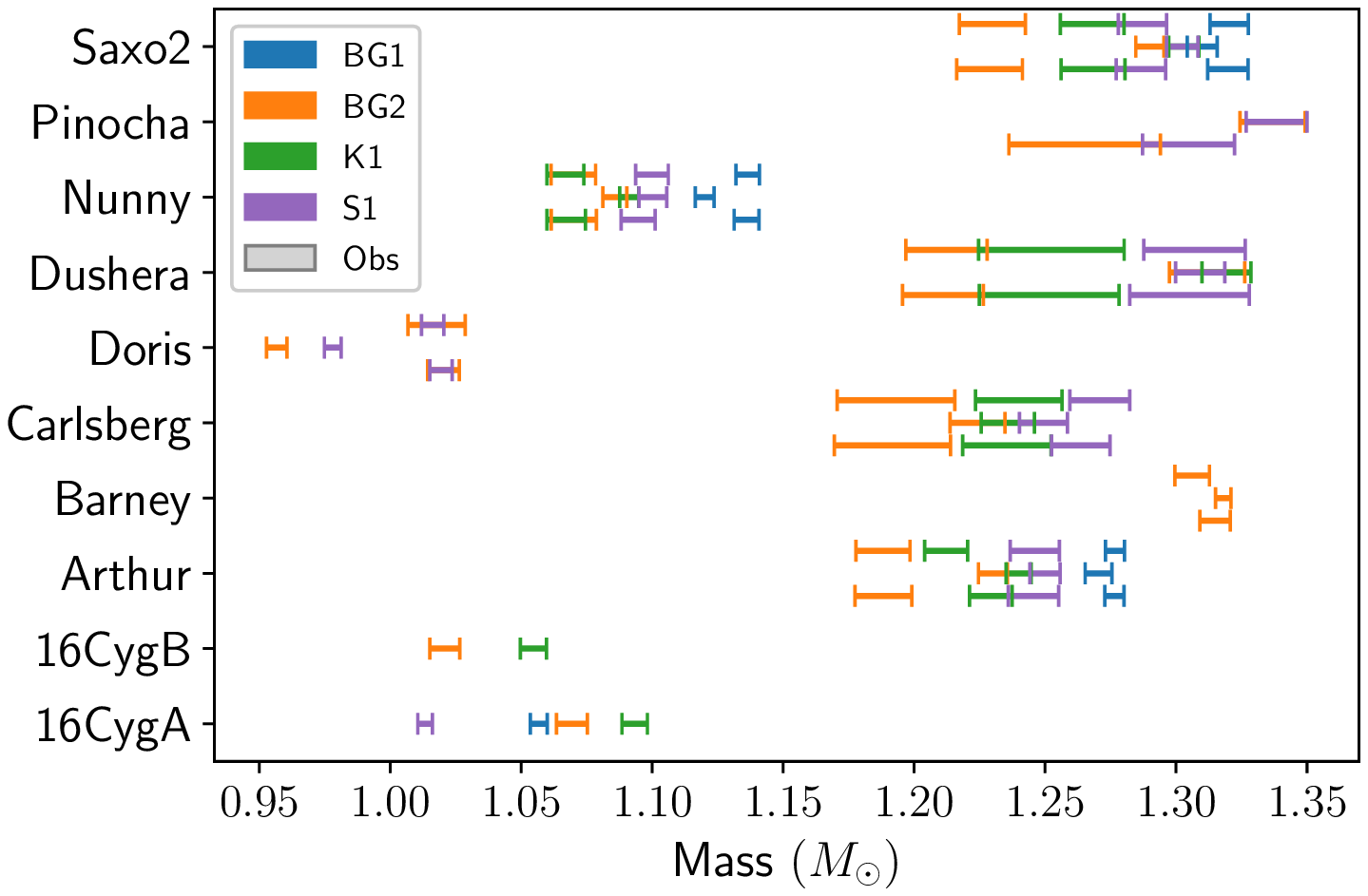}  
  \label{fig_LEGACY_frequencies_clever_mass}
\end{subfigure}
\begin{subfigure}[b]{.43\textwidth}
  \includegraphics[width=.99\linewidth]{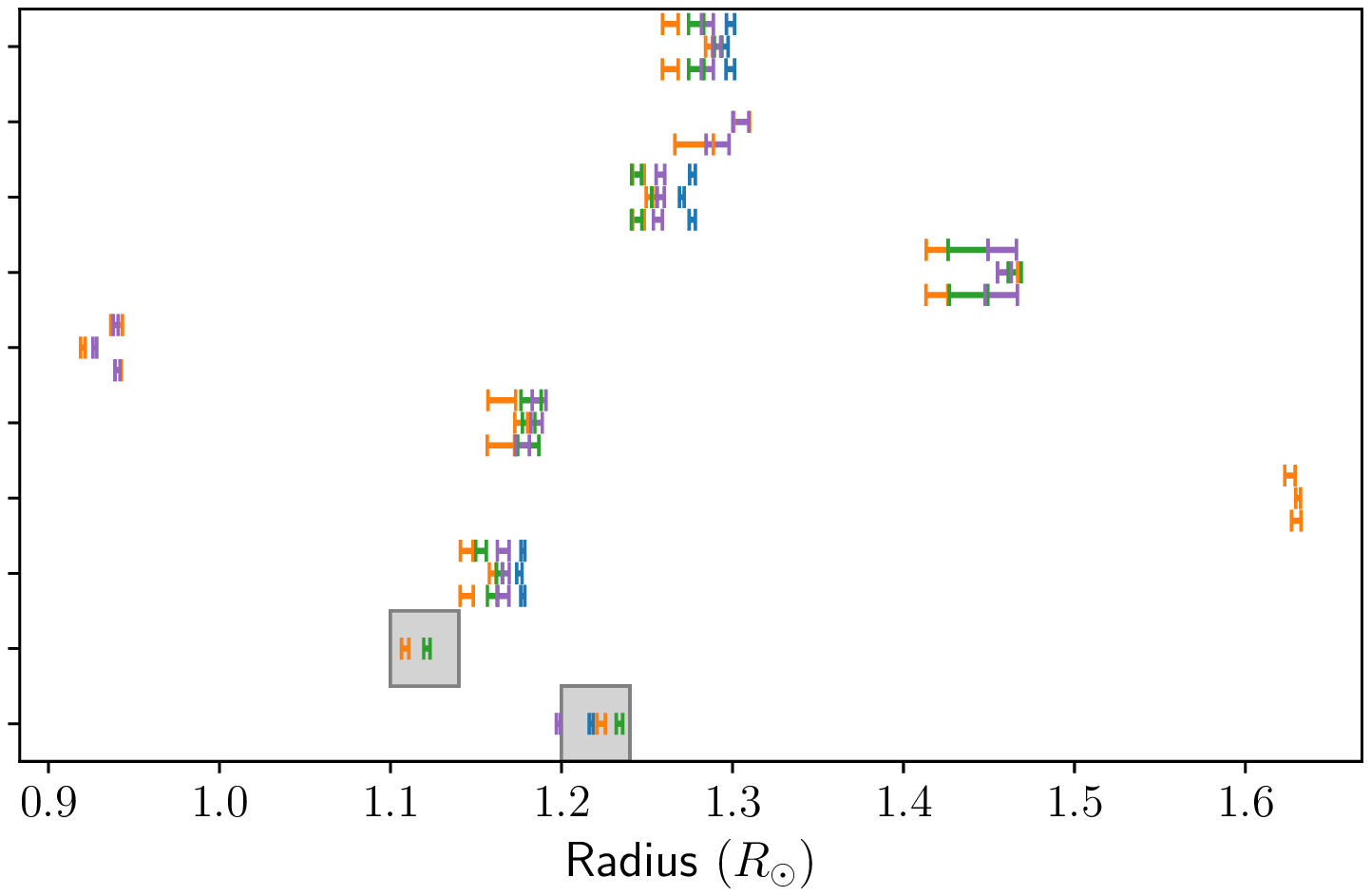}  
  \label{fig_LEGACY_frequencies_clever_radius}
\end{subfigure}
\begin{subfigure}[b]{.43\textwidth}
  \includegraphics[width=.99\linewidth]{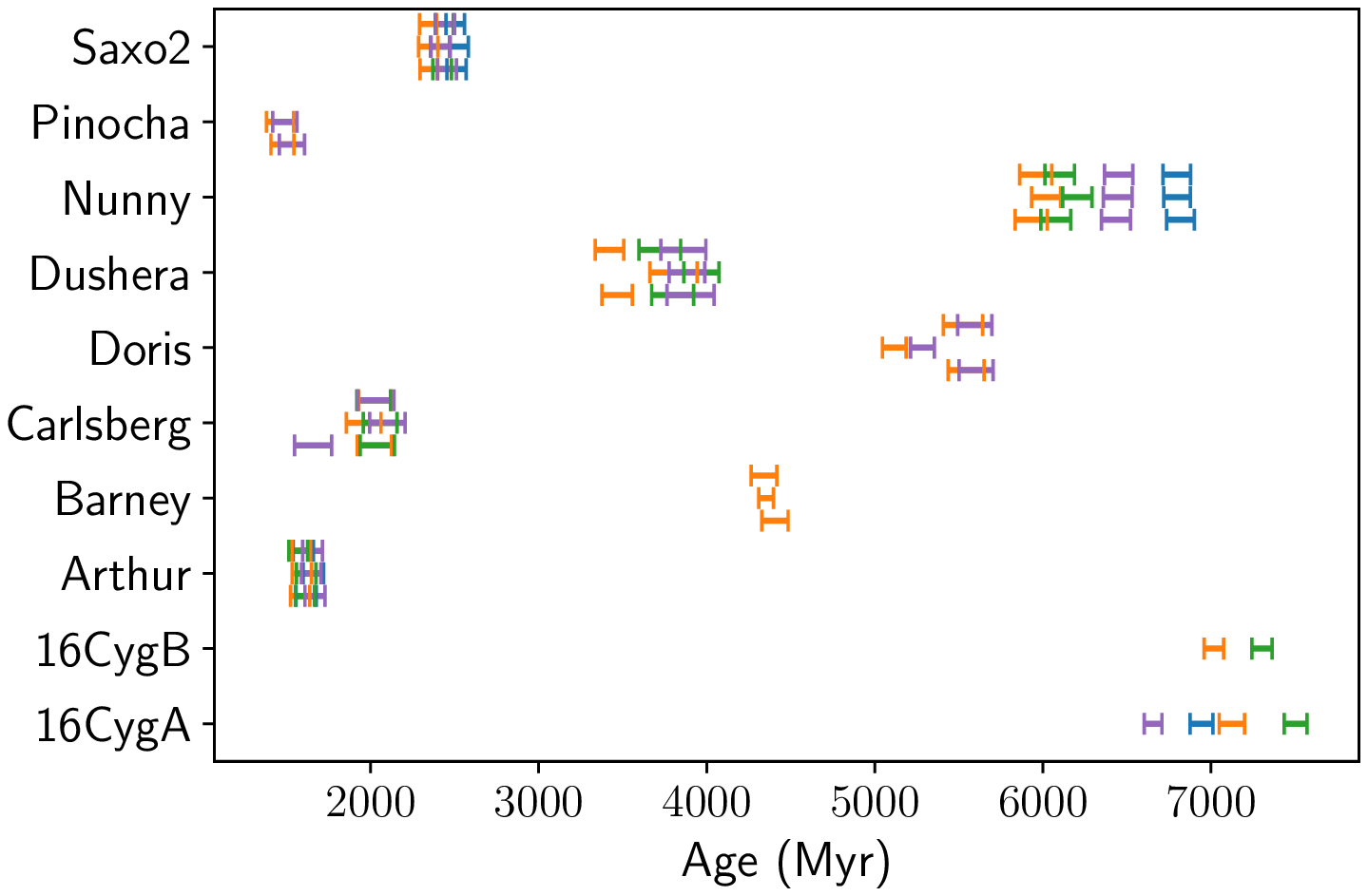}  
  \label{fig_LEGACY_frequencies_clever_age}
\end{subfigure}
\begin{subfigure}[b]{.43\textwidth}
  \includegraphics[width=.99\linewidth]{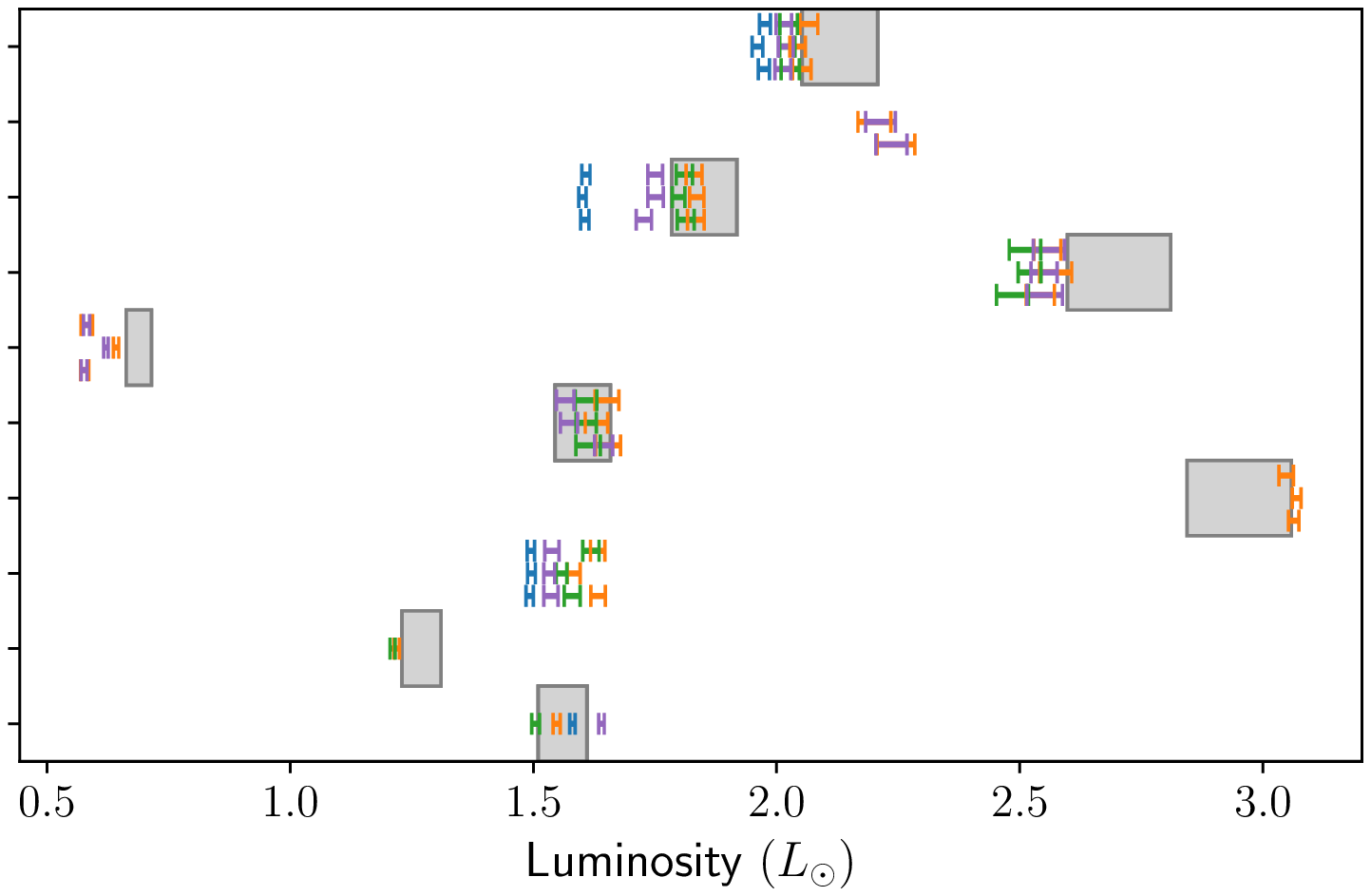}  
  \label{fig_LEGACY_frequencies_clever_luminosity}
\end{subfigure}
\caption{MCMC results for the \emph{LEGACY} targets, using different prescriptions of the surface effects described in Table \ref{tab_surface_prescriptions}. The runs with convergence issues were manually discarded. The modelling of Barney was more challenging, and MCMC runs converged successfully only with the \textit{BG2} surface effect prescription. For each target, three set of classical constraints were considered: set 1 (bottom line), set 2 (middle line), and set 3 (upper line). The grey boxes represent the observational constraints.}
\label{fig_LEGACY_frequencies_clever_full}
\end{figure*}

In Fig. \ref{fig_LEGACY_frequencies_clever_full}, we show the results of the fit of the individual frequencies and the different sets of classical constraints, by considering different prescriptions for the surface effects. We removed the models with convergence issues that result from the treatment of the surface effects. With the \textit{K2} and \textit{S2} prescriptions, the MCMC could not find optimal values for the free coefficients associated with the surface effects correction. For the other unsuccessful runs, the MCMC hit the grid boundaries by trying to compensate the other MCMC free parameters (mass, radius, and initial chemical composition with $X_0$ and $Z_0$) to force an inappropriate value for the free parameters of the surface effects. In comparison to the results of Sec. \ref{sec_individual_frequencies_as_constraints_takafumi} with the \citet{Sonoi2015} targets, more prescriptions lead to successful MCMC runs. This difference is most likely due to surface effects that are smaller with the \emph{LEGACY} targets and therefore are easier to reproduce. Although some of the results with the \textit{BG1} prescription did not show the usual convergence issues, they appear as outliers in Fig. \ref{fig_LEGACY_frequencies_clever_full}. We would recommend to consider them with caution because they failed to reproduce the high frequencies, which affects the estimate of the mass and radius.

Except for Arthur and Doris, the absolute luminosity estimated by the models is consistent with the observed value, independently of the set of classical constraints considered. As explained in the previous section, the luminosity of Arthur is flagged as unreliable, but because the fit is mainly driven by the seismic information, the results of the models that include the luminosity in the constraints or not are quasi-identical. This shows that except if the luminosity is very precisely constrained, it plays a small role on the final parameters. If the inverted mean density or/and the radial frequency of lower order are not included in the constraints, the situation may be different and the luminosity should only be included if reliable. Because the luminosity of Pinocha is poorly constrained, we did not consider the set 1 of classical constraints as it is equivalent to the set 3 in such conditions.

As illustrated in Fig. \ref{fig_LEGACY_frequencies_clever_full}, the systematic uncertainty due to the choice of the prescription for the surface effects is much larger than the individual uncertainties. Except for particular cases that are probably coincidental (e.g. ages of Arthur and Nunny), this systematic is several times larger than the statistical uncertainty. In addition, as for the \citet{Sonoi2015}, the numerical cost of each minimisation was high because we had to use 8000 steps of burn-in, which is very demanding from a pipeline perspective.

\subsection{Frequency separation ratios as constraints}
\label{sec_frequency_ratios_as_constraints_LEGACY}
As in Sec. \ref{sec_frequency_ratios_as_constraints_takafumi}, we fitted the frequency separation ratios along with the classical constraints and considering the same three prescriptions for the inclusion of the inverted mean density. The results are summarised in Table \ref{tab_ratios_LEGACY}. Just like with the \citet{Sonoi2015} targets, if the inverted mean density is part of the constraints, the precision of the stellar mass and radius is significantly improved. We found comparable precision by treating the mean density as a classical or seismic constraint. However, the convergence of the latter was less stable with real observations, favouring again the recommendation to keep using the mean density as a classical constraint. Moreover, we observed drifts with some of the models including the radial frequency of lower order. In those cases, we did not include this constraint, which resulted in ridges whose position is slightly less well reproduced. For the model of 16 Cyg A that treats the mean density as a seismic constraint, the estimated mass was too small and incompatible with the literature \citep[see e.g.][]{Buldgen2022b} or the other set of constraints that we tested. Even though the diagnostic plots did not show any issues, we consider this result as not reliable and probably due to an undetected shift during the minimisation linked with the lowest order radial frequency. In fact, these drifts are a recurrent disadvantage of including the lowest order radial frequency in the constraints. Even by assuming a more conservative uncertainty on that quantity, it does not prevent from such drifts to occur, and if the uncertainty is too large, it does not constrain the position of the ridges anymore. From a pipeline perspective, we would recommend to not use this quantity. Indeed, the detection of the drifts must be done manually and is sometimes difficult even for an experienced modeller. In addition, even though the inverted mean density may lead to an imperfect anchoring of the frequency ridges, it mostly occurs for the most complicated cases and the resulting slight bias on the stellar parameters are less significant than the bias due to a drift of the walkers.

\begin{table*}
\centering
\caption{Precision of the stellar parameters by fitting the frequency separation ratios for our selection of  \emph{LEGACY} targets.}
\resizebox{\linewidth}{!}{
\begin{tabular}{lccc|ccc|ccc}
\hline
List of & \multicolumn{3}{c}{Stellar mass} & \multicolumn{3}{c}{Stellar radius} & \multicolumn{3}{c}{Stellar age} \\ 
Constraints & no $\bar{\rho}_{\mathrm{inv}}$ & no corr. & incl. corr. & no $\bar{\rho}_{\mathrm{inv}}$ & no corr. & incl. corr. & no $\bar{\rho}_{\mathrm{inv}}$ & no corr. & incl. corr. \\ 
\hline \hline
\textit{16 Cyg A} \\
$[Fe/H],T_{\mathrm{eff}},R,r_{02},\nu_{n_{min},l=0}$ & 2.7\% & 2.7\% & 0.9\% & 0.9\% & 0.9\% & 0.3\% & 1.6\% & 1.5\% & 1.3\% \\ 
$[Fe/H],T_{\mathrm{eff}},R,r_{01},r_{02}$ & 2.4\% & 0.8\% & 0.5\% & 1.6\% & 0.4\% & 0.2\% & 1.4\% & 1.4\% & 1.4\% \\ 
\hline 
\textit{16 Cyg B} \\
$[Fe/H],T_{\mathrm{eff}},R,r_{02},\nu_{n_{min},l=0}$ & 2.4\% & 2.4\% & 2.4\% & 0.8\% & 0.8\% & 0.8\% & 1.8\% & 1.6\% & 1.4\% \\
\hline 
\textit{Arthur} \\
$[Fe/H],T_{\mathrm{eff}},L,r_{01},r_{02},\nu_{n_{min},l=0}$ & 3.1\% & 2.5\% & 2.9\% & 1.2\% & 1.0\% & 1.1\% & 7.5\% & 5.8\% & 6.3\% \\ 
\hline 
\textit{Barney} \\
$[Fe/H],T_{\mathrm{eff}},L,r_{01},r_{02}$ & 2.1\% & 1.1\% & 1.1\% & 1.9\% & 0.5\% & 0.4\% & 5.6\% & 4.3\% & 3.8\% \\ 
\hline 
\textit{Carlsberg} \\
$[Fe/H],T_{\mathrm{eff}},L,r_{01},r_{02},\nu_{n_{min},l=0}$ & 2.8\% & 2.5\% & 2.3\% & 1.0\% & 0.9\% & 0.9\% & 7.2\% & 6.6\% & 6.6\% \\
$[Fe/H],T_{\mathrm{eff}},L,r_{01},r_{02}$ & 3.7\% & 2.6\% & 2.8\% & 2.3\% & 0.9\% & 1.0\% & 6.8\% & 6.7\% & 6.9\% \\
\hline 
\textit{Doris} \\
$[Fe/H],T_{\mathrm{eff}},L,r_{01},r_{02}$ & 1.0\% & 0.7\% & 0.7\% & 1.5\% & 0.3\% & 0.2\% & 2.5\% & 2.2\% & 2.2\% \\
\hline 
\textit{Dushera} \\
$[Fe/H],T_{\mathrm{eff}},L,r_{01},r_{02},\nu_{n_{min},l=0}$ & 2.3\% & 2.4\% & 1.8\% & 0.8\% & 0.8\% & 0.7\% & 7.2\% & 5.5\% & 4.0\% \\ 
\hline 
\textit{Nunny} \\
$[Fe/H],T_{\mathrm{eff}},L,r_{01},r_{02},\nu_{n_{min},l=0}$ & 1.0\% & 1.0\% & 1.0\% & 0.4\% & 0.4\% & 0.4\% & 4.6\% & 3.4\% & 3.0\% \\ 
\hline 
\textit{Pinocha} \\
$[Fe/H],T_{\mathrm{eff}},L,r_{01},r_{02},\nu_{n_{min},l=0}$ & 2.5\% & 2.2\% & 2.3\% & 1.0\% & 0.8\% & 0.9\% & 8.0\% & 7.8\% & 7.9\% \\ 
\hline 
\textit{Saxo2} \\
$[Fe/H],T_{\mathrm{eff}},L,r_{01},r_{02},\nu_{n_{min},l=0}$ & 1.7\% & 1.5\% & 1.8\% & 0.7\% & 0.6\% & 0.7\% & 4.9\% & 3.3\% & 4.7\% \\ 
$[Fe/H],T_{\mathrm{eff}},L,r_{01},r_{02}$ & 2.7\% & 1.7\% & 1.8\% & 1.9\% & 0.6\% & 0.8\% & 6.6\% & 3.6\% & 5.1\% \\ 
\hline 
\end{tabular} 
}
{\par\small\justify\textbf{Notes.} We considered three prescriptions for the inclusion of the inverted mean density in the constraints: not including it (no $\bar{\rho}_{\mathrm{inv}}$), including it as a classical constraint (no corr.), or including it as a seismic constraint to account for the correlations with the ratios (incl. corr.). \par}
\label{tab_ratios_LEGACY}
\end{table*}

\subsection{Comparison and discussion}
\label{sec_comparison_LEGACY}

\begin{figure*}[htp!]
  \centering
\begin{subfigure}[b]{.45\textwidth}
  \includegraphics[width=.99\linewidth]{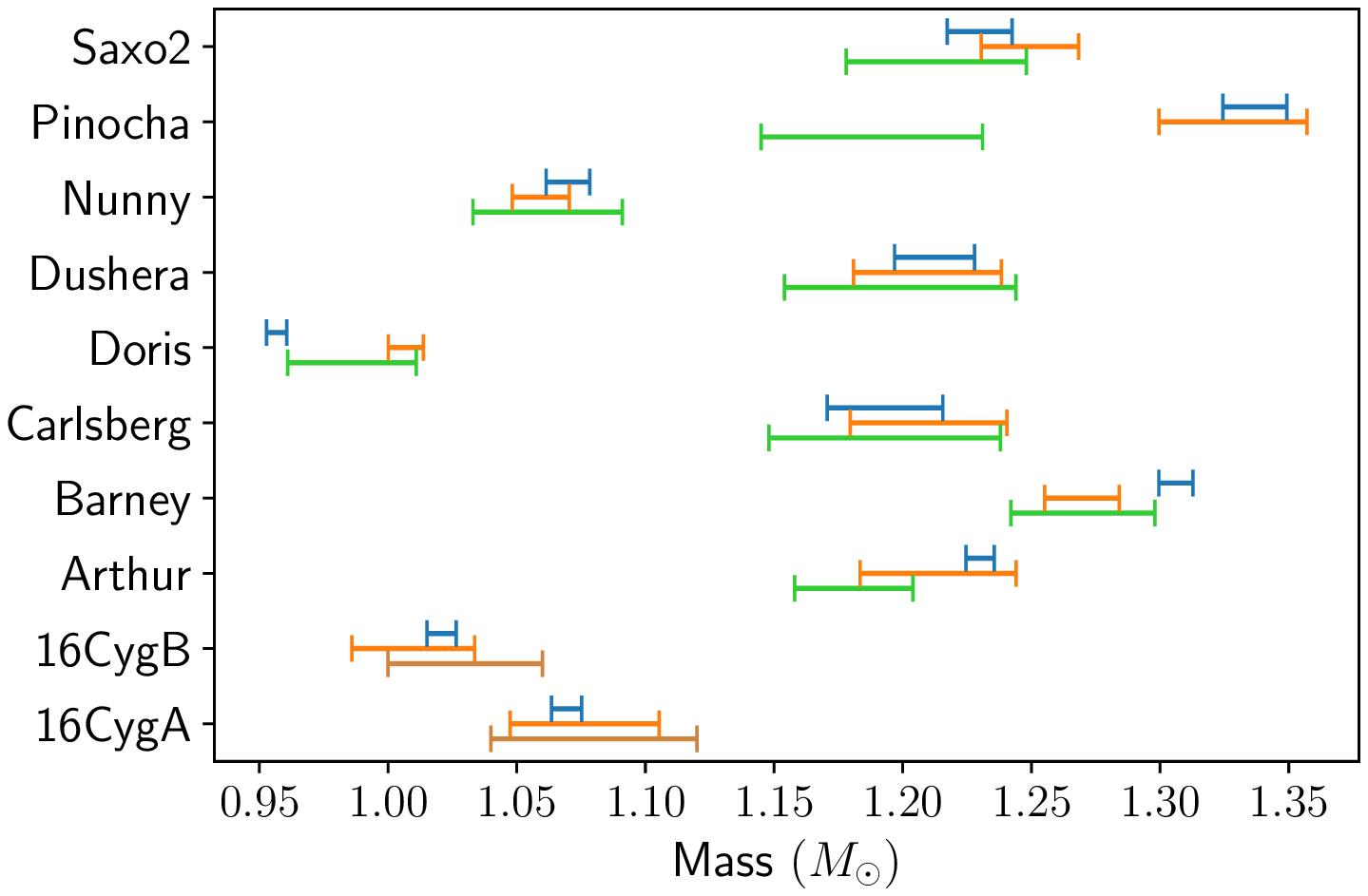}  
  \label{fig_LEGACY_VS_Mass}
\end{subfigure}
\begin{subfigure}[b]{.45\textwidth}
  \includegraphics[width=.99\linewidth]{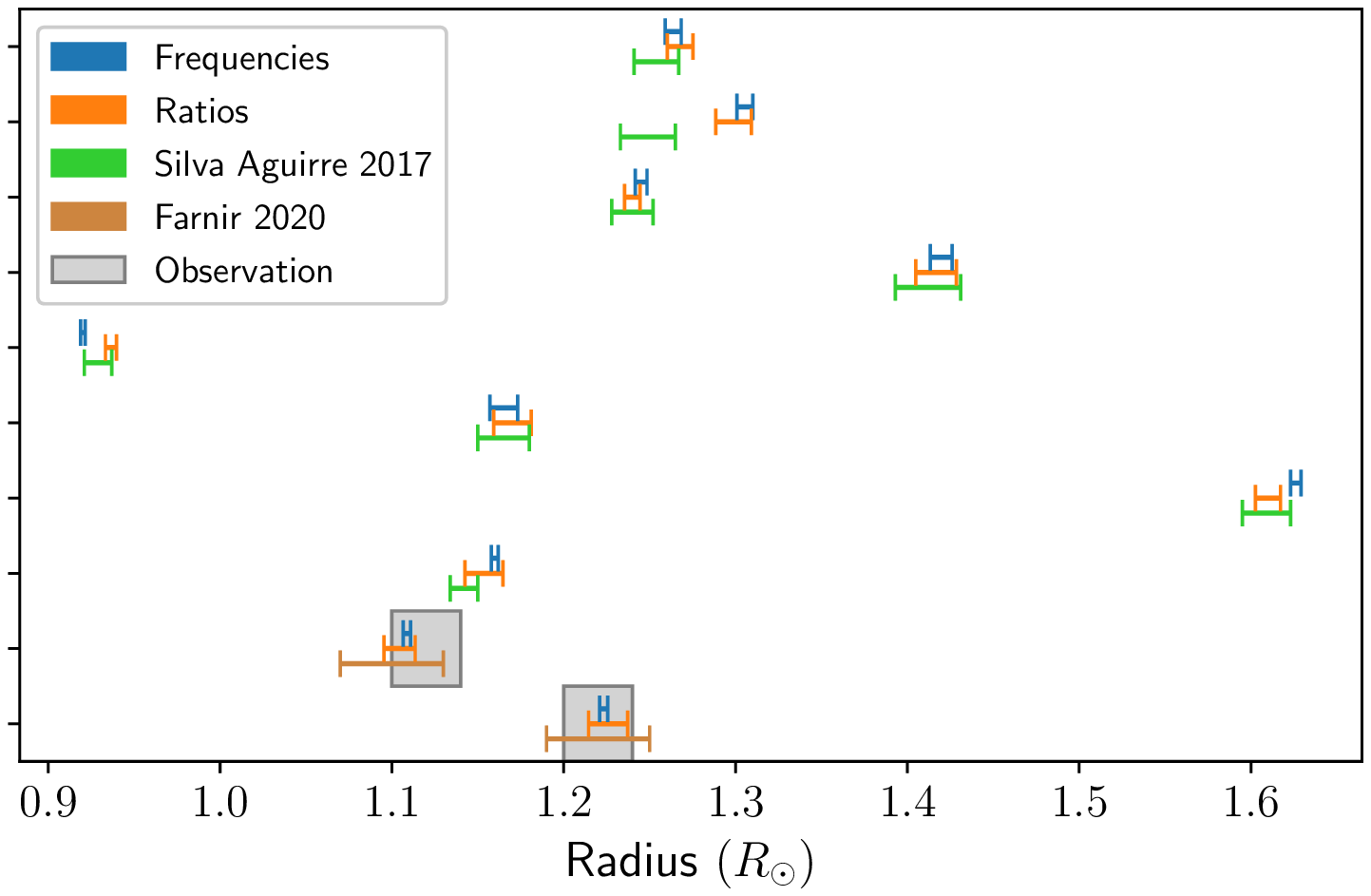}  
  \label{fig_LEGACY_VS_Radius}
\end{subfigure}
\begin{subfigure}[b]{.45\textwidth}
  \includegraphics[width=.99\linewidth]{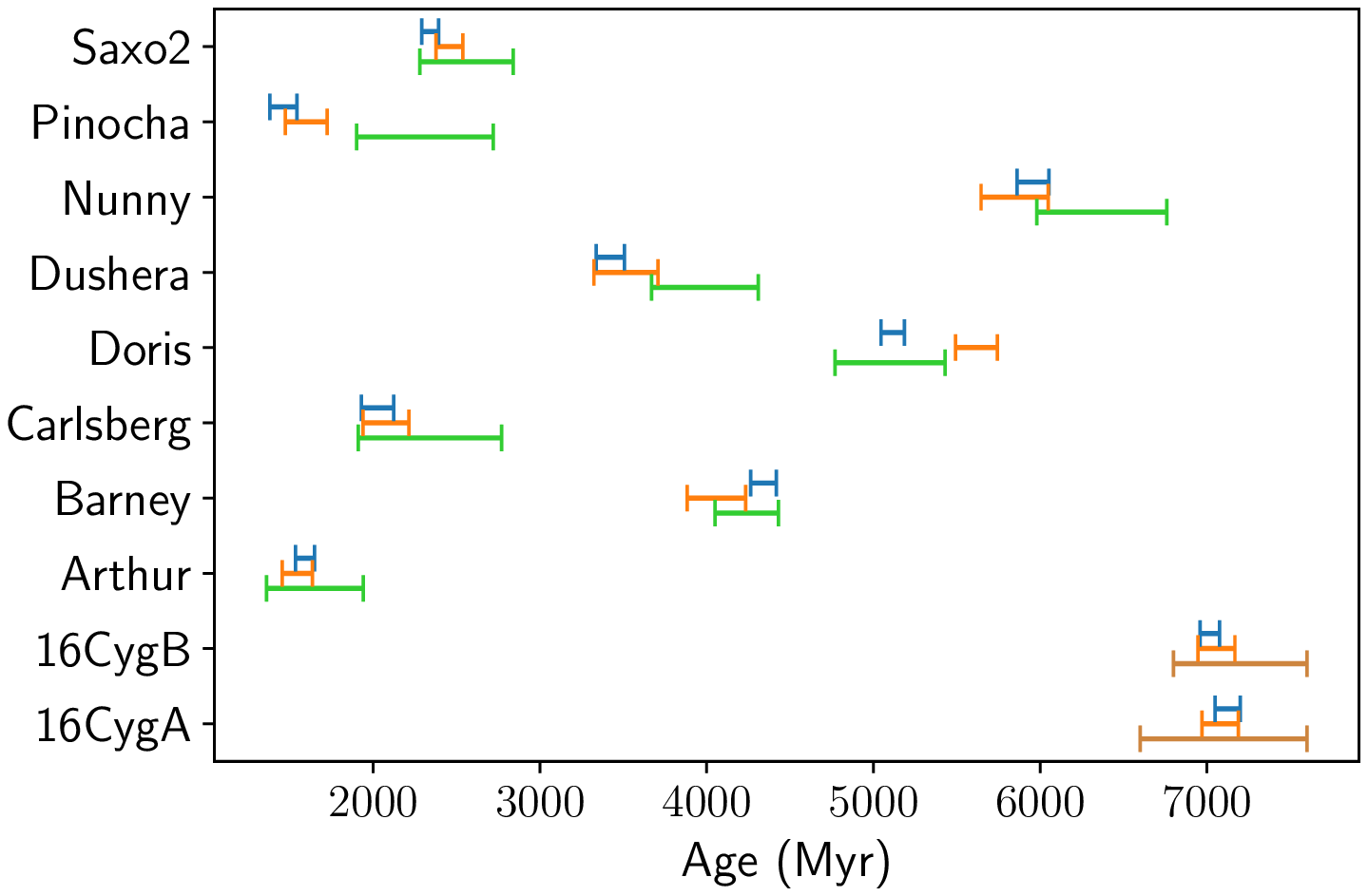}  
  \label{fig_LEGACY_VS_Age}
\end{subfigure}
\begin{subfigure}[b]{.45\textwidth}
  \includegraphics[width=.99\linewidth]{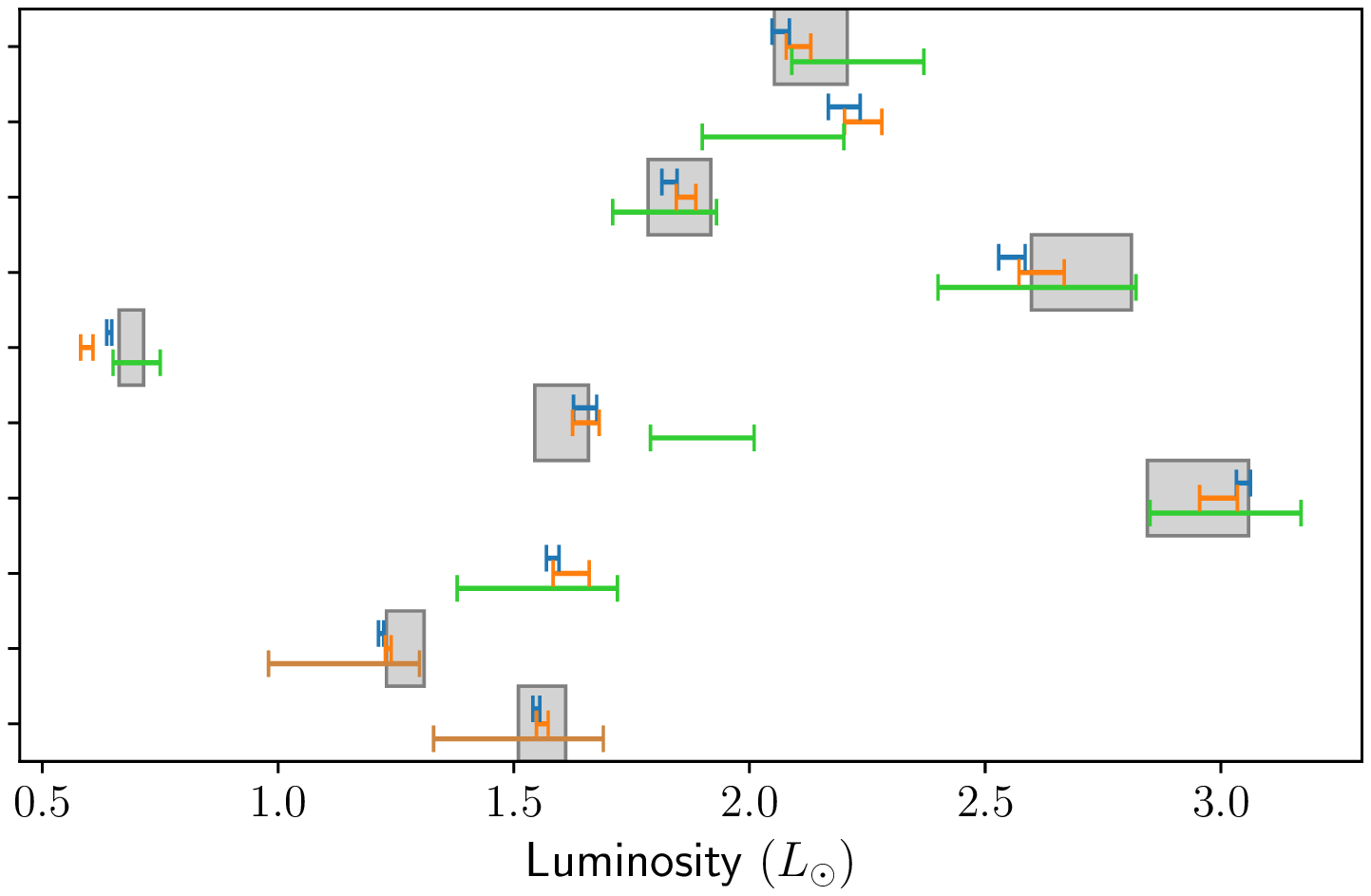}  
  \label{fig_LEGACY_VS_Luminosity}
\end{subfigure}
\caption{Comparison between the results of the modelling strategy fitting individual frequencies (in blue), the modelling strategy fitting frequency separations ratios and the inverted mean density (in orange), and the literature (in brown and green). The grey boxes represent the observational constraints.}
\label{fig_LEGACY_VS}
\end{figure*}

In Fig. \ref{fig_LEGACY_VS}, we compare the results between the fit of the individual frequencies, the fit of the frequency separation ratios, and the literature \citep{Silva-Aguirre2015,Silva-Aguirre2017,Farnir2020}. For the individual frequencies, we selected the models that include the absolute luminosity in the constraints, except for Arthur, Doris, and Pinocha. For these targets, the luminosity estimated with Eq. \eqref{eq_luminosity} is considered as unreliable, and we selected models that constrain the frequency of maximal power $\nu_{max}$ instead. For the fit of the ratios, we selected the models treating the mean density as a classical constraint, and the literature values come from \citet{Farnir2020} for 16 Cyg A \&~B and from the YMCM algorithm \citep{Silva-Aguirre2015,Silva-Aguirre2017} otherwise. We note that the latter authors used older references for some of the physical ingredients, especially they used the GS98 abundances \citep{Grevesse&Sauval1998} and the nuclear rates from \citet{Adelberger1998}. Hence, although our results are consistent with the literature, we can observe some slight differences due the differences in the physics of the models. In addition, as with the \citet{Sonoi2015} targets, the fit of the individual frequencies tends to overestimate the statistical precision of the stellar parameters. Finally, we note that we provided in Table \ref{tab_final_params_LEGACY} the optimal stellar parameters of the \emph{LEGACY} targets studied.

\section{Conclusions}
\label{sec_conclusions}

In this paper, we introduced in Sec. \ref{sec_spelaion_grid} a new high-resolution grid of stellar models of main-sequence stars between 0.8 and 1.6 solar masses. Then in Sec. \ref{sec_modelling_strategies}, we presented two modelling strategies focusing on a direct exploitation of the seismic information. We discussed the issues occurring with a fit of the individual frequencies, and presented a more elaborated modelling technique that combines mean density inversions and a fit of the frequency separation ratios to damp the surface effects and provide precisely and accurately constrained stellar parameters. We also discussed and compared three options for the inclusion of the inverted mean density in the constraints. In Sec. \ref{sec_modelling_strategies}, we applied the two modelling strategies to six synthetic targets from \citet{Sonoi2015}, but including a consistent treatment of non-adiabatic effects, and in Sec. \ref{sec_applications_LEGACY}, we conducted the same tests on a sample of ten \emph{Kepler LEGACY} targets.

The current treatment of the surface effects with semi-empirical prescriptions constitutes an important limiting factor in terms of precision, accuracy and numerical cost. This corroborates what was observed in previous studies for other targets \citep{Ball&Gizon2017,Nsamba2018,Jorgensen2020,Jorgensen2021}. The procedure that combines the mean density inversion and the ratios can significantly improve the precision and accuracy of the stellar parameters, especially the mass and the radius, but would benefit from an improvement in the understanding of surface effects as it would allow us to further improve the maximal precision and accuracy achievable. Regarding the treatment of the inverted mean density, we recommend applying it as a classical constraint and assuming a conservative precision. Further studies on benchmark targets would be welcomed to refine this conservative precision in certain mass ranges, and test if and how it is impacted by the chemical composition and overshooting. The treatment of the inverted mean density as a seismic constraint to account for the correlations with the ratios achieves a comparable precision but in a less stable manner and is therefore less recommended.

Putting this work in the context of PLATO, we showed that it was possible to get stellar parameters precise enough to meet the PLATO precision requirements for ten \emph{Kepler LEGACY} targets, thanks to the use of mean density inversions (cf. Table \ref{tab_ratios_LEGACY}). The numerical cost of the procedure will be challenging for a pipeline. The first step consists in fitting the individual frequencies, thus obtaining a reference model for the mean density inversion in order to circumvent the surface effects. In addition to a better understanding of these effects, PLATO would also benefit from a thorough characterisation of the systematics due to the choice of the physical ingredients, because it also impacts the maximal precision achievable for the stellar parameters \citep[see e.g.][]{Betrisey2022}. Finally, we would recommend using the following set of constraints if used in a pipeline: $r_{01}$, $r_{02}$, [Fe/H], $T_{\mathrm{eff}}$, $L$ if reliable, and $\bar{\rho}_{\mathrm{inv}}$. It was the most robust set and the benefits from the radial frequency of lowest order are too small in comparison to the biases that it may introduce.


\section*{Acknowledgements}
We would like to thank Takafumi Sonoi for providing the models and associated data from \citet{Sonoi2015}.

J.B and G.B. acknowledge funding from the SNF AMBIZIONE grant No 185805 (Seismic inversions and modelling of transport processes in stars). P.E. and G.M. have received funding from the European Research Council (ERC) under the European Union's Horizon 2020 research and innovation programme (grant agreement No 833925, project STAREX). M.F. acknowledges the support STFC consolidated grant ST/T000252/1. Finally, this work has benefited from financial support by CNES (Centre National des Etudes Spatiales) in the framework of its contribution to the PLATO mission.


\bibliography{bibliography.bib}

\begin{appendix}


\section{Mean density inversion - numerical treatment and results interpretation}
\label{appendix_inversion_coefficients}

For a more complete description of inversions, we invite the reader to look at \citet{Reese2012}, \citet{Betrisey2022}, \citet{Betrisey&Buldgen2022}, or \citet{Buldgen2022c}. The mean density inversion used in this work is based on the structure inversion equation, that relates the frequency perturbation directly to the structural perturbation \citep{Dziembowski1990},
\begin{equation}
\frac{\delta\nu^{n,l}}{\nu^{n,l}} = \int_{0}^{R} K_{\rho,\Gamma_1}^{n,l}\frac{\delta \rho}{\rho}dr + \int_{0}^{R} K_{\Gamma_1,\rho}^{n,l}\frac{\delta \Gamma_1}{\Gamma_1}dr + \mathcal{O}(\delta^2),
\label{eq_inversionGT}
\end{equation}
where $\nu$ is the oscillation frequency, $\rho$ is the density, $\Gamma_1=~\left(\frac{\partial\ln P}{\partial\ln \rho}\right)_{\mathrm{ad}}$ is the first adiabatic exponent, $P$ is the pressure, and $K_{\rho,\Gamma_1}^{n,l}$ and $K_{\Gamma_1,\rho}^{n,l}$ the corresponding structural kernels. We used the definition
\begin{equation}
\frac{\delta x}{x} = \frac{x_{\mathrm{obs}}-x_{\mathrm{ref}}}{x_{\mathrm{ref}}},
\end{equation}
where `ref' stands for reference and `obs' stands for observed. For a mean density inversion, the idea is then to combine the equations \eqref{eq_inversionGT} to compute a correction of the mean density of the reference model based on the observed frequency differences. In practice, the following cost function is minimised:
\begin{align}
\mathcal{J}_{\bar{\rho}}(c_i) &= \int_0^1 \big(\mathcal{K}_{\mathrm{avg}} - \mathcal{T}_{\bar{\rho}}\big)^2 dx 
                                        + \beta\int_0^1 \mathcal{K}_{\mathrm{cross}}^2dx + \lambda\left[2-\sum_i c_i\right] \nonumber\\
                                        &\quad+\tan\theta \frac{\sum_i (c_i\sigma_i)^2}{\langle\sigma^2\rangle}
                                        + \mathcal{F}_{\mathrm{Surf}}(\nu),
\label{eq_SOLA_cost_function}
\end{align}
where $x=r/R$, and the averaging kernel $\mathcal{K}_{\mathrm{avg}}$ and the cross-term kernel $\mathcal{K}_{\mathrm{cross}}$ are related to the structural kernels,
\begin{align}
\mathcal{K}_{\mathrm{avg}} &= \sum_i c_i K_{\rho,\Gamma_1}^{i} \\
\mathcal{K}_{\mathrm{cross}} &= \sum_i c_i K_{\Gamma_1,\rho}^{i}.
\end{align}
The balance between the amplitudes of the different terms during the fitting is adjusted with trade-off parameters, $\beta$ and $\theta$. The idea is to obtain a good fit of the target function, in our case $\mathcal{T}_{\bar{\rho}}(x)=4\pi x^2 \frac{\rho}{\rho_R}$ with $\rho_R=\frac{M}{R^3}$, while reducing the contribution from the cross-term and of the observational errors on the individual frequencies $\sigma_i$. An accurate inversion result is ensured by a good fit of the target function by the averaging kernel. In addition, we defined $\langle\sigma^2\rangle = \sum_i^N\sigma_i^2$, where $N$ is the number of observed frequencies. The $\lambda$ symbol is a Lagrange multiplier and the coefficients $c_i$ are the inversion coefficients. The surface term is denoted by $\mathcal{F}_{\mathrm{Surf}}(\nu)$, and is implemented using Eq. \eqref{eq_BG2} for the \citet{Ball&Gizon2014} prescription and the linearised version of Eq. \eqref{eq_S2} for the \citet{Sonoi2015} prescription.

The first term in Eq. \eqref{eq_SOLA_cost_function} is the main term, the equivalent of the usual least squares term in other minimisations techniques. The second term is related to the second structural variable. Indeed, the structural kernels are based on a structural pair, while we are only interested in one of the variables, in our case the density. Hence, the idea is the ensure that the contribution of this cross term is as small as possible. The third term is a normalisation term to ensure that the coefficients give the correct result for a homologous transformation, and the fourth term accounts for the observational uncertainty. As for the cross term, the idea is to ensure that its contribution is as small as possible. Finally, the last term should be treated with caution, because it allows us to take the surface effects into account, but at the expense of the fit of the target function. Indeed, asteroseismology is dealing with a limited number of frequencies (about 50 for high quality targets) compared to helioseismology (a few thousands). Hence, the seismic information may be completely used by the additional free variables introduced with the surface term and no structural differences can be extracted by the inversion. In such a case, the target function is poorly reproduced by the averaging kernel, and the inversion coefficients tend to take high amplitudes with large variations between two consecutive coefficients. The target function is also poorly reproduced if the data quality is low, either due to high observational uncertainties or due to a too small number of observed frequencies.

\begin{figure*}[htp!]
  \centering
\begin{subfigure}[b]{.33\textwidth}
  \includegraphics[width=.99\linewidth]{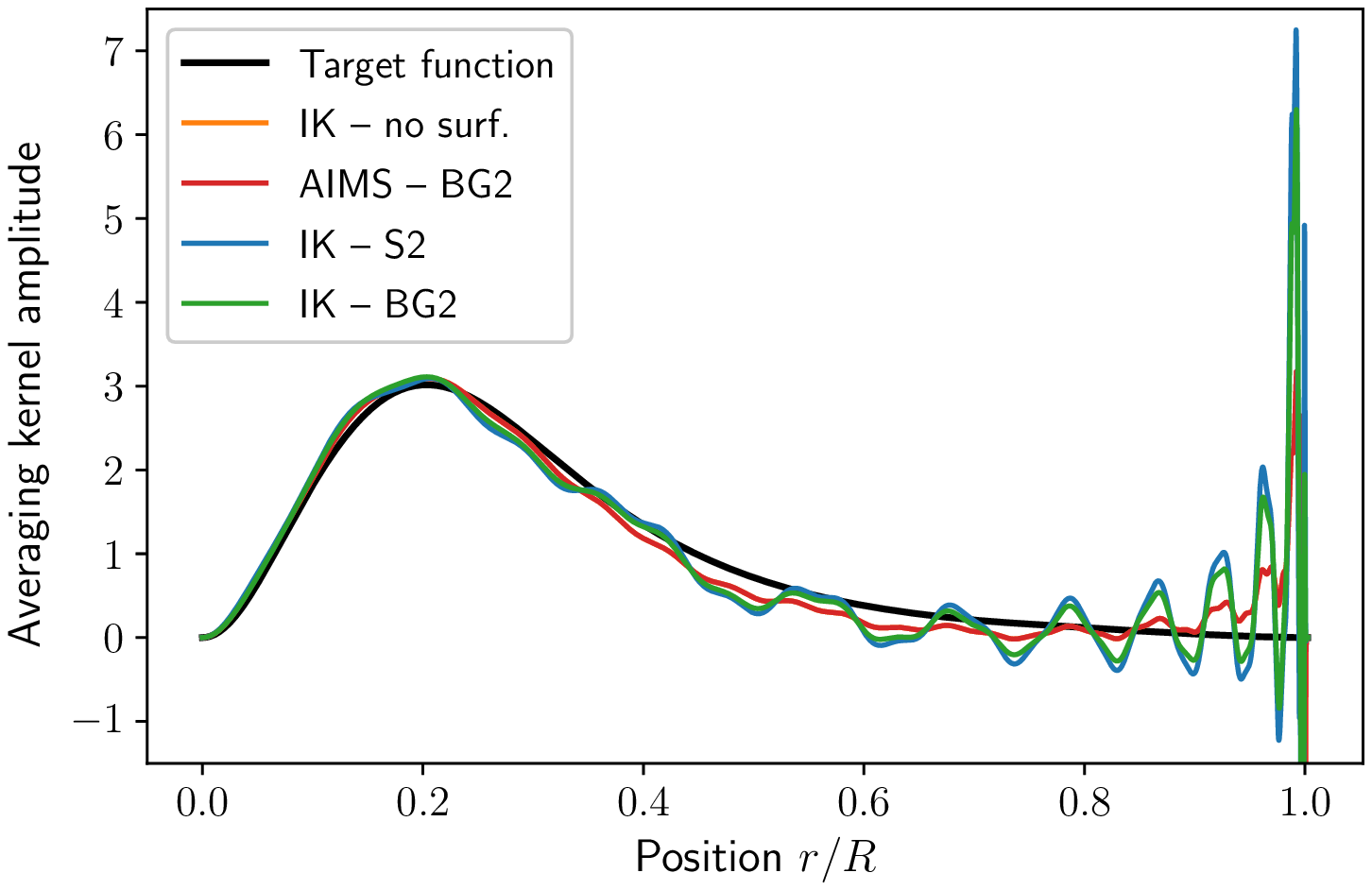}  
  \caption{\centering Averaging kernel of model A}
  \label{fig_appendix_comp_Kavg_modelA}
\end{subfigure}
\begin{subfigure}[b]{.33\textwidth}
  \includegraphics[width=.99\linewidth]{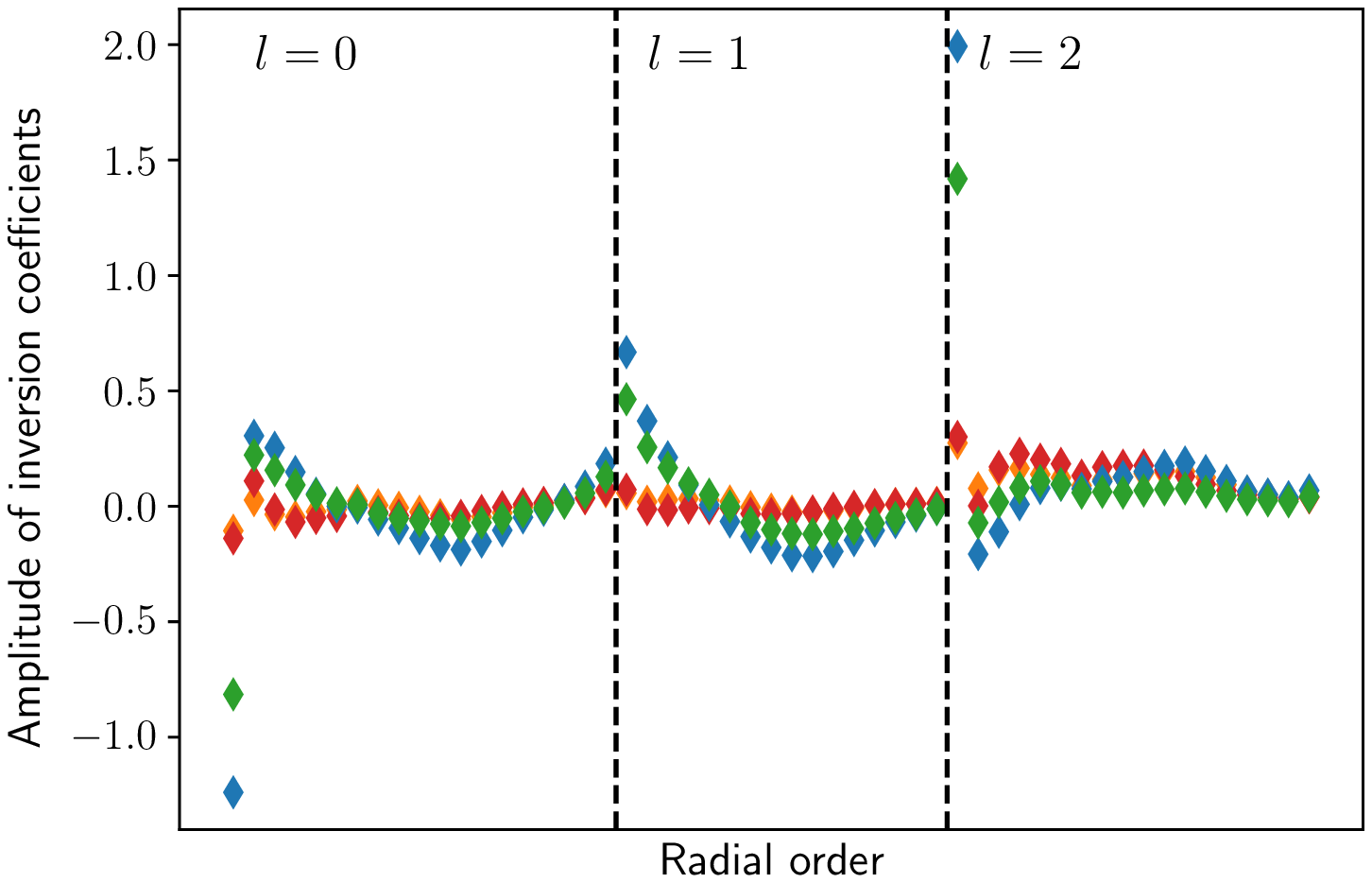} 
  \caption{\centering Inversion coefficients of model A} 
  \label{fig_appendix_comp_ci_modelA}
\end{subfigure}
\begin{subfigure}[b]{.33\textwidth}
  \includegraphics[width=.99\linewidth]{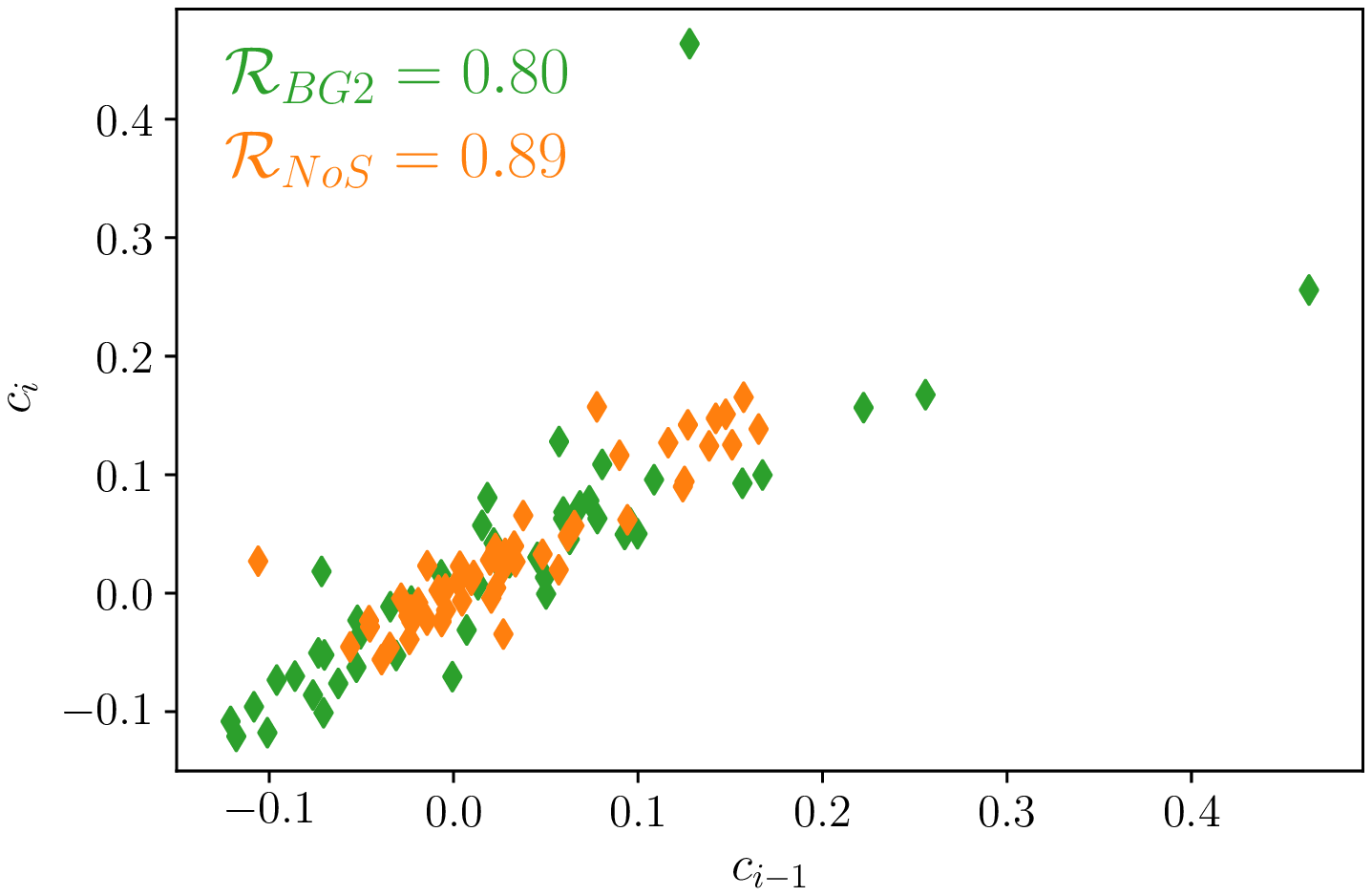} 
  \caption{\centering Lag plot of inversion coefficients of model A} 
  \label{fig_appendix_lag_plot_modelA}
\end{subfigure}
\begin{subfigure}[b]{.33\textwidth}
  \includegraphics[width=.99\linewidth]{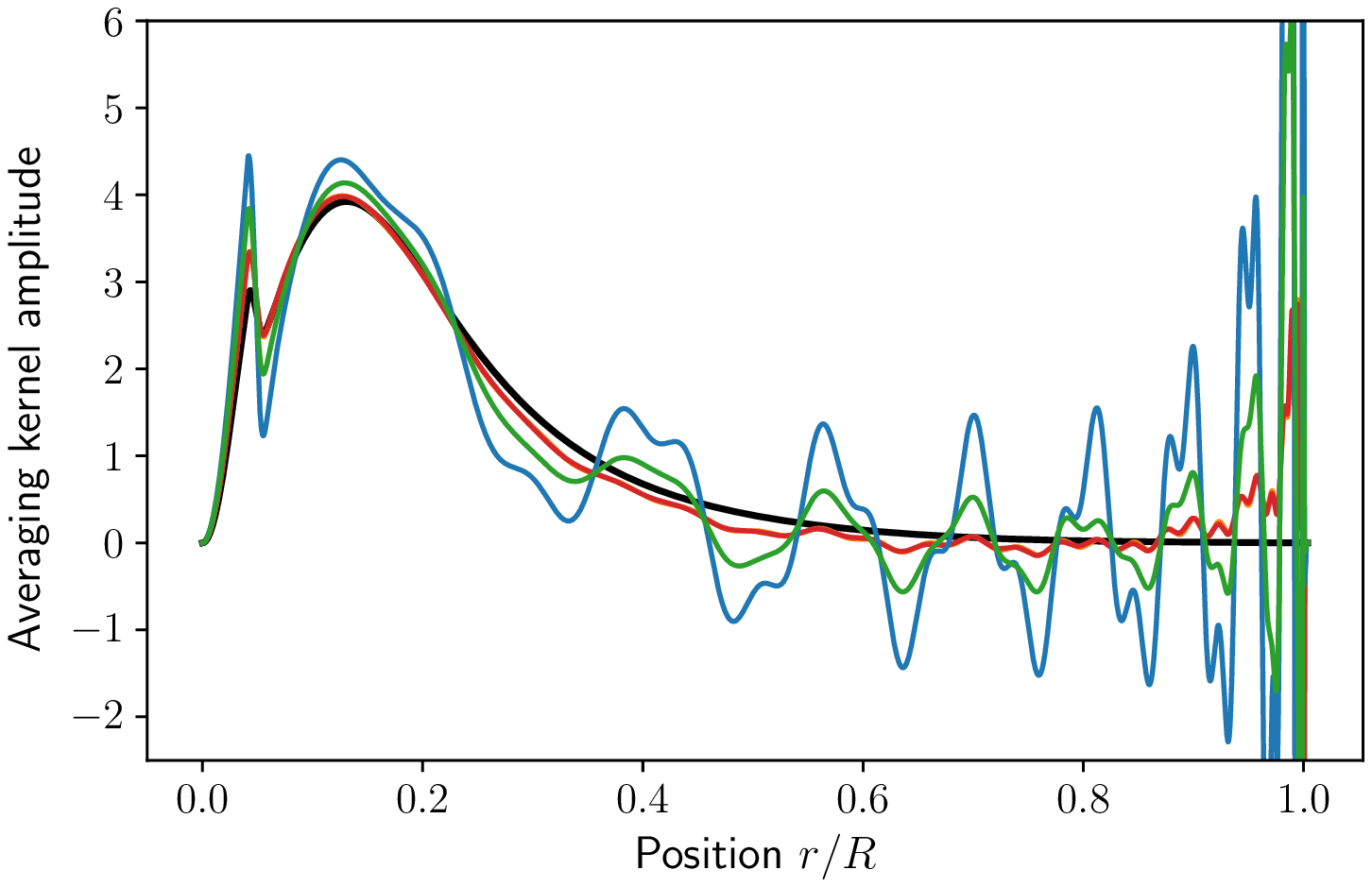} 
  \caption{\centering Averaging kernel of model E} 
  \label{fig_appendix_comp_Kavg_modelE}
\end{subfigure}
\begin{subfigure}[b]{.33\textwidth}
  \includegraphics[width=.99\linewidth]{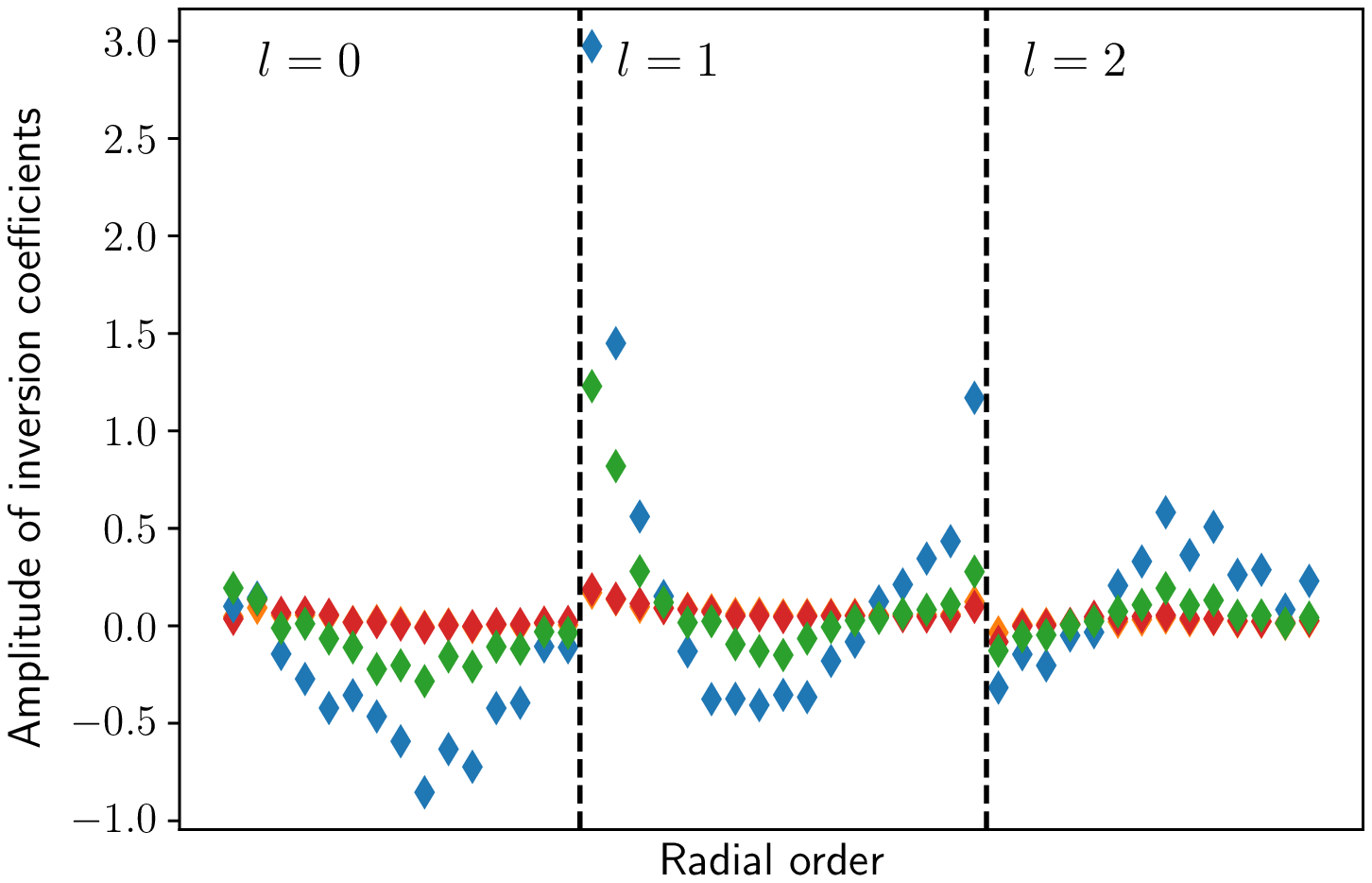}  
  \caption{\centering Inversion coefficients of model E} 
  \label{fig_appendix_comp_ci_modelE}
\end{subfigure}
\begin{subfigure}[b]{.33\textwidth}
  \includegraphics[width=.99\linewidth]{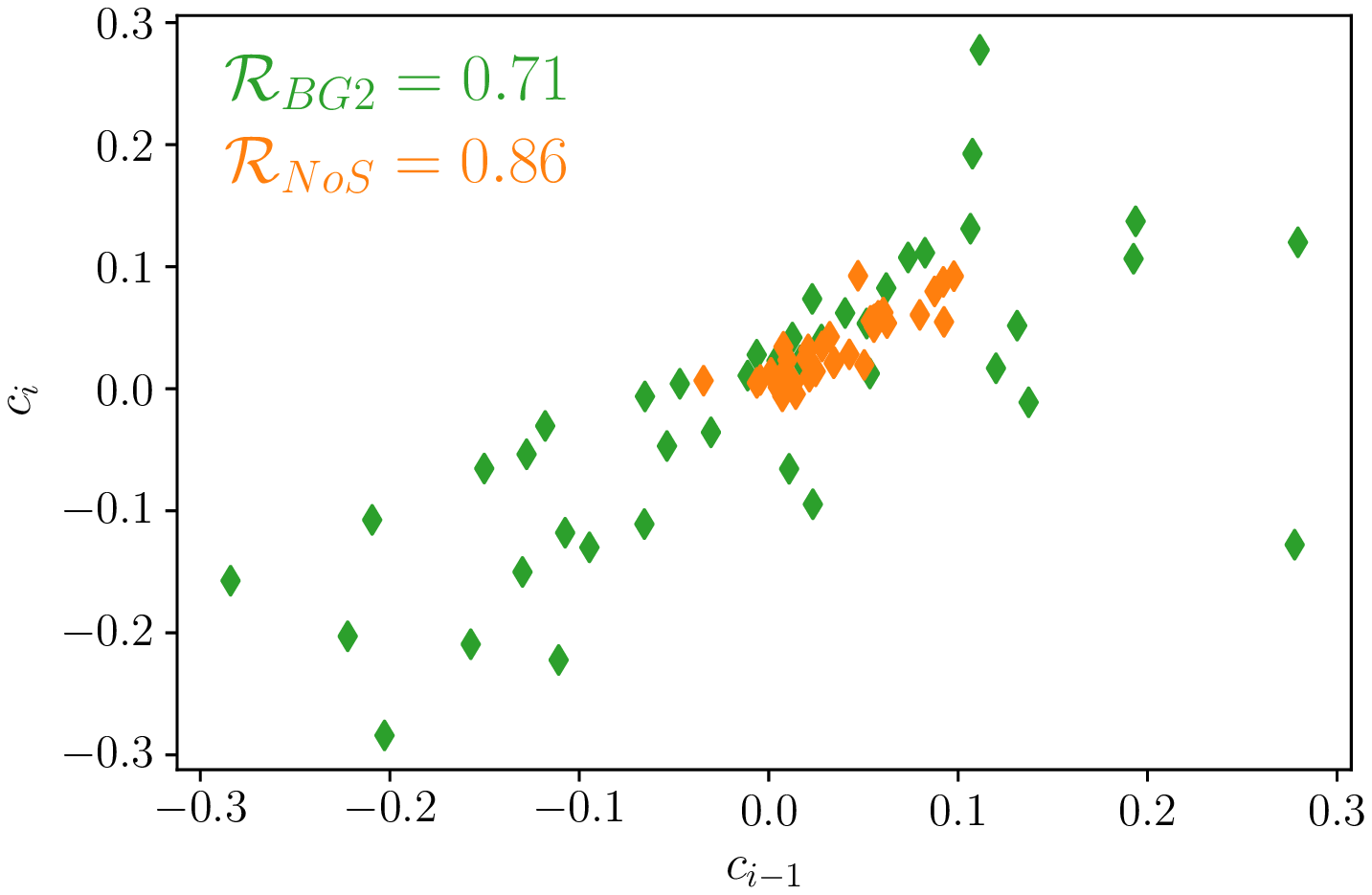} 
  \caption{\centering Lag plot of inversion coefficients of model E} 
  \label{fig_appendix_lag_plot_modelE}
\end{subfigure}
\begin{subfigure}[b]{.33\textwidth}
  \includegraphics[width=.99\linewidth]{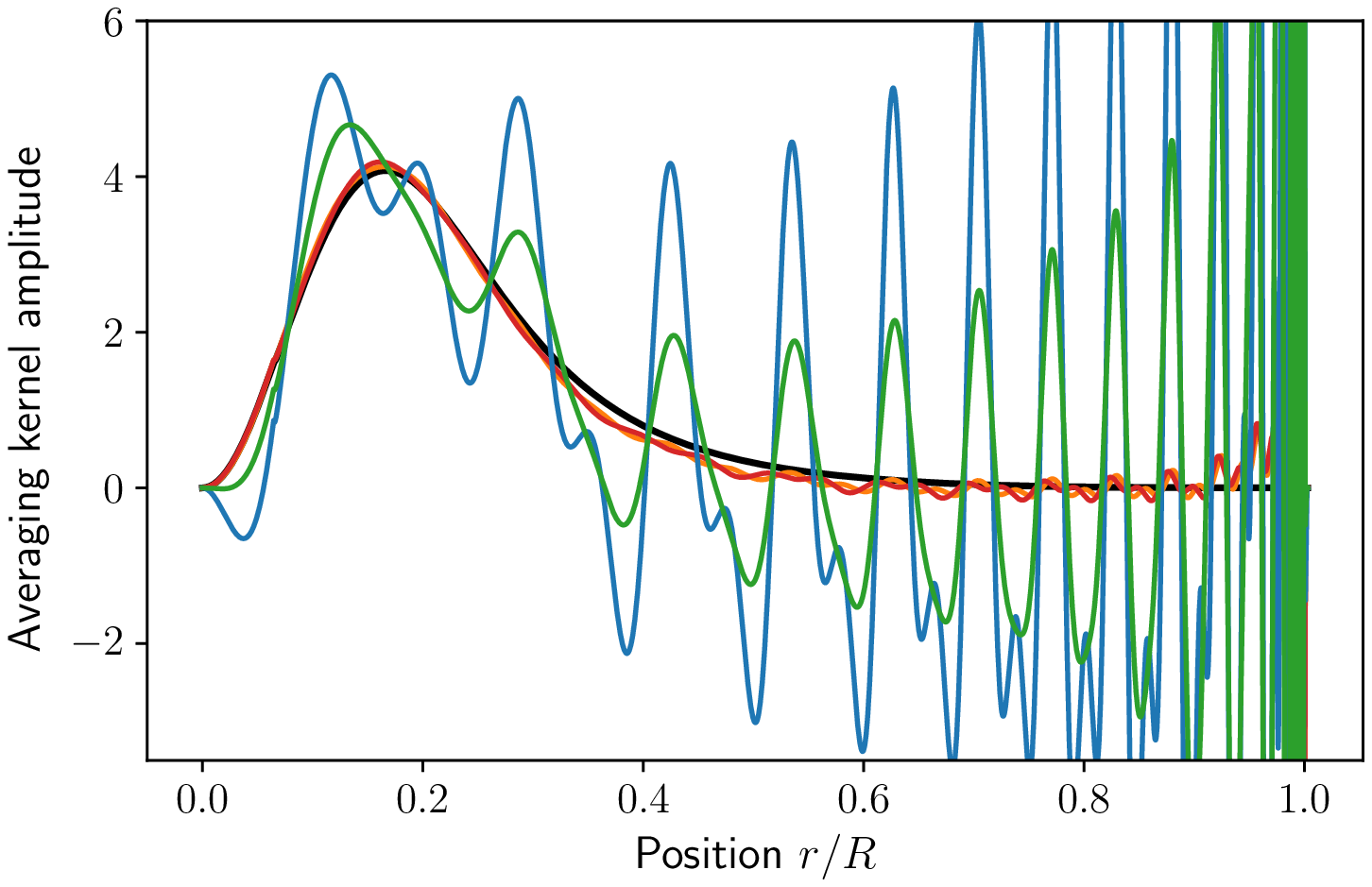}  
  \caption{\centering Averaging kernel of model B}
  \label{fig_appendix_comp_Kavg_modelB}
\end{subfigure}
\begin{subfigure}[b]{.33\textwidth}
  \includegraphics[width=.99\linewidth]{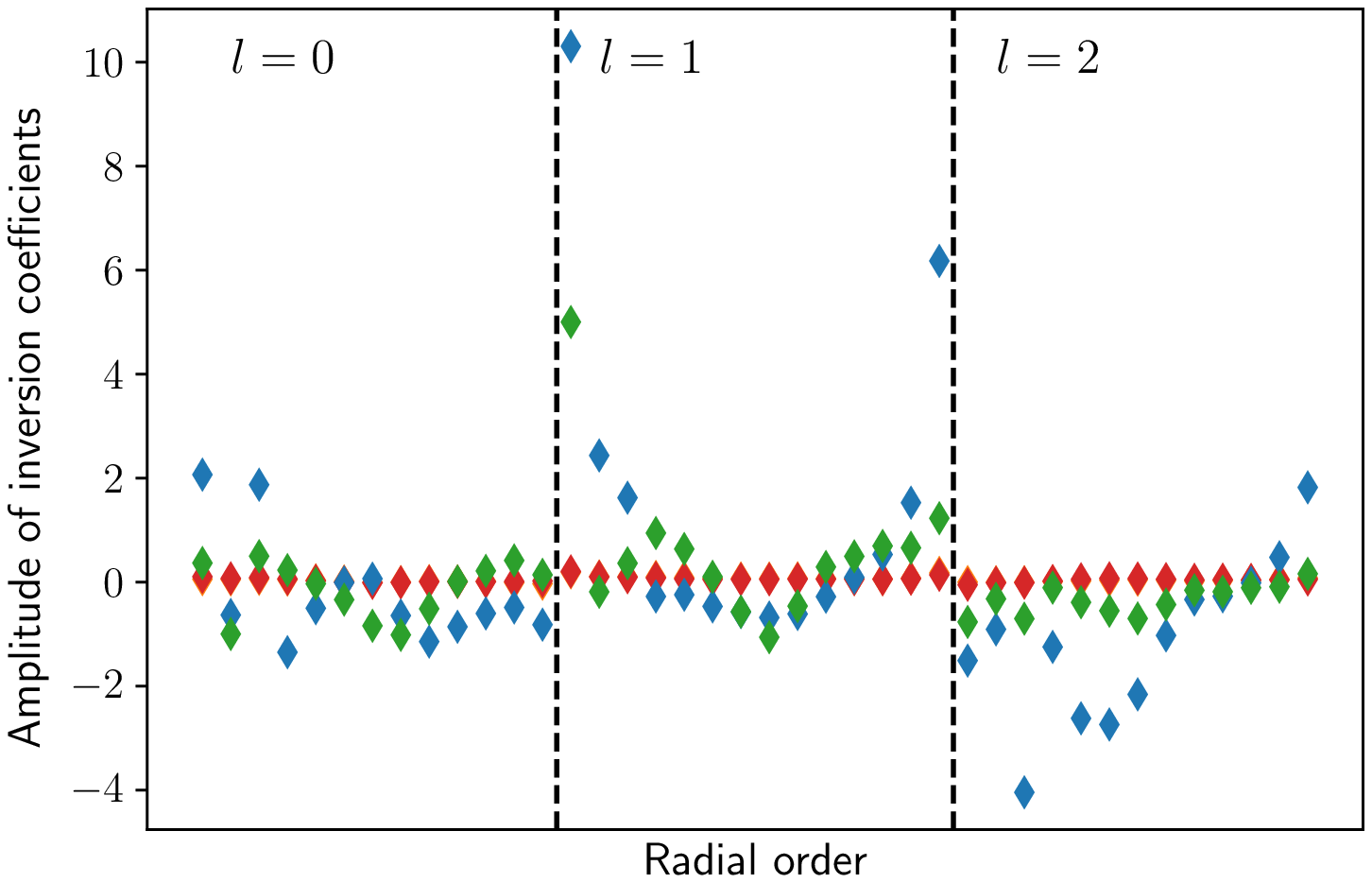}  
  \caption{\centering Inversion coefficients of model B} 
  \label{fig_appendix_comp_ci_modelB}
\end{subfigure}
\begin{subfigure}[b]{.33\textwidth}
  \includegraphics[width=.99\linewidth]{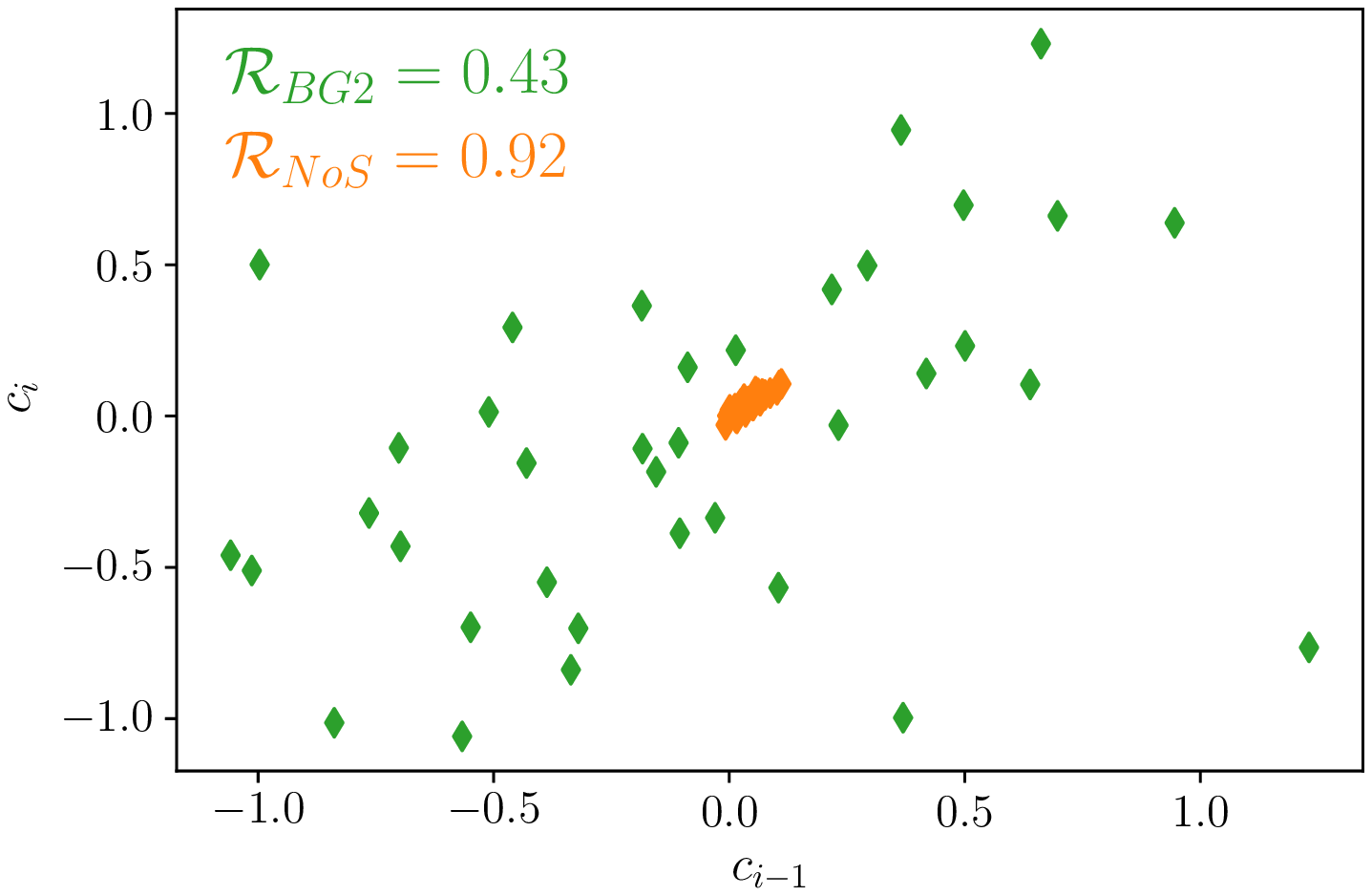} 
  \caption{\centering Lag plot of inversion coefficients of model B} 
  \label{fig_appendix_lag_plot_modelB}
\end{subfigure}
\begin{subfigure}[b]{.33\textwidth}
  \includegraphics[width=.99\linewidth]{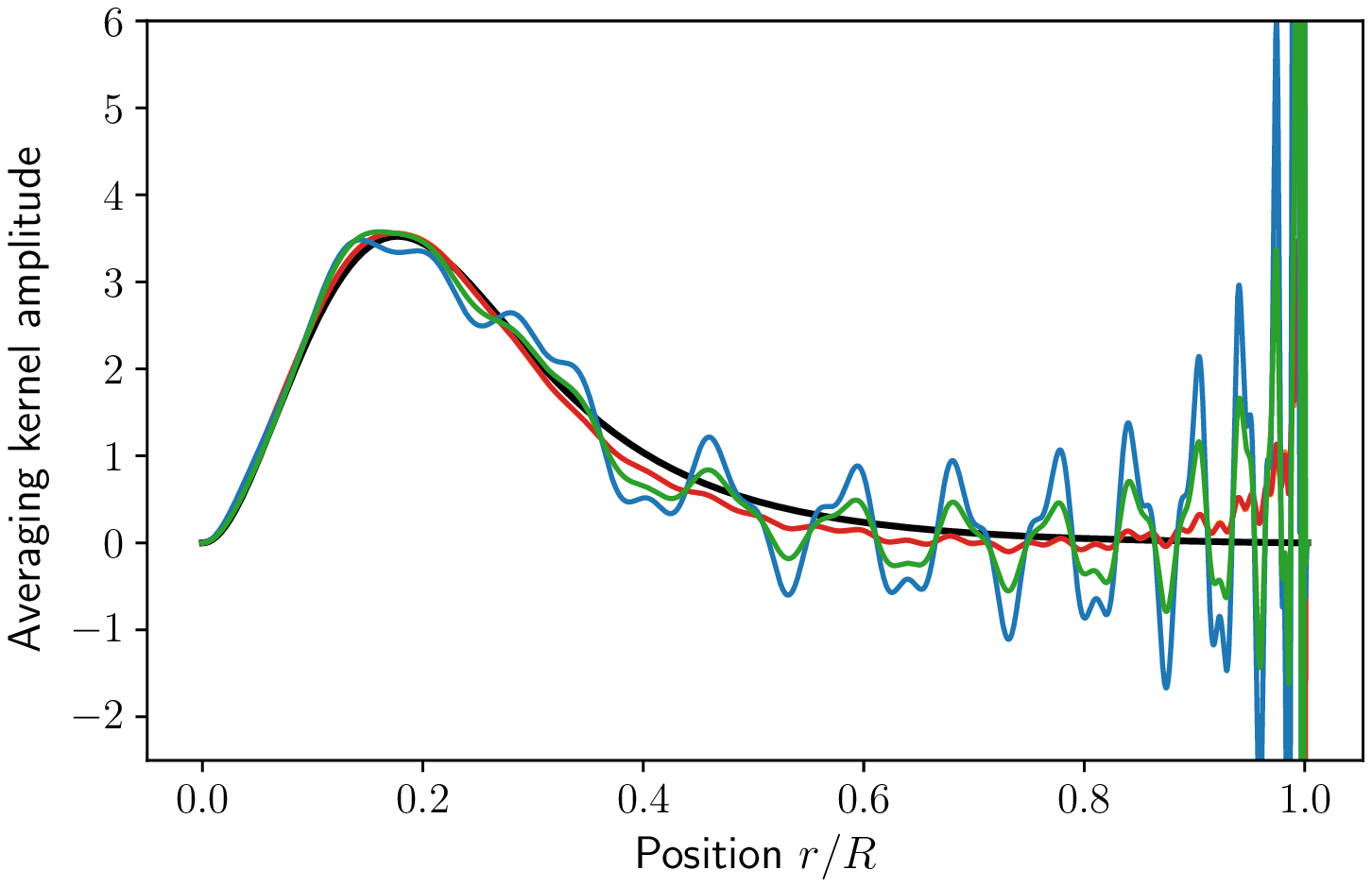}  
  \caption{\centering Averaging kernel of Saxo2}
  \label{fig_appendix_comp_Kavg_Saxo2}
\end{subfigure}
\begin{subfigure}[b]{.33\textwidth}
  \includegraphics[width=.99\linewidth]{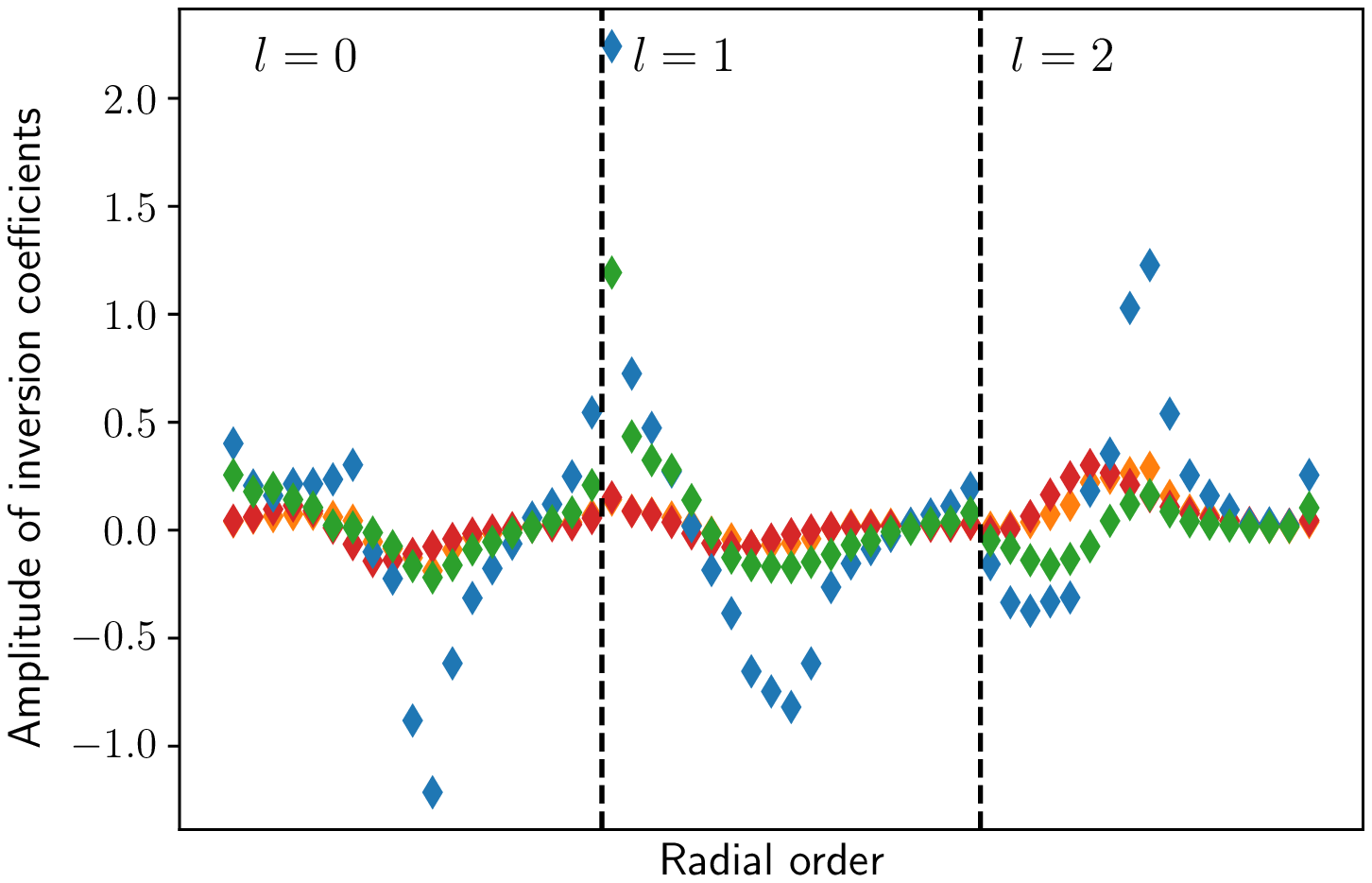}  
  \caption{\centering Inversion coefficients of Saxo2} 
  \label{fig_appendix_comp_ci_Saxo2}
\end{subfigure}
\begin{subfigure}[b]{.33\textwidth}
  \includegraphics[width=.99\linewidth]{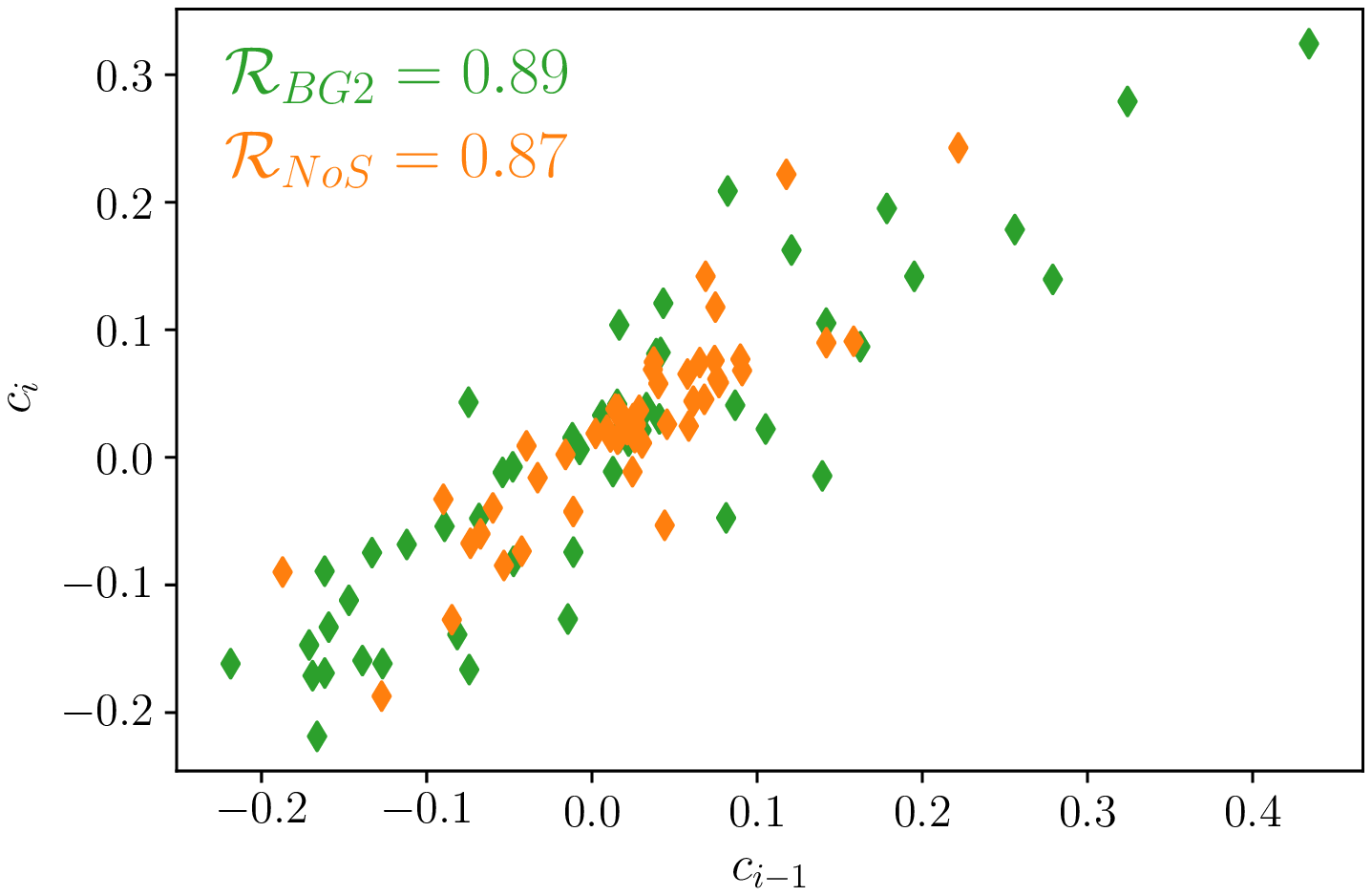} 
  \caption{\centering Lag plot of inversion coefficients of Saxo2} 
  \label{fig_appendix_lag_plot_Saxo2}
\end{subfigure}
\begin{subfigure}[b]{.33\textwidth}
  \includegraphics[width=.99\linewidth]{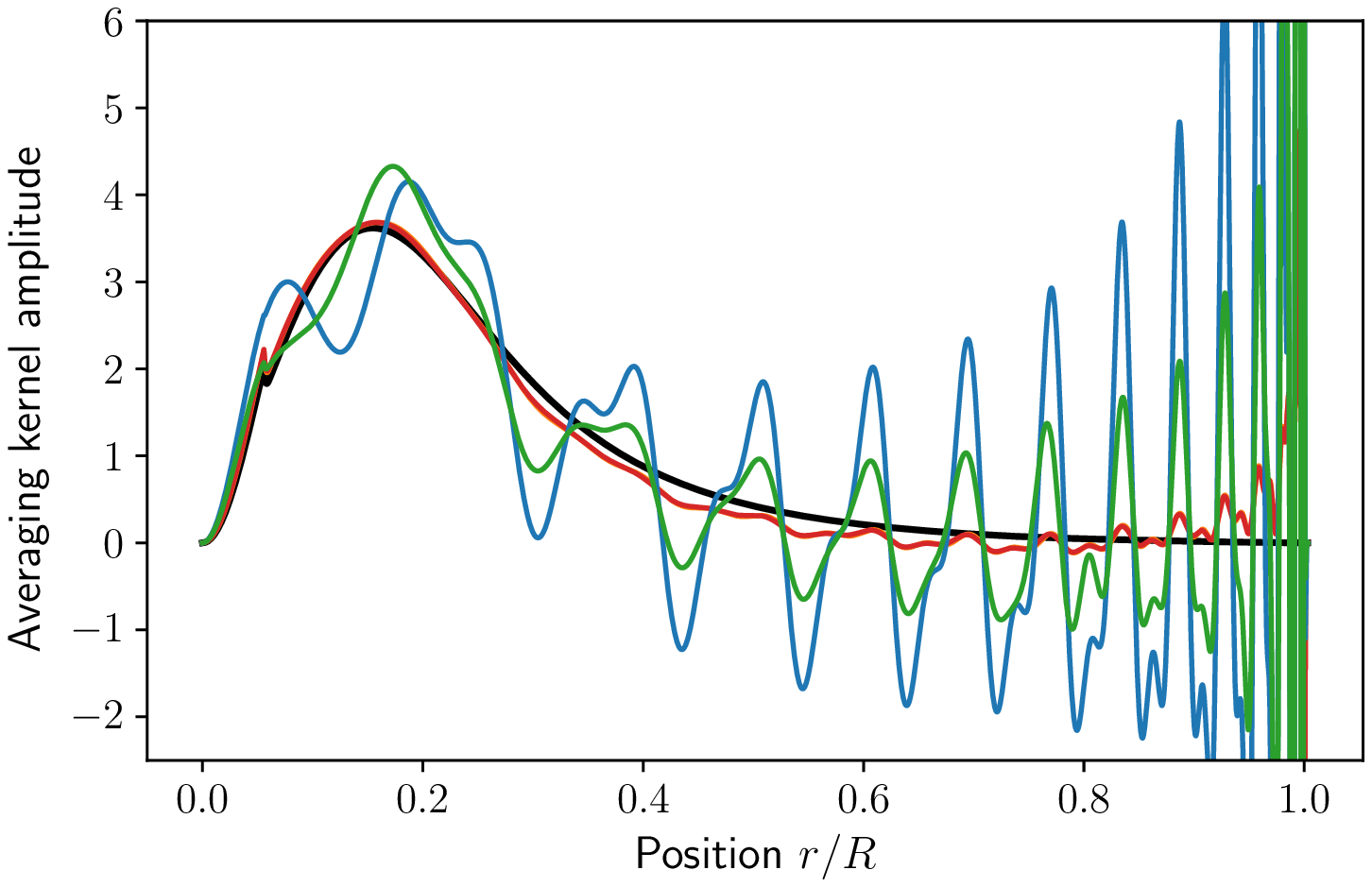}  
  \caption{\centering Averaging kernel of Dushera}
  \label{fig_appendix_comp_Kavg_Dushera}
\end{subfigure}
\begin{subfigure}[b]{.33\textwidth}
  \includegraphics[width=.99\linewidth]{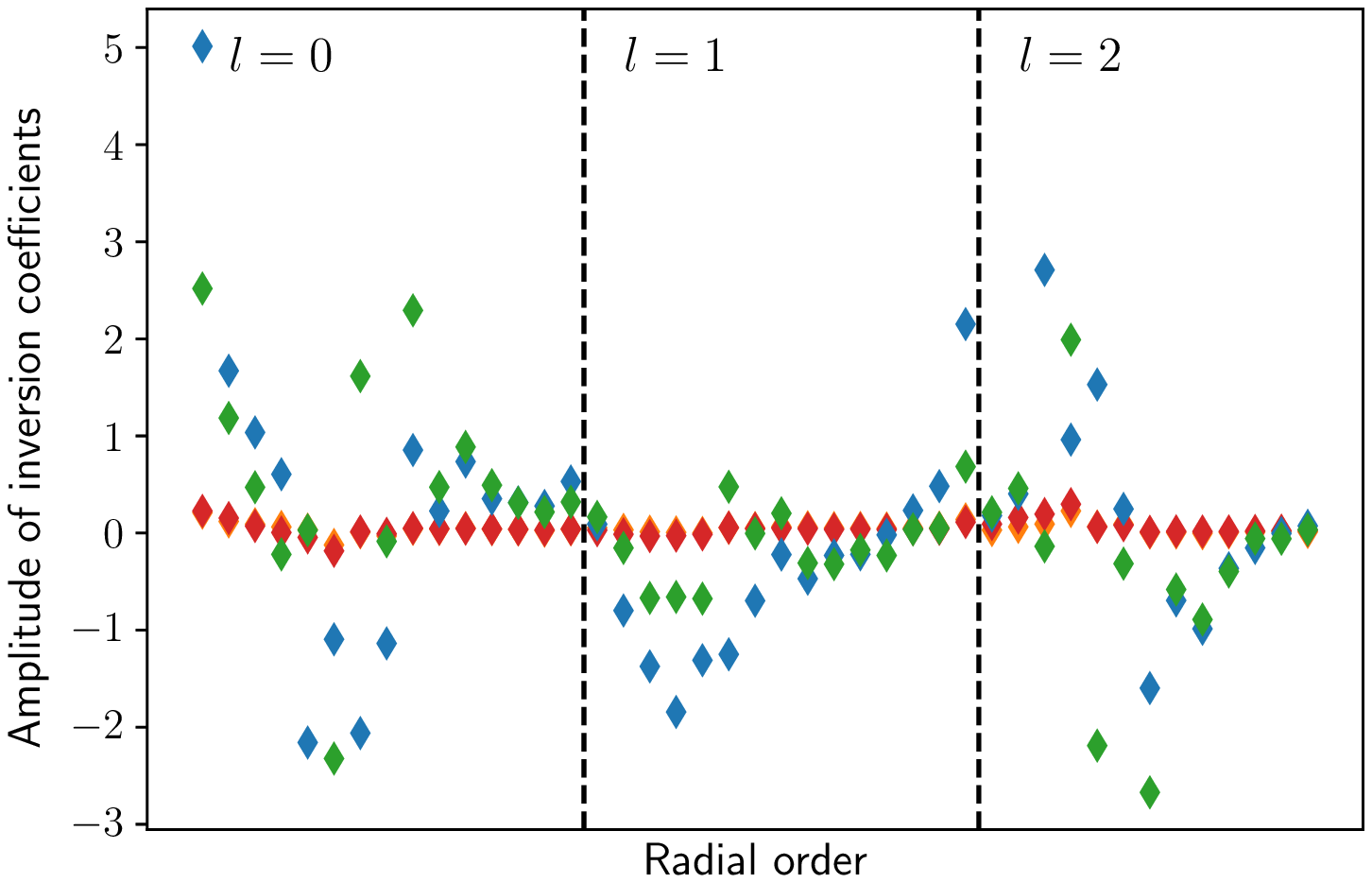}  
  \caption{\centering Inversion coefficients of Dushera} 
  \label{fig_appendix_comp_ci_Dushera}
\end{subfigure}
\begin{subfigure}[b]{.33\textwidth}
  \includegraphics[width=.99\linewidth]{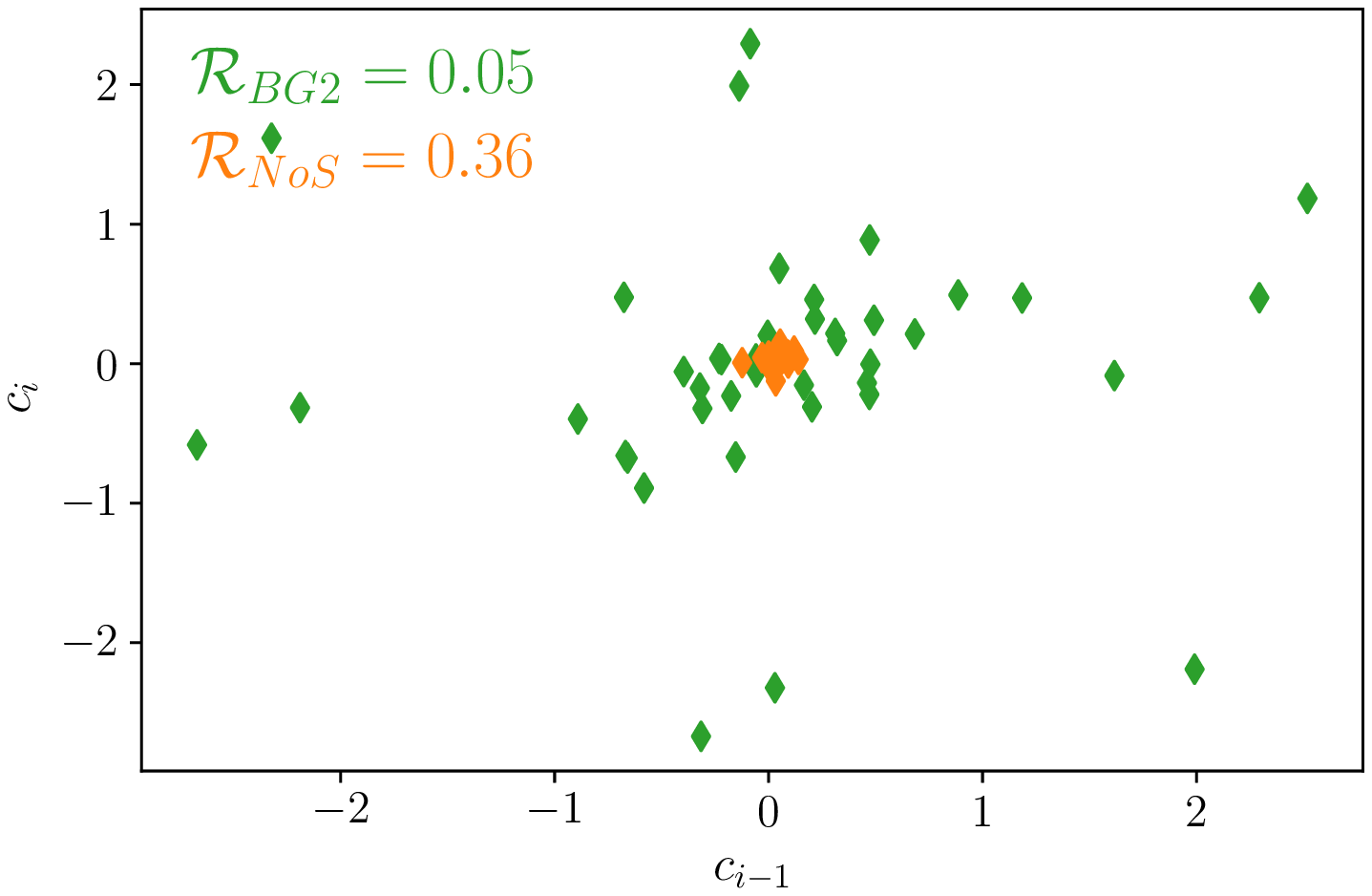} 
  \caption{\centering Lag plot of inversion coefficients of Dushera} 
  \label{fig_appendix_lag_plot_Dushera}
\end{subfigure}
\caption{Comparison of the averaging kernels (left column), inversion coefficients (middle column), and lag plots (right column) of models A, E and B, Saxo2, and Dushera by considering different implementations for the surface effects in the inversion. In orange, the surface effects are neglected (abbreviated as \textit{NoS}), in green and blue the surface effects are treated as free variables in the inversion in InversionKit (IK), respectively with the \textit{BG2} and \textit{S2} prescriptions, and in red, the frequencies are corrected before the inversion with the optimized coefficients from AIMS.}
\label{fig_appendix_comp_ci&Kavg}
\end{figure*}

Hence, verifying how the target function is reproduced by the averaging kernels constitutes a good visual test to assess how the inversion is behaving. In an effort of automation, one could be tempted to assess the quality of the inversion with the $L_2$ norm similarly to \citet{Backus&Gilbert1968}, \citet{Backus&Gilbert1970}, \citet{Pijpers&Thompson1994}, \citet{Rabello-Soares1999}, \citet{Reese2012}, or \citet{Buldgen2015b}:
\begin{align}
||K_{\mathrm{avg}}||_2^2 = \int_0^R \left(\mathcal{K_{\mathrm{avg}}} - \mathcal{T}_{\bar{\rho}} \right)^2 dr.
\end{align}
However, such an approach can only be trusted for inversions with reference models that have a target function with a similar amplitude. This condition was fulfilled in the papers that we quoted, but in our study, we analysed targets spreading across a large mass span. The amplitudes of the targets functions are therefore not comparable (see e.g. Fig. \ref{fig_appendix_comp_Kavg_modelA} and Fig. \ref{fig_appendix_comp_Kavg_modelE}) and the absolute value of the $L_2$ norm cannot be used as a quality indicator of the inversion. Hence, we constructed a new test based on the inversion coefficients. If the inversion behaves optimally, the coefficients form smooth structures, as illustrated in Fig. \ref{fig_appendix_comp_Kavg_modelA}. Hence, it is relevant to look at the autocorrelation of the coefficients because an instability in the inversion tends to destroy these structures. In such conditions, the coefficients seem to be more randomly distributed. To measure the autocorrelation, we produced the lag plot of the coefficients (with $\mathrm{lag}=1$), where the coefficients present a linear correlation, as illustrated in the third column of Fig. \ref{fig_appendix_comp_ci&Kavg}. Physically, we interpret this behaviour as a consequence that the frequencies are not fully independent. Indeed, the frequencies follow an asymptotic behaviour within the same harmonic degree, which implies that the seismic information contained in the frequencies may be redundant. This affects the inversion that picks a bit of the same seismic information in multiple frequencies, thus generating the smooth structures in the inversion coefficients. The linear correlation of the coefficients observed with the lag plot is likely related to the linear formalism at the basis of the inversion. To quantify the degree of instability of the inversion, we used the Pearson correlation coefficient $\mathcal{R}$, a low value corresponding to a high degree of instability. We identified three instability regimes, high ($\mathcal{R}<0.5$), intermediate ($0.5<\mathcal{R}<0.75$), and low ($0.75<\mathcal{R}$). If $\mathcal{R}<0.5$ further investigations are required. The boundaries of the different regimes are empirical, and were determined based on our limited sample of 16 targets, on the analysis of the averaging kernels, on the lag plots, and on our experience of inversions. We also point out that these regimes were identified for mean density inversions and that further investigations should be conducted for other types of inversions. From a pipeline perspective, we would however recommend to define a unique threshold below which we reject the result of the inversion, in our case at $\mathcal{R}\sim 0.6$, and refine this threshold with a larger statistics. 

In Fig. \ref{fig_appendix_comp_ci&Kavg}, we illustrate models representative of the different instability regimes. The three first rows correspond to results for synthetic models, respectively models A, E, and B, while the last two rows corresponds to \emph{LEGACY} targets, respectively Saxo2 and Dushera. Model A and Saxo2 are representative of a robust inversion. The instability is low and the target function is well reproduced independently of the surface prescription that we considered. Model E is representative of an inversion in the intermediate regime. The target function is less well reproduced, especially at the surface, but the main features of the central regions are still captured by the inversion. We note that the \textit{S2} prescription is more unstable than the \textit{BG2} prescription. Finally, model B and Dushera are representative of high-instability inversions. The target function is poorly reproduced, the central features are missed, and the amplitude of the averaging kernel diverges at the surface. In the lag plot, the coefficients including surface effects (in green) are significantly different from the coefficients that do not include them (in orange). In such conditions, the inversion could not see the structural differences and therefore did not correct the reference mean density. Hence, using the results of an unstable inversion amounts admitting that the mean density of the reference model is robust, which is not an unreasonable assumption since it comes from an MCMC run in a grid. However, some caution should be considered because an unstable inversion can also provide a non-negligible correction of the mean density, which would in that case be a numerical artefact resulting from the poor fit of the target function by the averaging kernel. In our study, we identified four targets in the high-instability regime, models B, D, F, and Dushera, and a numerical artefact only affected the results of model F. To test the impact of using inversion coefficients in the high instability regime, we did not discard these targets, and point out that the conservative precision that we adopted when treating the inverted mean density as a classical constraint accounts, at least partially, for this kind of systematics.

\begin{figure}[t]
  \includegraphics[width=.45\textwidth]{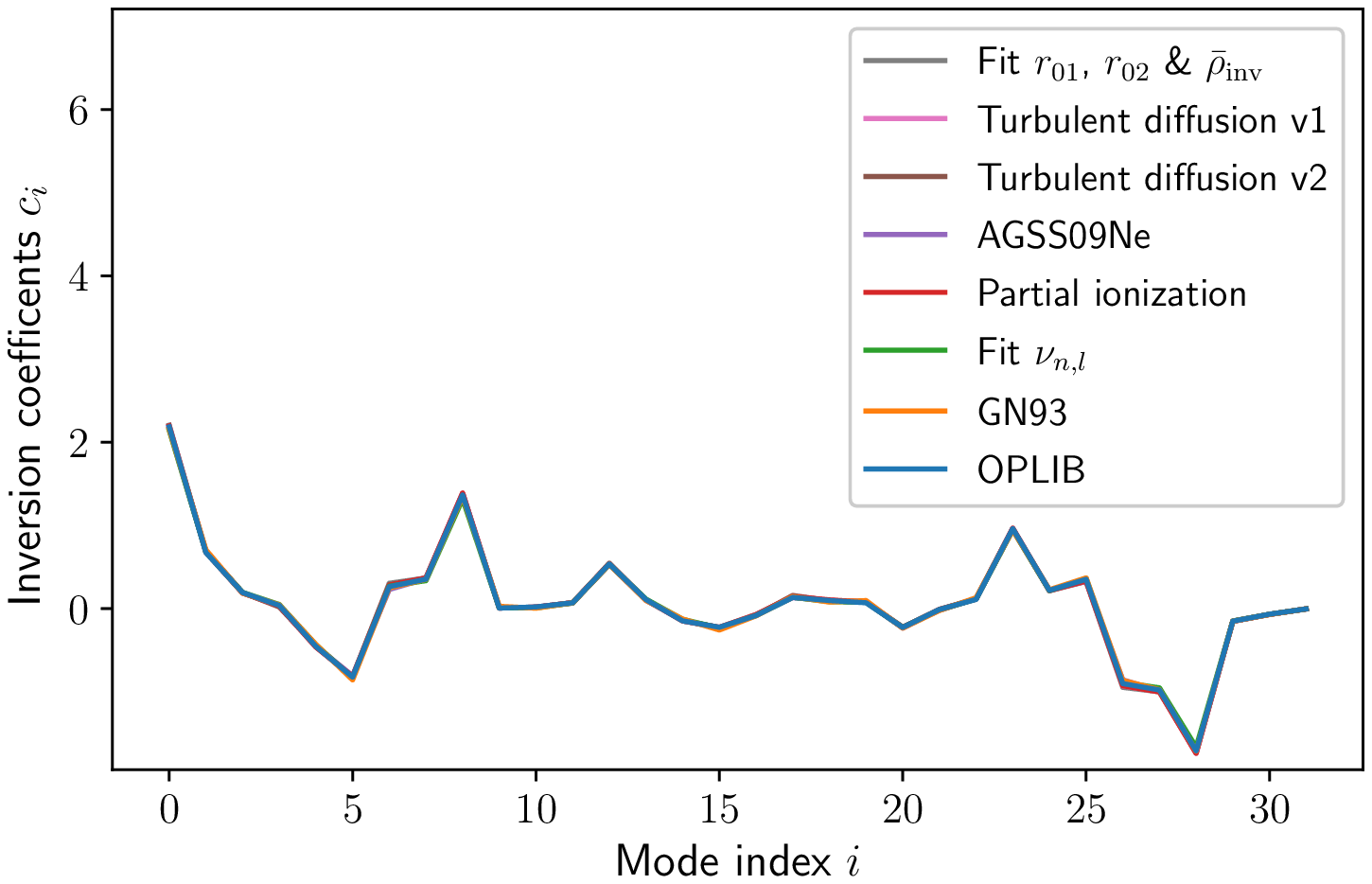}  
  \caption{Inversion coefficients of the models of Kepler-93 from \citet{Betrisey2022}, for which various physical ingredients were considered. The differences between the coefficients of the different models are very small. Hence, the lines in this figure are nearly-indistinguishable.}
   \label{fig_inversion_coefficients_Kepler93}
\end{figure}

In Fig. \ref{fig_inversion_coefficients_Kepler93}, we show the inversion coefficients of the models of Kepler-93 from \citet{Betrisey2022}, which include different sets of physical ingredients. Changing the physics shifts slightly the position of the global minimum in the parameter space. Hence, these models generate a scatter of close models in a confined region of the parameter space. The MCMC generates a scatter of close models in a confined region of the parameter space with a random walk algorithm. Although the form of the scatter is different in these two cases, the assumption of constant coefficients only requires the models to be close enough in the parameter space to be valid. As illustrated in Fig. \ref{fig_inversion_coefficients_Kepler93}, the variations between the coefficients of Kepler-93 are negligible. It is therefore reasonable to assume constant coefficients for all the MCMC steps. In Table~\ref{tab_tests_correlations}, we show the exact and approximated correlations of the toy model, computed with Eqs. \eqref{eq_s_exact} and \eqref{eq_s_approx} respectively. The differences are very small, about 0.3\% on average, and are therefore negligible compared to other sources of uncertainty. If the set of constraints is changed, it would invalidate our assumption. Indeed, a different frequency set, even by removing or adding only one frequency, would change significantly the inversion coefficients. This could happen if one of the observed frequency is not in some of the precomputed frequency sets of the grid. To avoid this issue, we computed extended frequency ranges in our grid, including low and high order modes that are currently not observable.

\begin{table}[t!]
\centering
\caption{Approximate and exact correlations between the inverted mean density and the $r_{02}$ ratios for the toy model.}
\begin{tabular}{lcc}
\hline 
$n$ & $\mathrm{corr}_{\bar{\rho}_{\mathrm{inv}},r_{02}(n)}^{	\mathrm{approx}}$ & $\mathrm{corr}_{\bar{\rho}_{\mathrm{inv}},r_{02}(n)}^{\mathrm{exact}}$ \\ 
\hline \hline
11 &  0.111971 &  0.112293 \\
12 &  0.075914 &  0.076130 \\
13 &  0.072055 &  0.072263 \\
14 &  0.047814 &  0.047955 \\
15 &  0.009646 &  0.009666 \\
16 & -0.004499 & -0.004491 \\
17 & -0.026609 & -0.026703 \\
18 & -0.032226 & -0.032323 \\
19 & -0.064974 & -0.065117 \\
20 & -0.075464 & -0.075570 \\
21 & -0.085204 & -0.085319 \\
22 & -0.098082 & -0.098188 \\
23 & -0.073656 & -0.073620 \\
24 & -0.069512 & -0.069540 \\
25 & -0.052201 & -0.052204 \\
26 & -0.021957 & -0.021940 \\
27 & -0.017685 & -0.017678 \\
28 & -0.013870 & -0.013856 \\
29 & -0.005165 & -0.005139 \\
30 &  0.003003 &  0.002978 \\
31 &  0.009338 &  0.009505 \\
32 &  0.004755 &  0.004769 \\
\hline
\end{tabular} 
{\par\small\justify\textbf{Notes.} The toy model is Model~S from \citet{JCD1996} and observational data is from \citet{Lazrek1997}. \par}
\label{tab_tests_correlations}
\end{table}

\section{Supplementary data for the \citet{Sonoi2015} targets}
\label{appendix_additional_data_takafumi}

\begin{figure}[t!]
\centering
\includegraphics[scale=0.4]{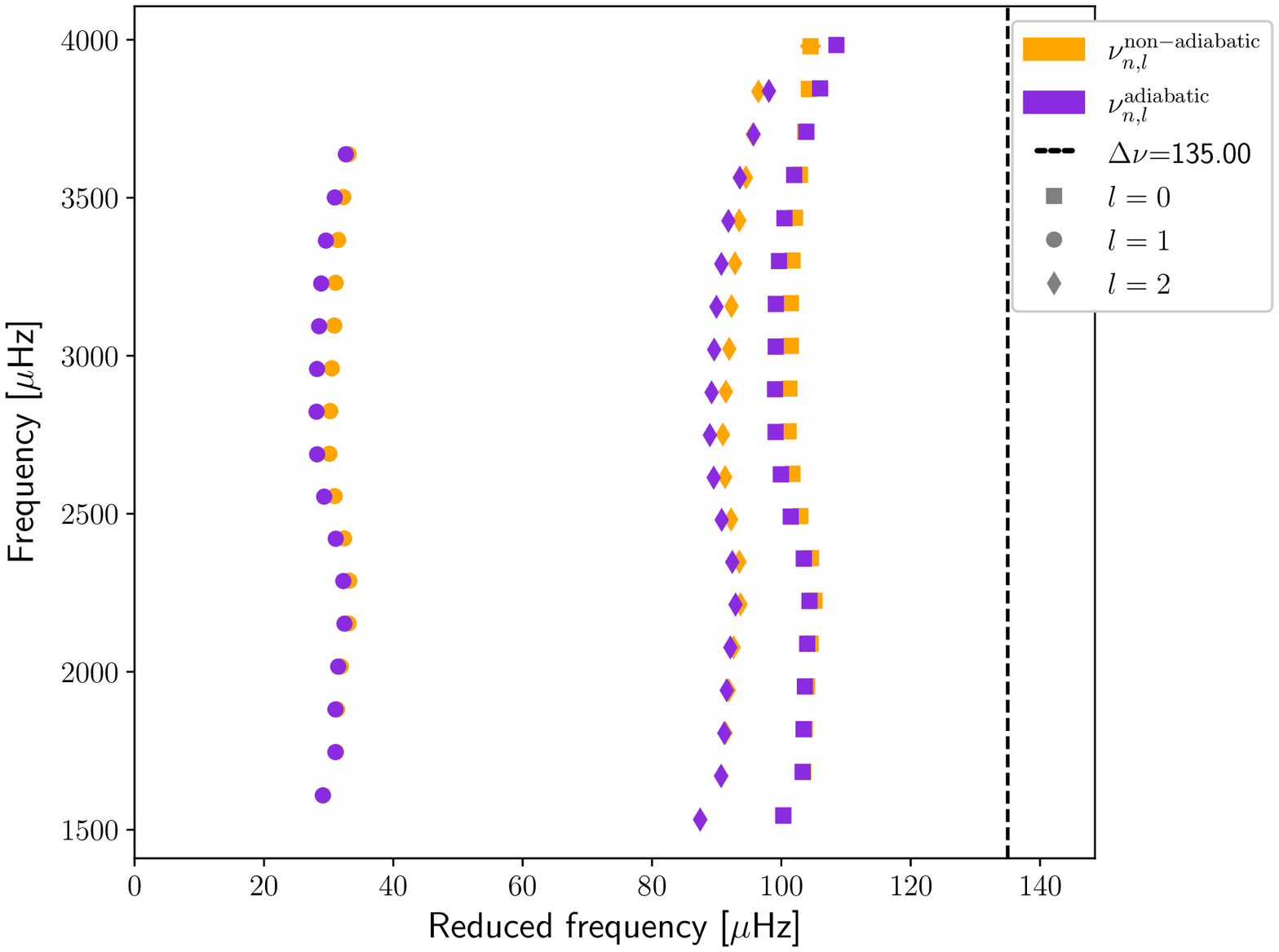}
\caption{Impact of the non-adiabatic effects on the individual frequencies of model A.}
\label{fig_ModelA_nad}
\end{figure}

\begin{figure}[t!]
\centering
\includegraphics[scale=0.4]{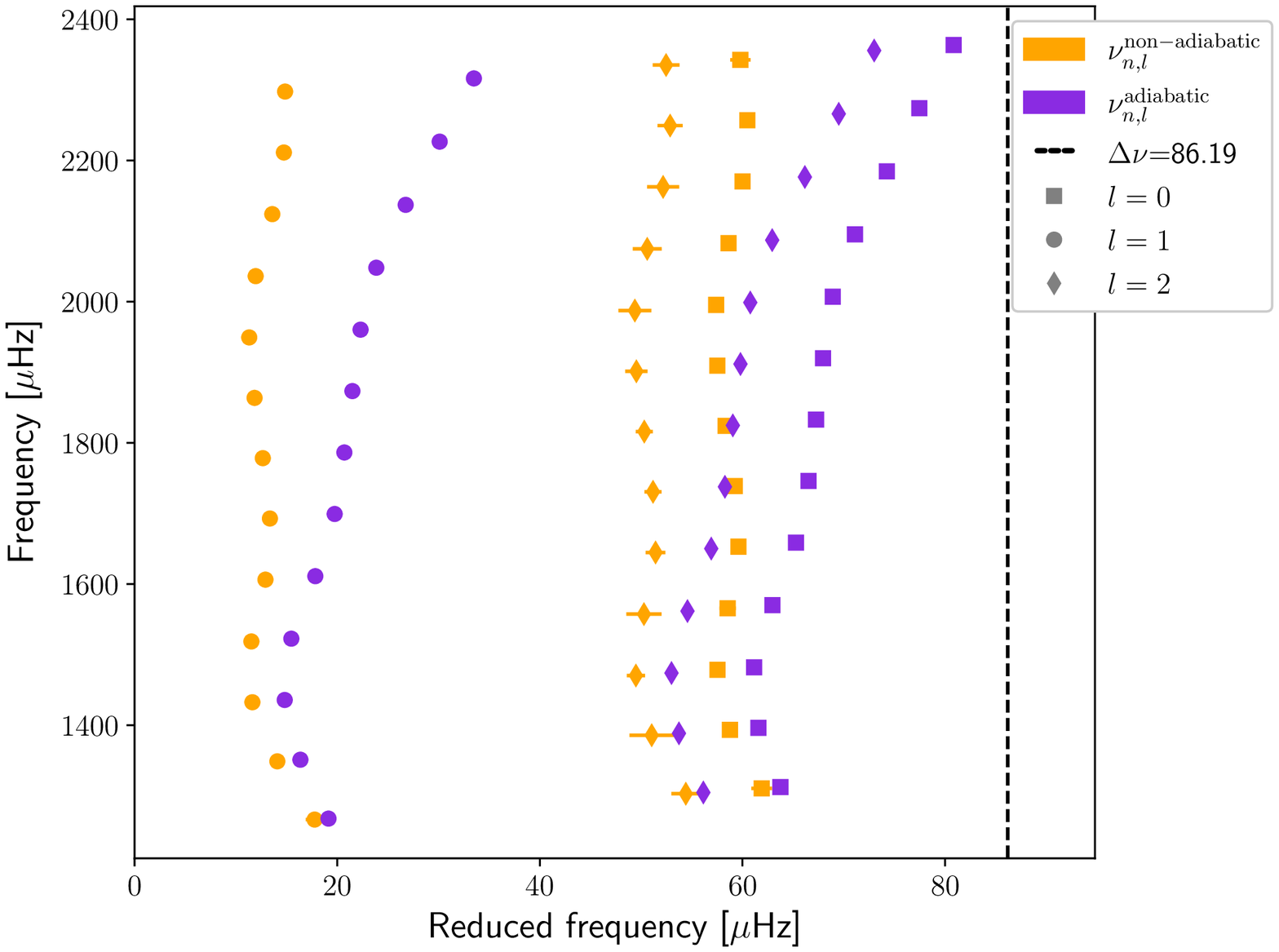}
\caption{Impact of the non-adiabatic effects on the individual frequencies of model B.}
\label{fig_ModelB_nad}
\end{figure}

In Figs. \ref{fig_ModelA_nad} and \ref{fig_ModelB_nad}, we show how the non-adiabatic effects impact the individual frequencies. The adiabatic part of the frequencies comes from a 3D simulation of the upper stellar layers patching a 1D model. For the solar model (model A), the non-adiabatic correction estimated by MAD is small, of the order of a few $\mu$Hz, while for model B, which is a higher mass star, the impact is significant, up to 20 $\mu$Hz for the highest order frequencies. Models C, D, E and F have corrections with magnitudes similar to model B. For all the models, the non-adiabatic correction is significantly larger than the observational uncertainties, up to more than an order of magnitude larger in the most extreme cases. We checked that the large separation was correctly estimated by the non-adiabatic oscillation code. Whether such large corrections are physically realistic is beyond the scope of this study and would require further investigations.

\begin{figure}[t!]
\centering
\includegraphics[scale=0.4]{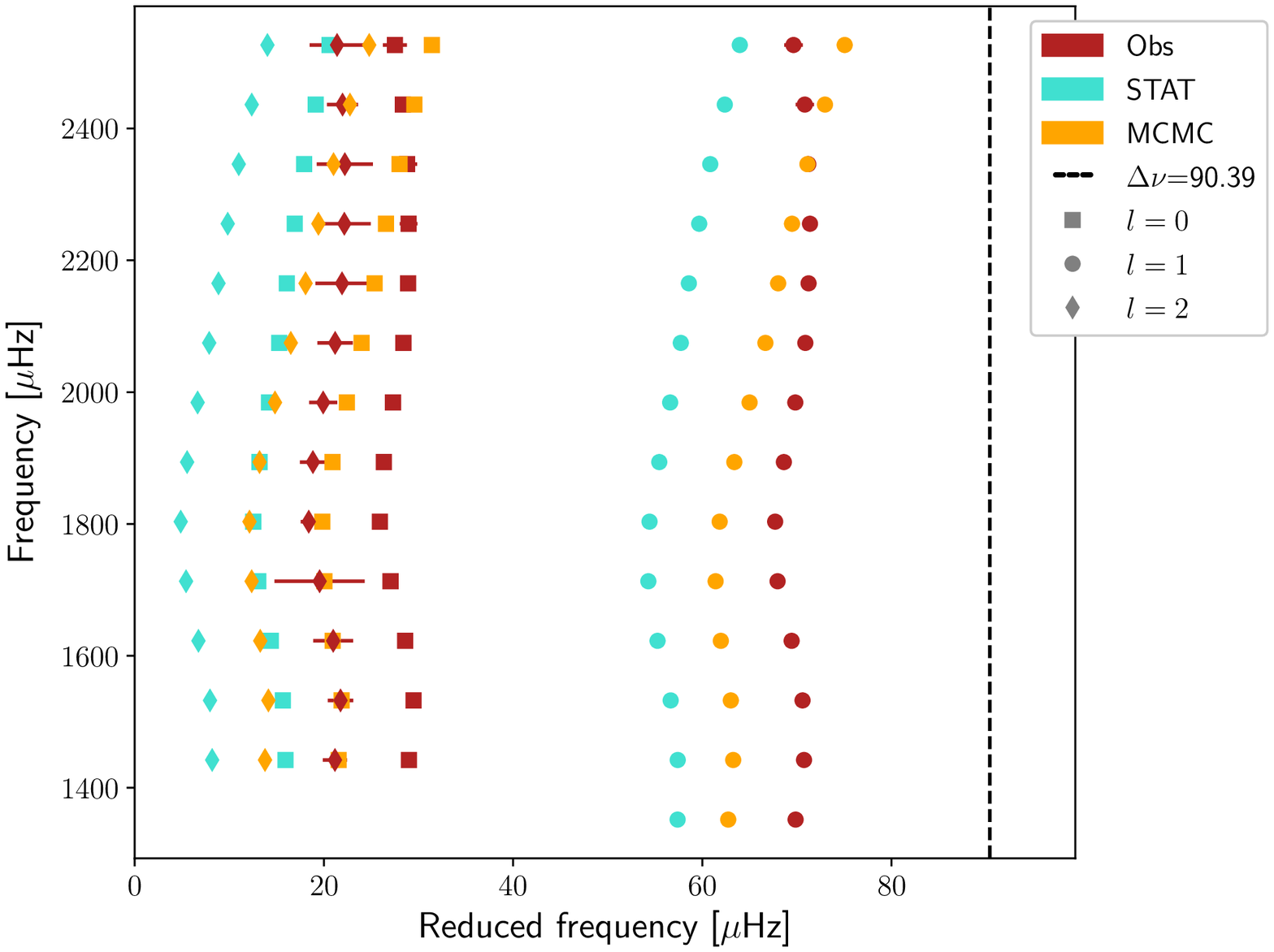}
\caption{Illustration of an imperfect anchoring of the frequency ridges. The observed frequencies are shown in red. The cyan frequencies correspond to the model based on the median of the posterior distributions of the MCMC run and the orange model to the best MCMC model, which minimises the $\chi^2$.}
\label{fig_ModelD_anchoring}
\end{figure}

In Fig. \ref{fig_ModelD_anchoring}, we show an illustration of an imperfect anchoring. In this example, the inverted mean density was part of the constraints but proved to be insufficient to perfectly anchor the frequency ridges. Such an offset in the échelle diagram implies that the stellar parameters are slightly biased. In this example, the offset is small and the stellar parameters are therefore not significantly affected. We note that adding the frequency of lowest radial order in that case leads to walker drifts that were more problematic than the imperfect anchoring.

\begin{figure*}[htp!]
  \centering
\begin{subfigure}[b]{.43\textwidth}
  \includegraphics[width=.99\linewidth]{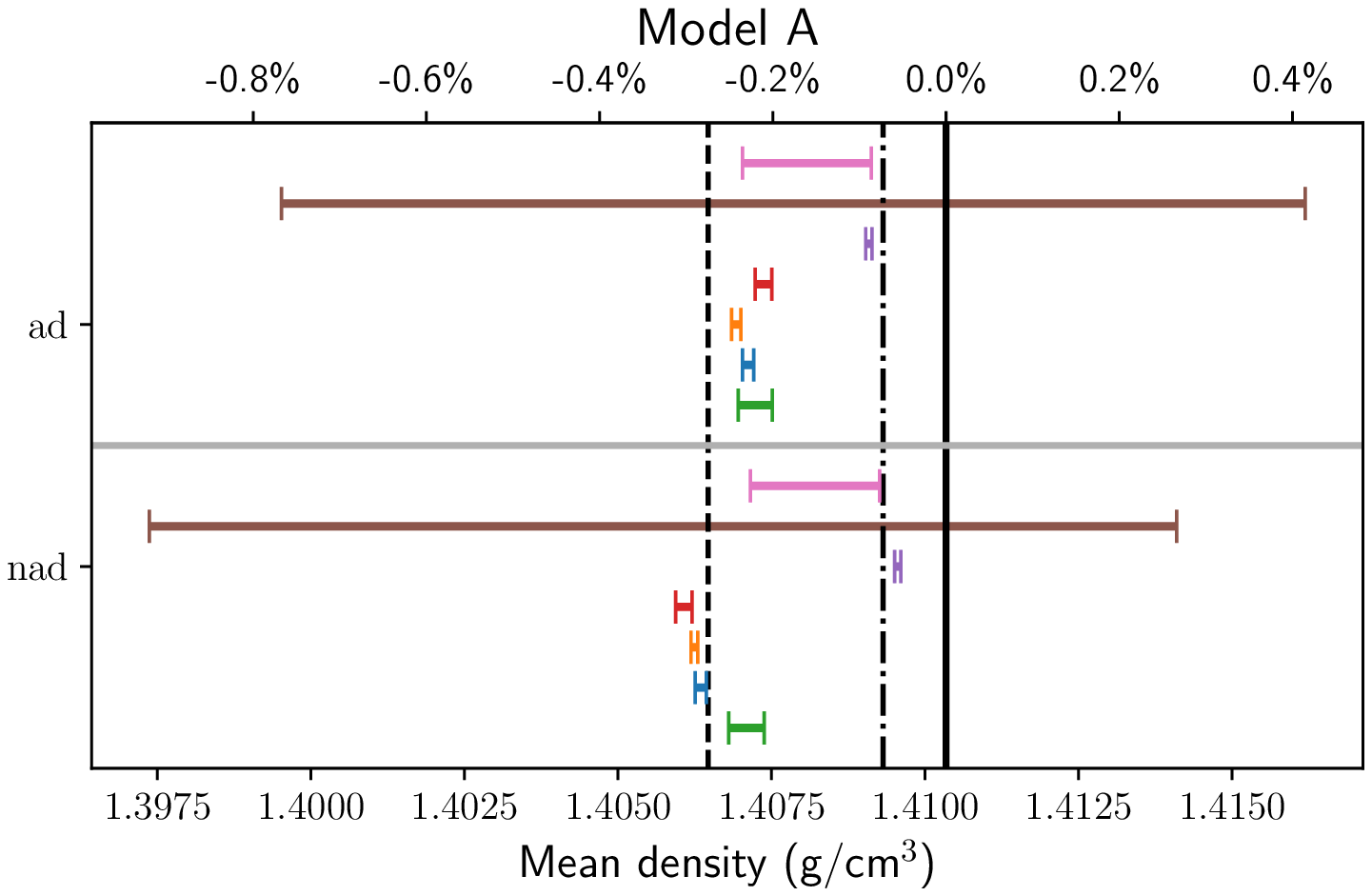}  
  \label{fig_appendix_Takafumi_comp_mean_density_modelA}
\end{subfigure}
\begin{subfigure}[b]{.43\textwidth}
  \includegraphics[width=.99\linewidth]{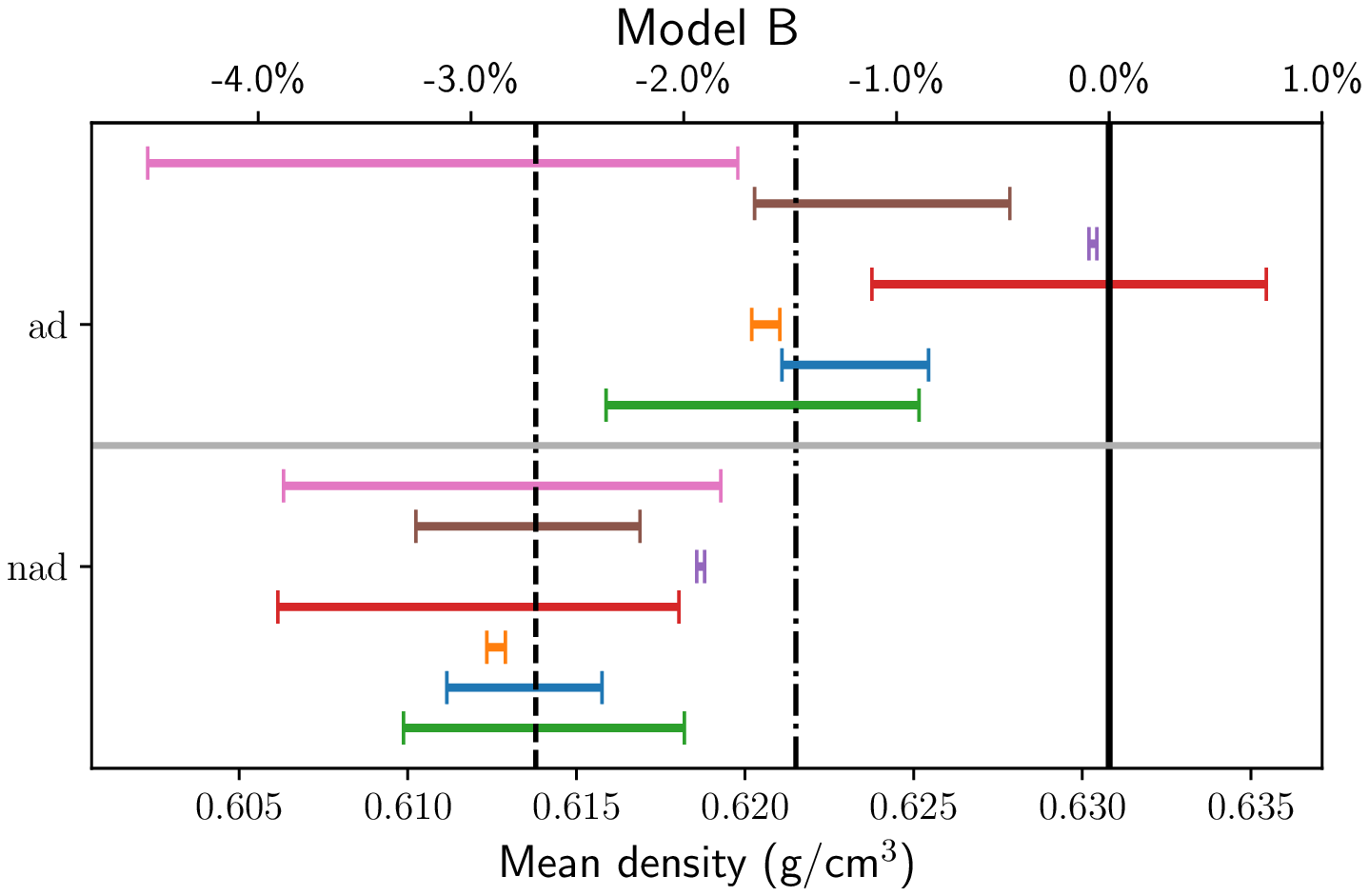} 
  \label{fig_appendix_Takafumi_comp_mean_densityi_modelB}
\end{subfigure}
\begin{subfigure}[b]{.43\textwidth}
  \includegraphics[width=.99\linewidth]{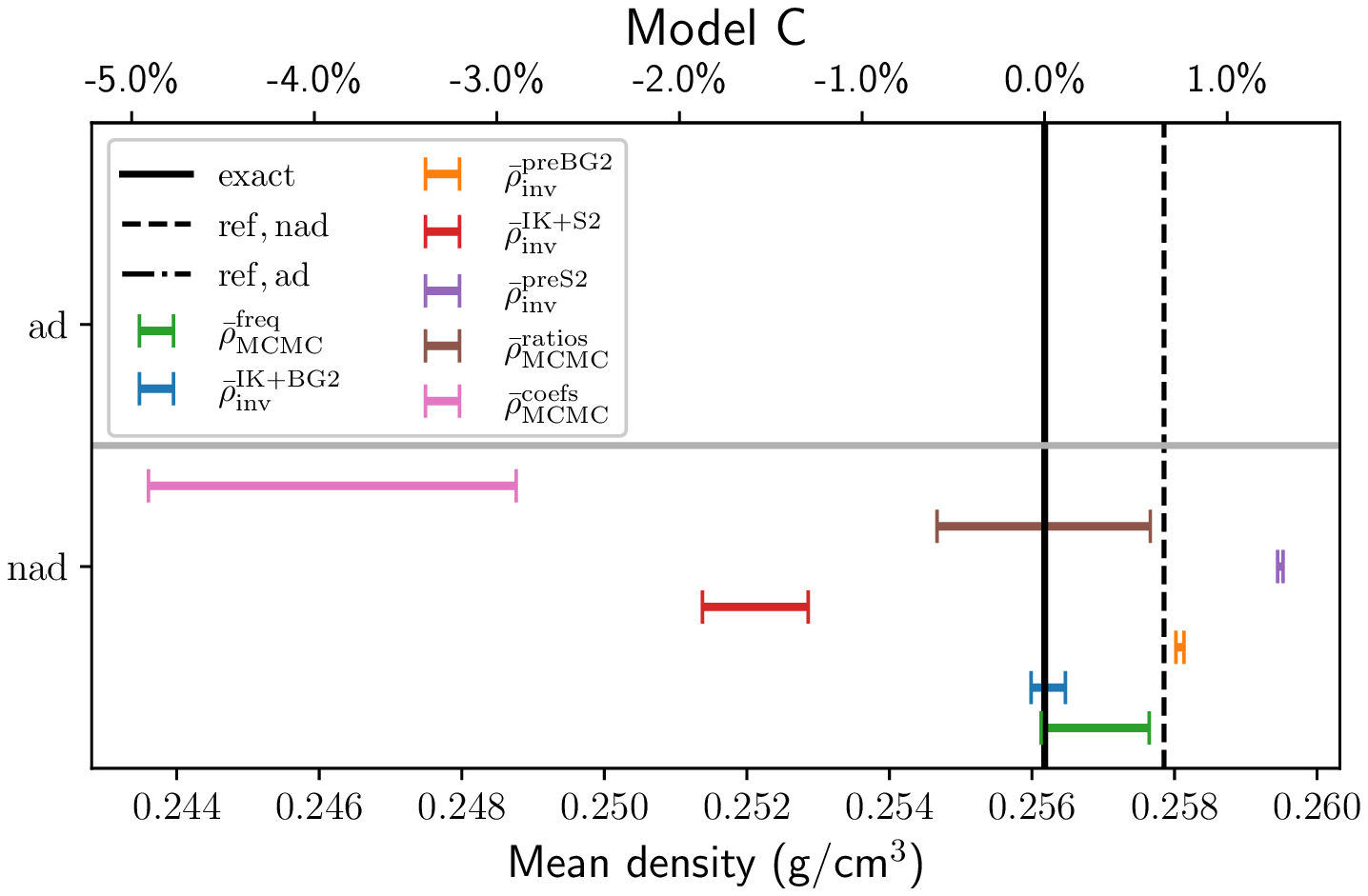} 
  \label{fig_appendix_Takafumi_comp_mean_density_modelC}
\end{subfigure}
\begin{subfigure}[b]{.43\textwidth}
  \includegraphics[width=.99\linewidth]{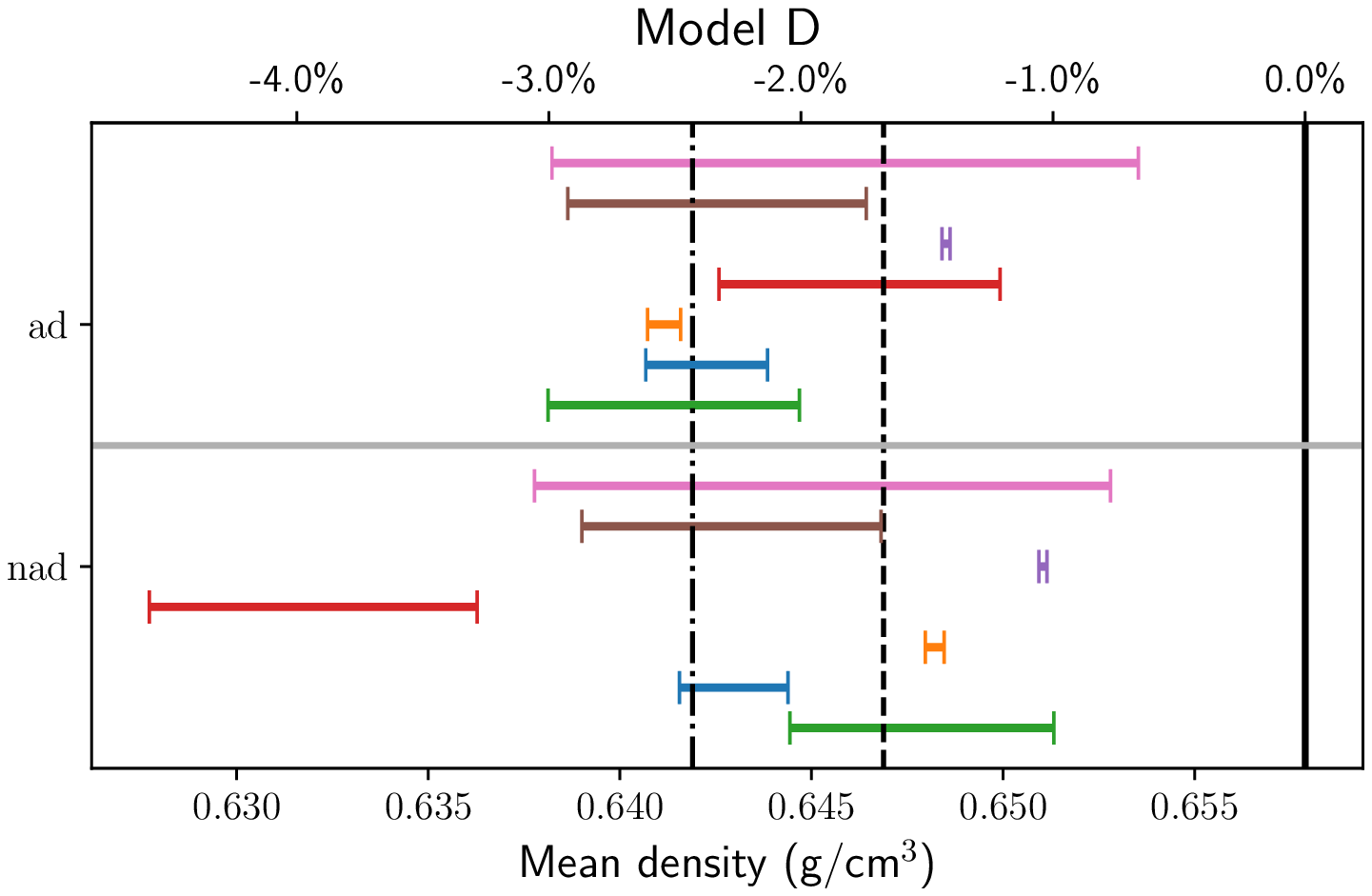}  
  \label{fig_appendix_Takafumi_comp_mean_density_modelD}
\end{subfigure}
\begin{subfigure}[b]{.43\textwidth}
  \includegraphics[width=.99\linewidth]{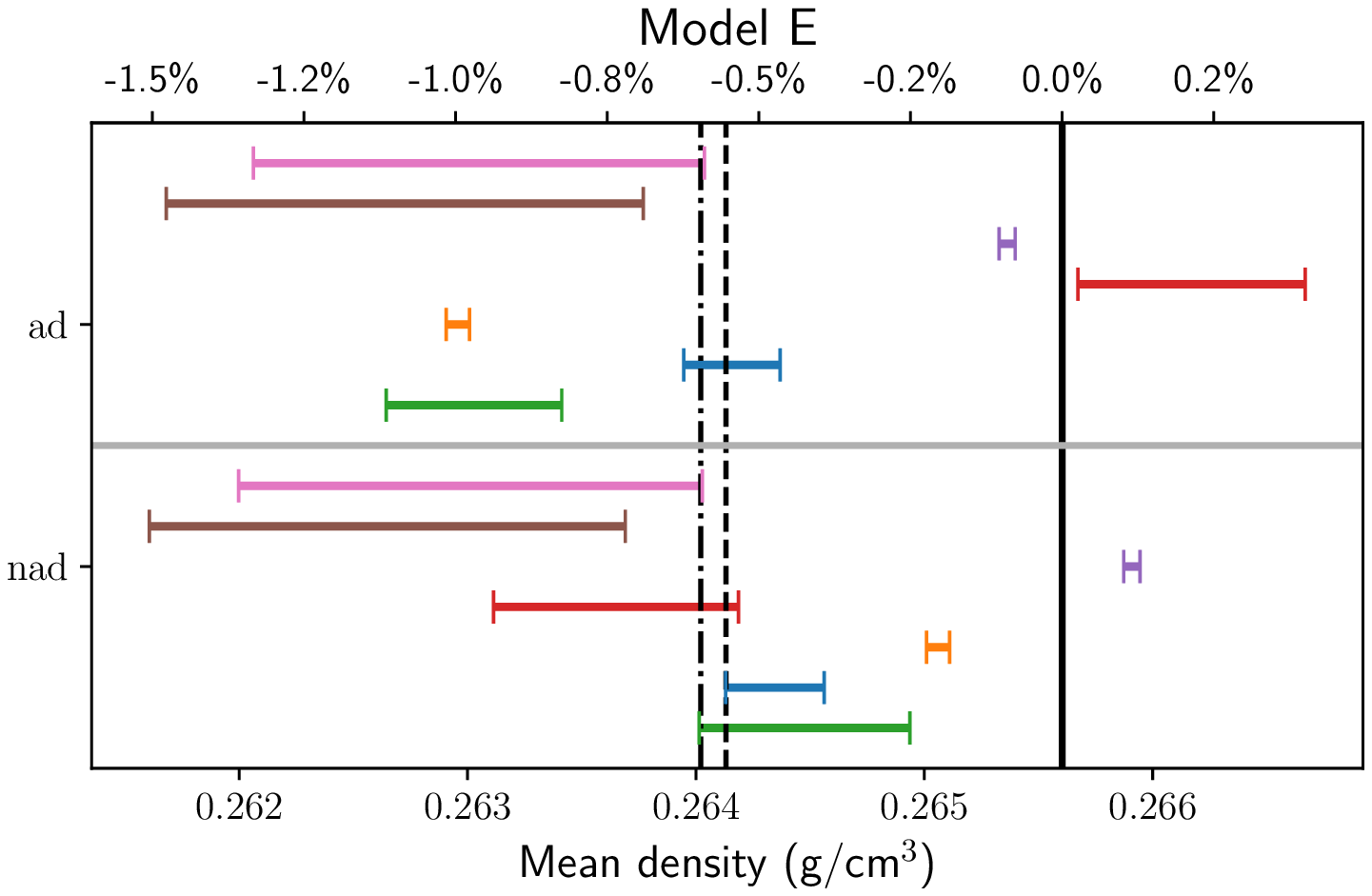}  
  \label{fig_appendix_Takafumi_comp_mean_density_modelE}
\end{subfigure}
\begin{subfigure}[b]{.43\textwidth}
  \includegraphics[width=.99\linewidth]{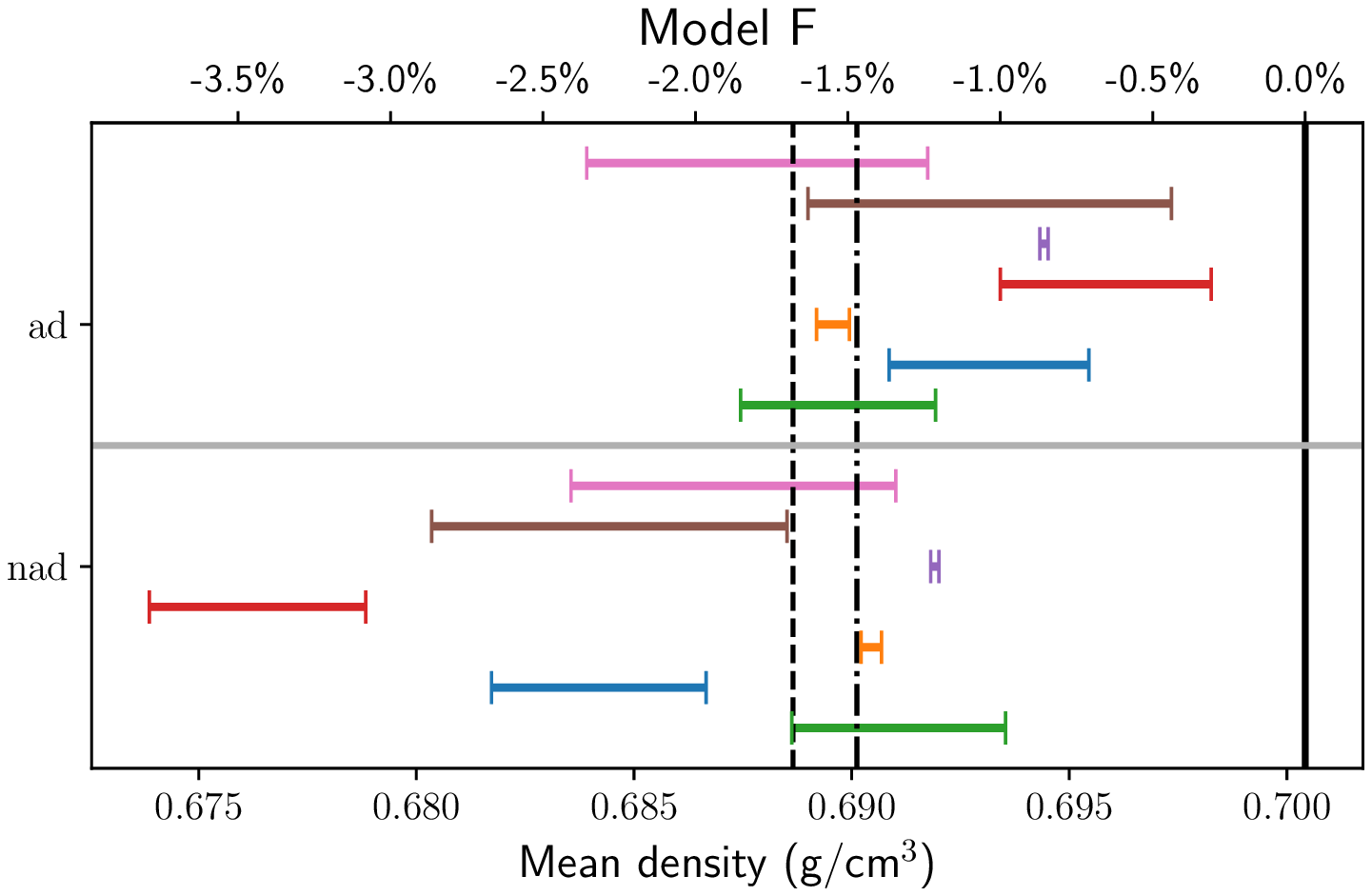}  
  \label{fig_appendix_Takafumi_comp_mean_density_modelF}
\end{subfigure}
\caption{Mean density of the \citet{Sonoi2015} targets estimated using different techniques to account for the surface effects or damp them. The black dashed and dot-dashed lines correspond to the mean density of the reference model including the non-adiabatic correction or not, respectively. The exact mean density is shown by the solid black line. Each panel is divided in two parts, separated by a solid grey line. The lower part shows the results using the frequencies that include the non-adiabatic correction (labelled as \textit{nad}), while the upper part is based on the frequencies that do not include this correction (labelled as \textit{ad}). For model C, there are no \textit{ad} results because the MCMC that provides the reference model did not converge successfully with this set of frequencies.}
\label{fig_appendix_Takafumi_comp_mean_density}
\end{figure*}

In Fig. \ref{fig_appendix_Takafumi_comp_mean_density}, we show the impact of different surface effect prescriptions on the mean density. Theoretically, the mean density should not be affected, independently of the prescription. In practice however, the choice of the prescription affects the modelled frequencies, and therefore the large separation and the mean density. If the mean density is poorly reproduced, it implies that the underlying surface effects prescription performs poorly. For this test, we considered two sets of frequencies, by including the non-adiabatic correction (labelled as `nad') or excluding it (labelled as `ad'). For each frequency set, we considered the following ways of determining the mean density:
\begin{enumerate}
\item \textit{BG2} prescription whose coefficients are optimised with AIMS by fitting the individual frequencies (in green)
\item \textit{BG2} prescription whose coefficients are optimised within the mean density inversion (in blue)
\item \textit{BG2} prescription whose  coefficients are optimised with AIMS by fitting the individual frequencies and a mean density inversion is conducted based on relative difference between the corrected frequencies and the observed frequencies (in orange)
\item \textit{S2} prescription whose coefficients are optimised within the mean density inversion (in red)
\item \textit{S2} prescription whose coefficients are derived with the scaling relations of \citet{Sonoi2015} (Eqs. 10 and 11). The individual frequencies from the reference models are corrected before carrying out the mean density inversion.
\item Damping surface effects with AIMS by fitting frequency separation ratios and treating the inverted mean density as a classical constraint (in purple)
\item Damping surface effects with AIMS by fitting frequency separation ratios and treating the inverted mean density as a seismic constraint (in brown)
\end{enumerate}

The synthetic targets fall into four categories. The first one is composed of model A, where all the estimated mean densities are consistent, all falling within $\sim 0.2\%$, which is the precision that we would expect for this kind of star \citep[model A is similar to Kepler-93,][]{Betrisey2022}. The mean density obtained with the fit of the individual frequencies is already very accurate and the inversion confirms this value, as well as the fit of the ratios. There is no significant difference by including the non-adiabatic effects or not. The second category is composed of models E and F. These models have consistent mean densities when including the non-adiabatic effects or not. The dispersion of the mean densities is larger than what we would expect for an actual observed target. This raises the question to what extend synthetic models are representative of actual observed targets, and calls into question the performance of the 3D patching. The third category is composed of models B and D. There is a significant difference when the non-adiabatic effects are included or not. We interpret this difference as a hint that the non-adiabatic correction is incompatible with the surface effects prescriptions. These results are not surprising since these prescriptions were not designed to describe corrections as strong as the one predicted by the non-adiabatic effects. Further investigations and statistics would be require testing whether the limitations lie in the formalism of the non-adiabatic effects, in the surface effect prescriptions, or in both of them. The last category is composed of model C. The MCMC of fitting the individual frequencies excluding the non-adiabatic correction failed to converge because it hits the grid boundaries. This behaviour was unexpected since the actual stellar parameters of model C should fall within the grid, which raises the question whether there is an issue with the 3D patching. This issue could also originate from differences in the physical ingredients, especially in the mixing-length parameter. Finally, we point out that fitting frequency separation ratios efficiently damp these issues, as shown by the purple and brown results.

The surface effects can be accounted for directly in the inversion or by correcting the frequencies before carrying out the inversion. Both versions performed equivalently with the \textit{BG2} prescription, but not with the \textit{S2} prescription. This latter performs poorly with the pre-corrected frequencies because it overestimated significantly the frequencies differences, resulting in a shift towards the left in the HR diagram, which is then interpreted by the inversion as a reference mean density that is too small. The inversion therefore wrongly corrected the model towards a larger mean density. Accounting for the \textit{S2} prescription directly in the inversion also performs poorly. The inversion cannot determine robustly the prescription coefficients, and a non-negligible correction would be the result of the poor fit of the target function and not of a physical structure difference.

\section{Supplementary data for the \emph{LEGACY} targets}
\label{appendix_additional_data_LEGACY}

In Table \ref{tab_final_params_LEGACY}, we provided the optimal stellar parameters of our subsample of \emph{LEGACY} targets, determined with the procedure coupling mean density inversions and frequency separation ratios. For these fits, we treated the inverted mean density as a classical constraint.

\begin{table*}[h!]
\centering
\caption{Stellar parameters of the targets selected from the \emph{LEGACY} sample.}
\resizebox{\linewidth}{!}{
\begin{tabular}{lccccccccc}
\hline 
 & $M$ & $R$ & Age & $\bar{\rho}$ & $Y$ & $Z$ & $T_{\mathrm{eff}}$ & $L$ & $\nu_{max}$ \\
 & $(M_\odot)$ & $(R_\odot)$ & (Myr) & (g/cm$^3$) & (\%) & (\%) & (K) & $(L_\odot)$ & ($\mu$Hz) \\
\hline \hline 
16 Cyg A & $1.08 \pm 0.03$ & $1.226 \pm 0.011$ & $7079 \pm 109$ & $0.824 \pm 0.001$ & $0.26 \pm 0.02$ & $0.019 \pm 0.001$ & $5827 \pm 21$ & $1.56 \pm 0.01$ & $2203.9 \pm 22.6$ \\ 
16 Cyg B & $1.01 \pm 0.02$ & $1.105 \pm 0.009$ & $7057 \pm 111$ & $1.057 \pm 0.001$ & $0.27 \pm 0.02$ & $0.017 \pm 0.001$ & $5788 \pm 27$ & $1.23 \pm 0.01$ & $2555.2 \pm 24.5$ \\ 
Arthur & $1.21 \pm 0.03$ & $1.154 \pm 0.011$ & $1545 \pm 90$ & $1.115 \pm 0.004$ & $0.24 \pm 0.01$ & $0.018 \pm 0.003$ & $6065 \pm 62$ & $1.62 \pm 0.04$ & $2750.7 \pm 29.7$ \\ 
Barney & $1.27 \pm 0.01$ & $1.610 \pm 0.007$ & $4059 \pm 176$ & $0.429 \pm 0.003$ & $0.30 \pm 0.02$ & $0.027 \pm 0.002$ & $5985 \pm 17$ & $3.00 \pm 0.04$ & $1487.3 \pm 7.7$ \\ 
Carlsberg & $1.21 \pm 0.03$ & $1.170 \pm 0.011$ & $2076 \pm 138$ & $1.065 \pm 0.003$ & $0.24 \pm 0.02$ & $0.019 \pm 0.002$ & $6050 \pm 47$ & $1.65 \pm 0.03$ & $2668.1 \pm 27.5$ \\ 
Doris & $1.01 \pm 0.01$ & $0.936 \pm 0.003$ & $5617 \pm 125$ & $1.728 \pm 0.011$ & $0.26 \pm 0.00$ & $0.030 \pm 0.002$ & $5237 \pm 31$ & $0.59 \pm 0.01$ & $3726.0 \pm 24.5$ \\ 
Dushera & $1.21 \pm 0.03$ & $1.417 \pm 0.012$ & $3516 \pm 192$ & $0.600 \pm 0.002$ & $0.28 \pm 0.02$ & $0.019 \pm 0.002$ & $6170 \pm 32$ & $2.62 \pm 0.05$ & $1802.1 \pm 14.6$ \\ 
Nunny & $1.06 \pm 0.01$ & $1.240 \pm 0.004$ & $5847 \pm 201$ & $0.783 \pm 0.001$ & $0.26 \pm 0.01$ & $0.012 \pm 0.001$ & $6058 \pm 17$ & $1.87 \pm 0.02$ & $2079.3 \pm 7.9$ \\ 
Pinocha & $1.33 \pm 0.03$ & $1.299 \pm 0.010$ & $1599 \pm 125$ & $0.855 \pm 0.004$ & $0.24 \pm 0.01$ & $0.022 \pm 0.003$ & $6197 \pm 47$ & $2.24 \pm 0.04$ & $2349.1 \pm 22.3$ \\ 
Saxo2 & $1.25 \pm 0.02$ & $1.268 \pm 0.008$ & $2457 \pm 80$ & $0.864 \pm 0.003$ & $0.24 \pm 0.01$ & $0.018 \pm 0.002$ & $6174 \pm 34$ & $2.10 \pm 0.03$ & $2323.6 \pm 14.3$ \\ 
\hline 
\end{tabular} 
}
{\par\small\justify\textbf{Notes.} These are the results of the fit of the frequency separation ratios, the classical constraints (metallicity and effective temperature for all the targets, and luminosity except for Arthur, Doris, and Pinocha), and the inverted mean density treated as a classical constraint. \par}
\label{tab_final_params_LEGACY}
\end{table*}

\section{Supplementary diagnostic plots for AIMS convergence}
\label{appendix_AIMS_convergence}
\begin{figure*}[h!]
  \centering
\begin{subfigure}[b]{.45\textwidth}
  \includegraphics[width=.99\linewidth]{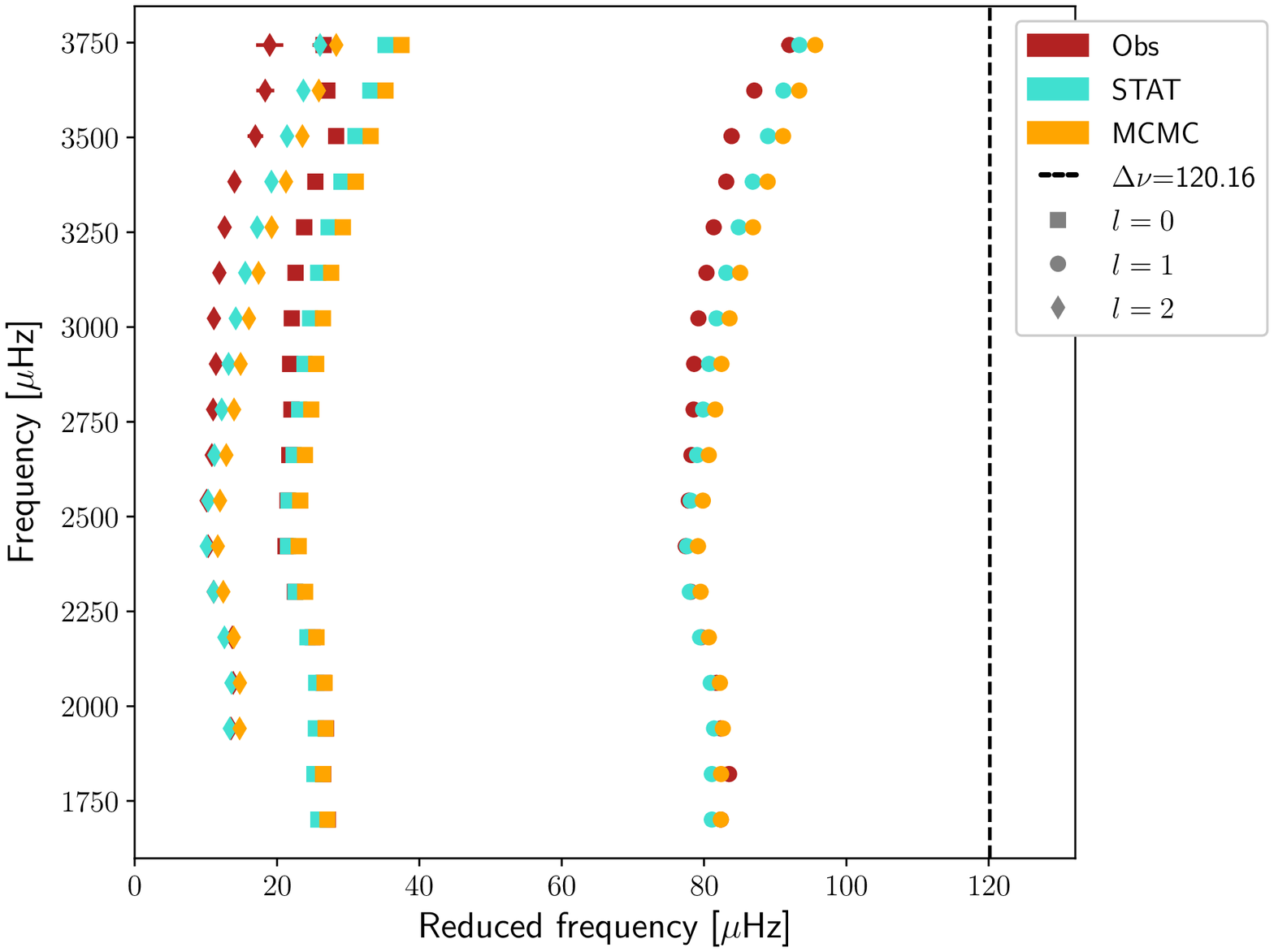}  
  \caption{\centering Echelle diagram}
  \label{fig_appendix_successful_convergence_echelle}
\end{subfigure}
\begin{subfigure}[b]{.47\textwidth}
  \includegraphics[width=.99\linewidth]{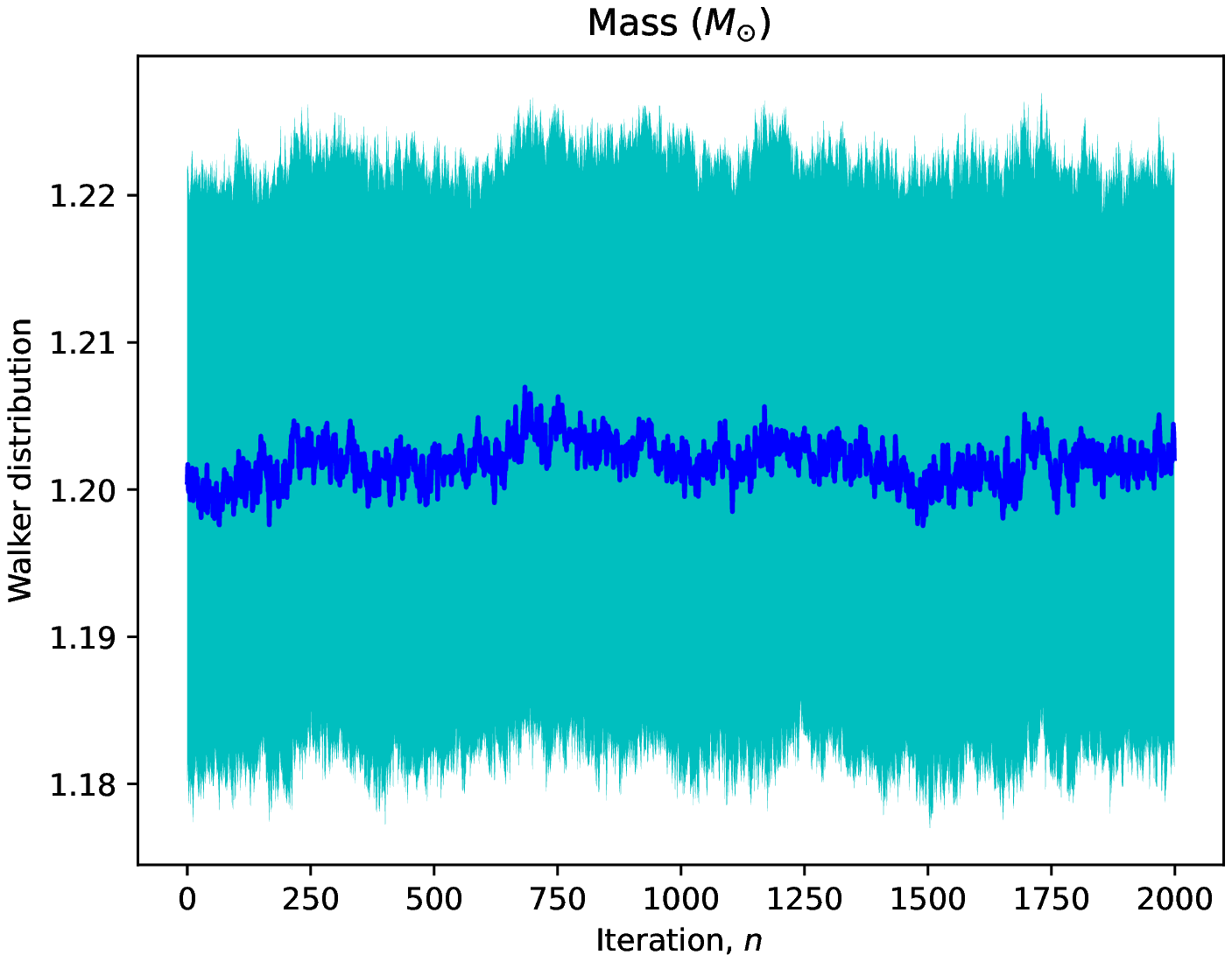} 
  \caption{\centering Distribution of the walkers as a function of the iteration} 
  \label{fig_appendix_successful_convergence_iter}
\end{subfigure}
\begin{subfigure}[b]{.82\textwidth}
  \includegraphics[width=.99\linewidth]{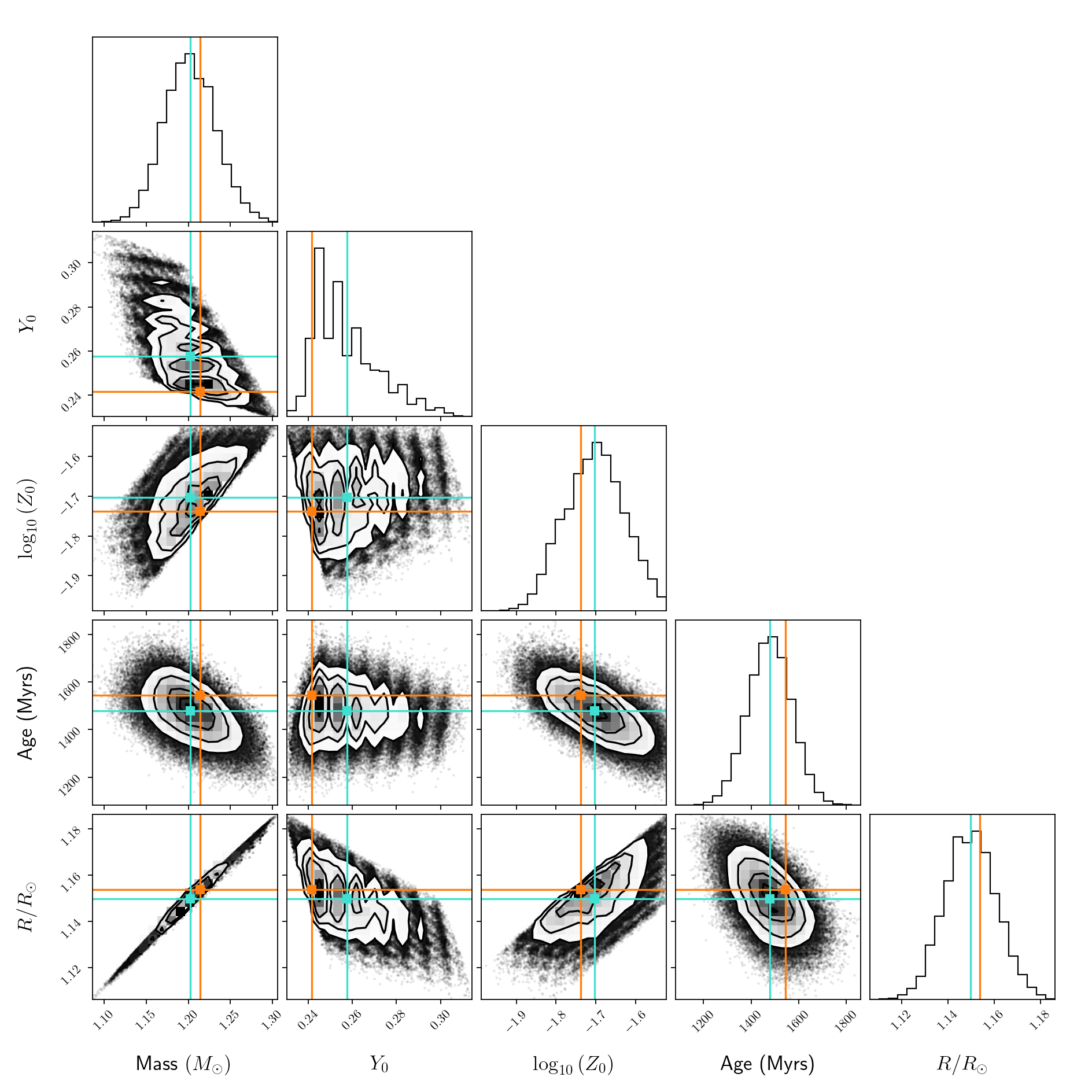} 
  \caption{\centering Triangle plot} 
  \label{fig_appendix_successful_convergence_triangle}
\end{subfigure}
\caption{Diagnostic plots of a MCMC run with successful convergence. The median parameters are denoted in cyan, and the best MCMC model, for which the $\chi^2$ is minimal, is denoted in orange.}
\label{fig_appendix_successful_convergence}
\end{figure*}

In Fig. \ref{fig_appendix_successful_convergence}, we show an illustration of the diagnostic plots of a successful convergence with AIMS. The échelle diagram is consistent, the temporal evolution of the walkers is flat, indicating that the burn-in phase was successful and that the walkers reached the global minimum in the parameter space. The triangle plot of the radius and optimised variables (mass, chemical composition, and age) is consistent and the posterior distributions show uni-modal distributions, which shows that the MCMC found the global minimum.

\begin{figure*}[h!]
  \centering
\begin{subfigure}[b]{.45\textwidth}
  \includegraphics[width=.99\linewidth]{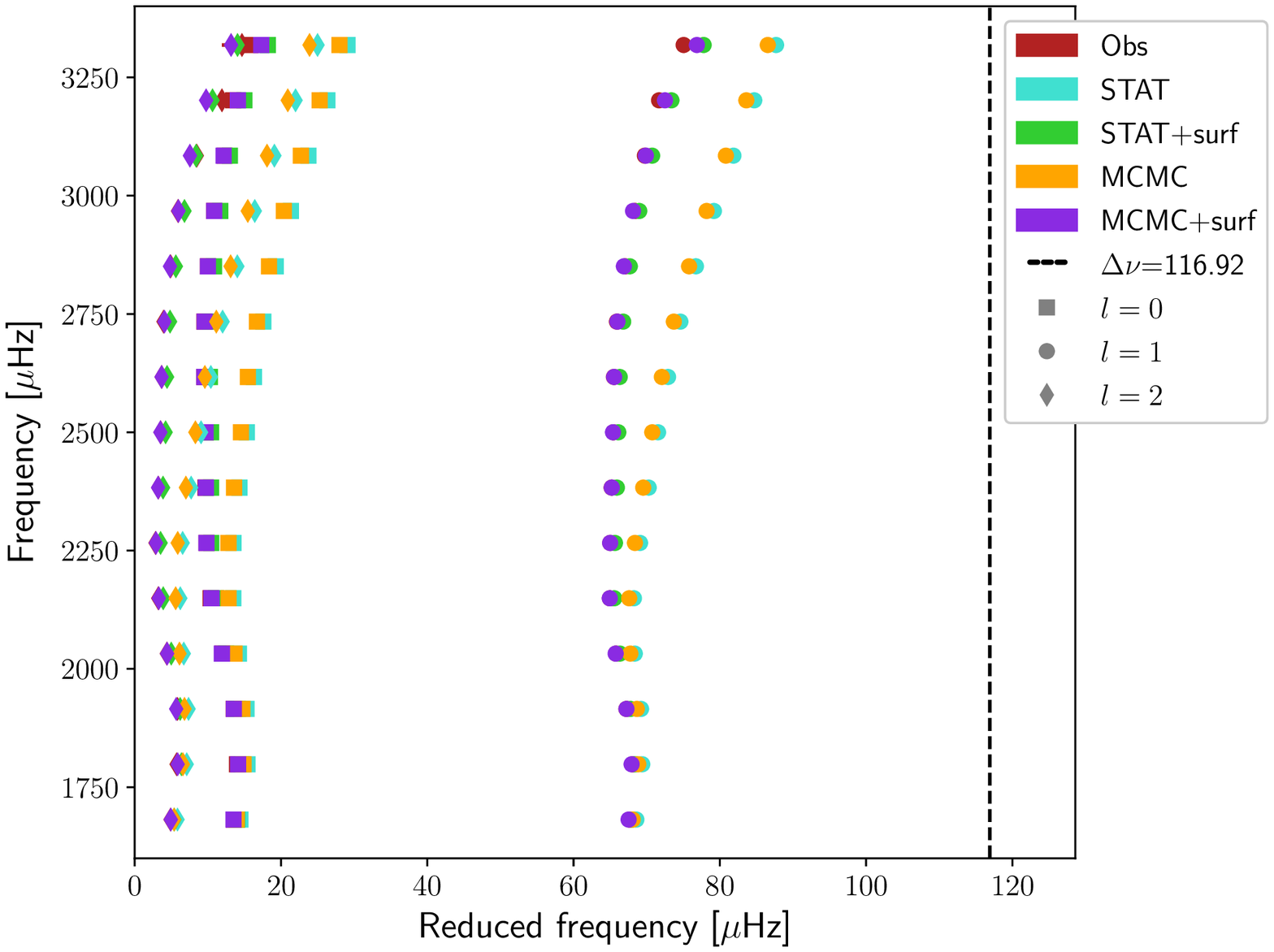}  
  \caption{\centering Echelle diagram}
  \label{fig_appendix_drift_walkers_echelle}
\end{subfigure}
\begin{subfigure}[b]{.47\textwidth}
  \includegraphics[width=.99\linewidth]{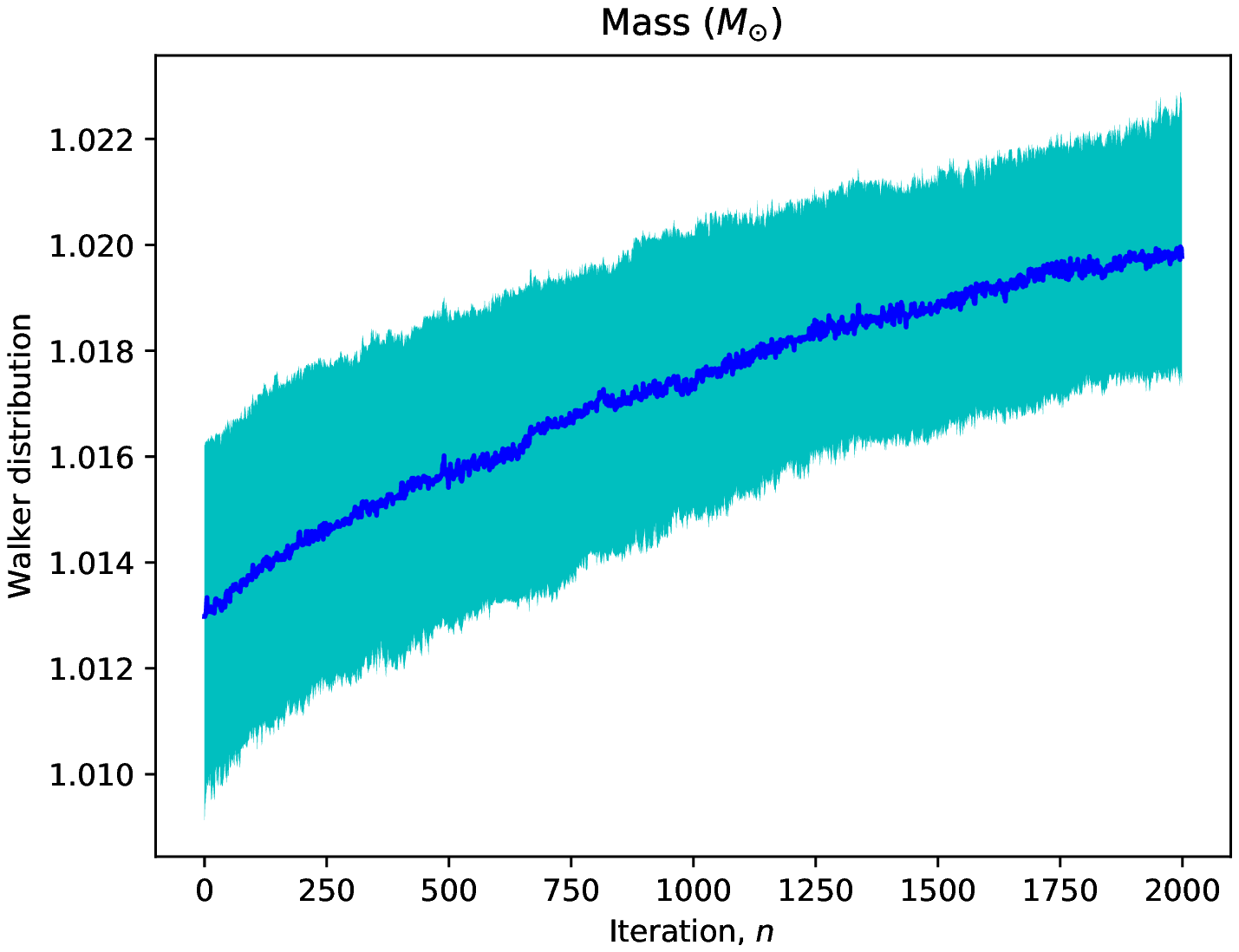} 
  \caption{\centering Distribution of the walkers as a function of the iteration} 
  \label{fig_appendix_drift_walkers_iter}
\end{subfigure}
\begin{subfigure}[b]{.82\textwidth}
  \includegraphics[width=.99\linewidth]{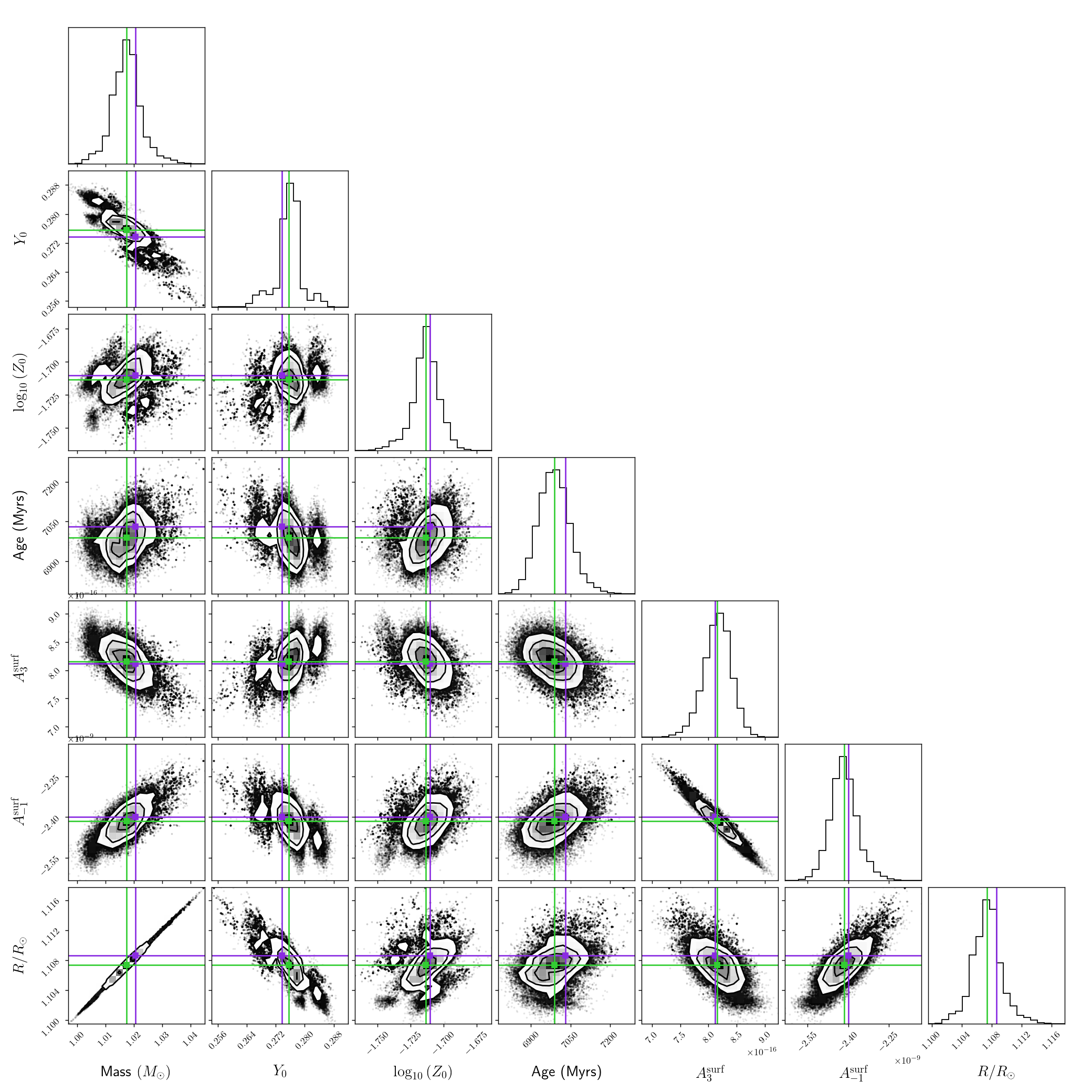} 
  \caption{\centering Triangle plot} 
  \label{fig_appendix_drift_walkers_triangle}
\end{subfigure}
\caption{Illustration of walkers drifting during the MCMC iterations. In this case, the walkers are still drifting after a burn-in of 2000 steps. The median parameters are denoted in green, and the best MCMC model, for which the $\chi^2$ is minimal, is denoted in purple.}
\label{fig_appendix_drift_walkers}
\end{figure*}

In Fig. \ref{fig_appendix_drift_walkers}, we show an illustration of an unsuccessful convergence due to a drift of the walkers during the iterations of the MCMC. This issue means that the MCMC is still in the burn-in phase.

\begin{figure*}[h!]
  \centering
\begin{subfigure}[b]{.45\textwidth}
  \includegraphics[width=.99\linewidth]{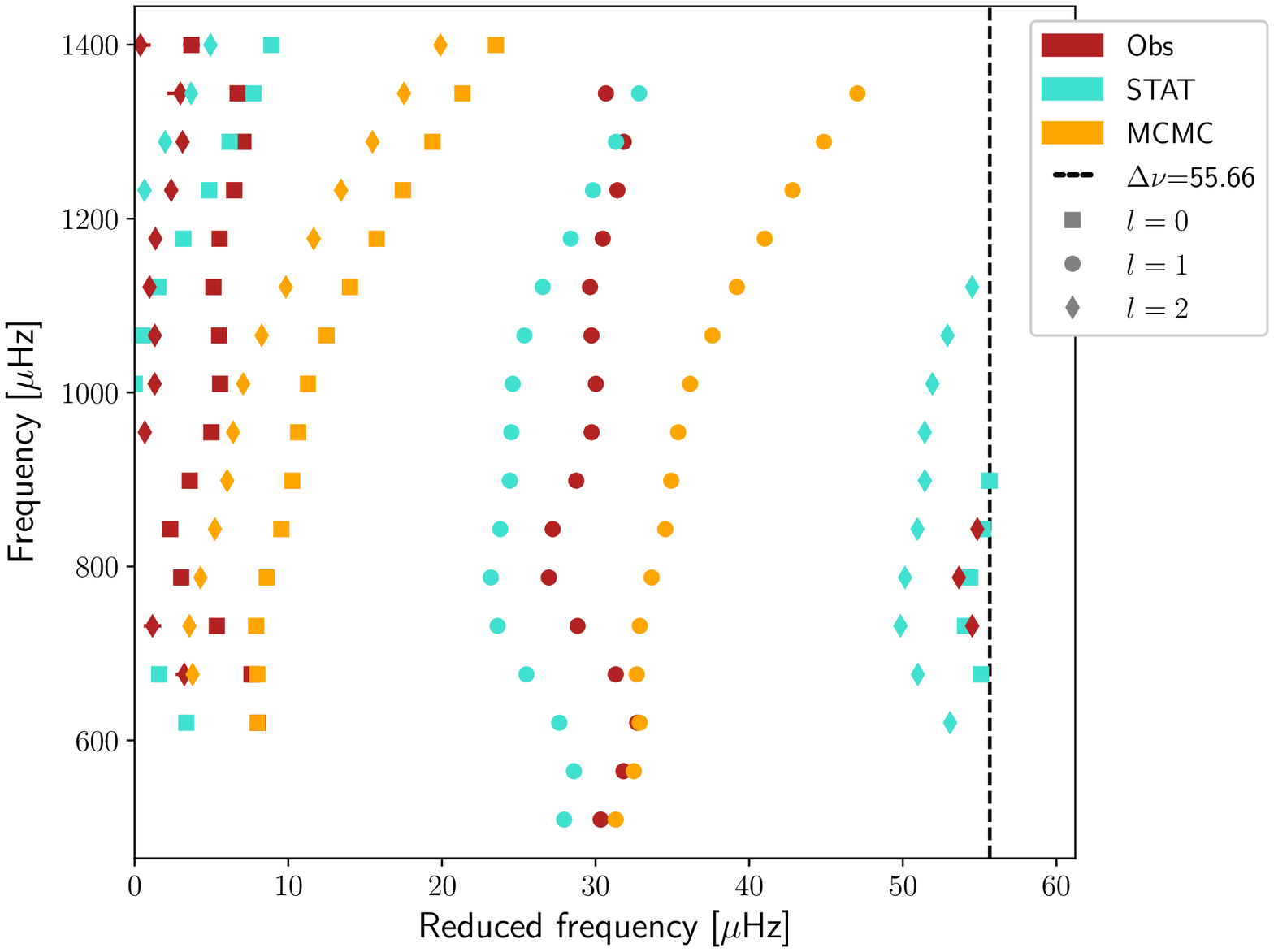}  
  \caption{\centering Echelle diagram}
  \label{fig_appendix_issue_numin0_echelle}
\end{subfigure}
\begin{subfigure}[b]{.47\textwidth}
  \includegraphics[width=.99\linewidth]{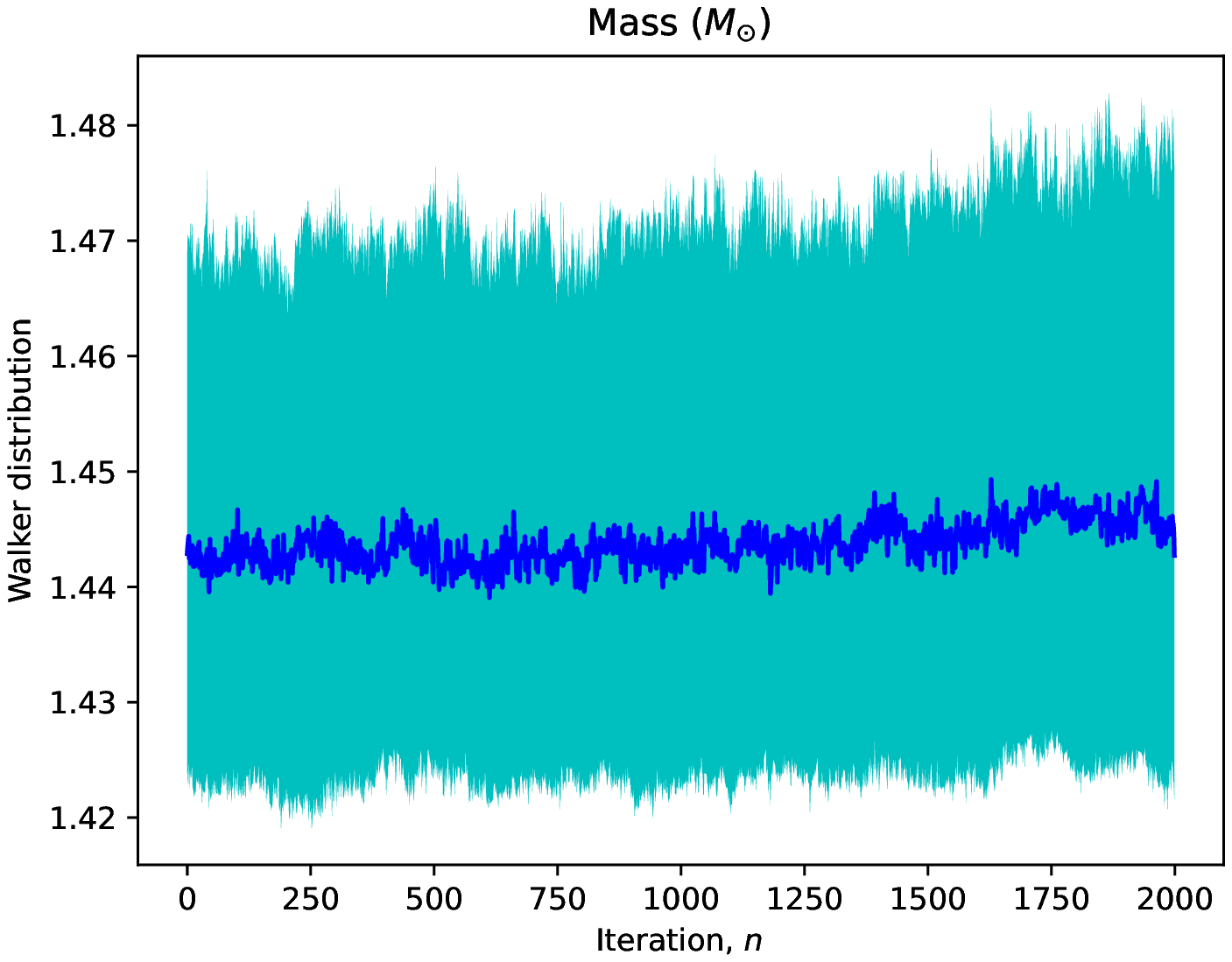} 
  \caption{\centering Distribution of the walkers as a function of the iteration} 
  \label{fig_appendix_issue_numin0_iter}
\end{subfigure}
\begin{subfigure}[b]{.82\textwidth}
  \includegraphics[width=.99\linewidth]{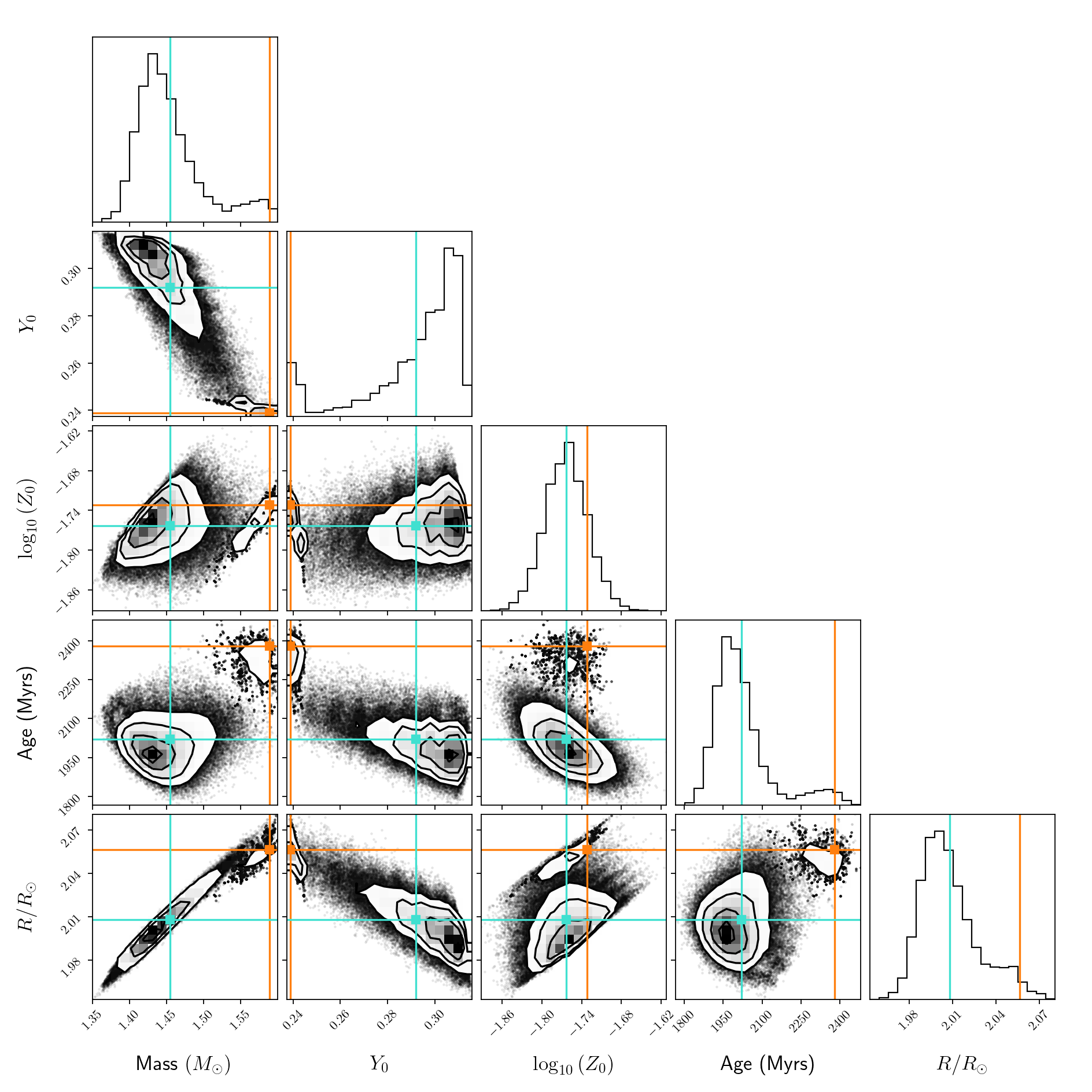} 
  \caption{\centering Triangle plot} 
  \label{fig_appendix_issue_numin0_triangle}
\end{subfigure}
\caption{Illustration of an issue that occurs while trying to fit the lowest order radial frequency. The MCMC sees a second local minimum and traps the walkers in it. The median parameters are denoted in cyan, and the best MCMC model, for which the $\chi^2$ is minimal, is denoted in orange.}
\label{fig_appendix_issue_numin0}
\end{figure*}

In Fig. \ref{fig_appendix_issue_numin0}, we show an illustration of an unsuccessful convergence due to an issue occurring while trying to fit the lowest order radial frequency. The MCMC sees a second suspicious local minimum and traps the walkers in it, thus biasing the stellar parameters.

\begin{figure*}[h!]
  \centering
\begin{subfigure}[b]{.45\textwidth}
  \includegraphics[width=.99\linewidth]{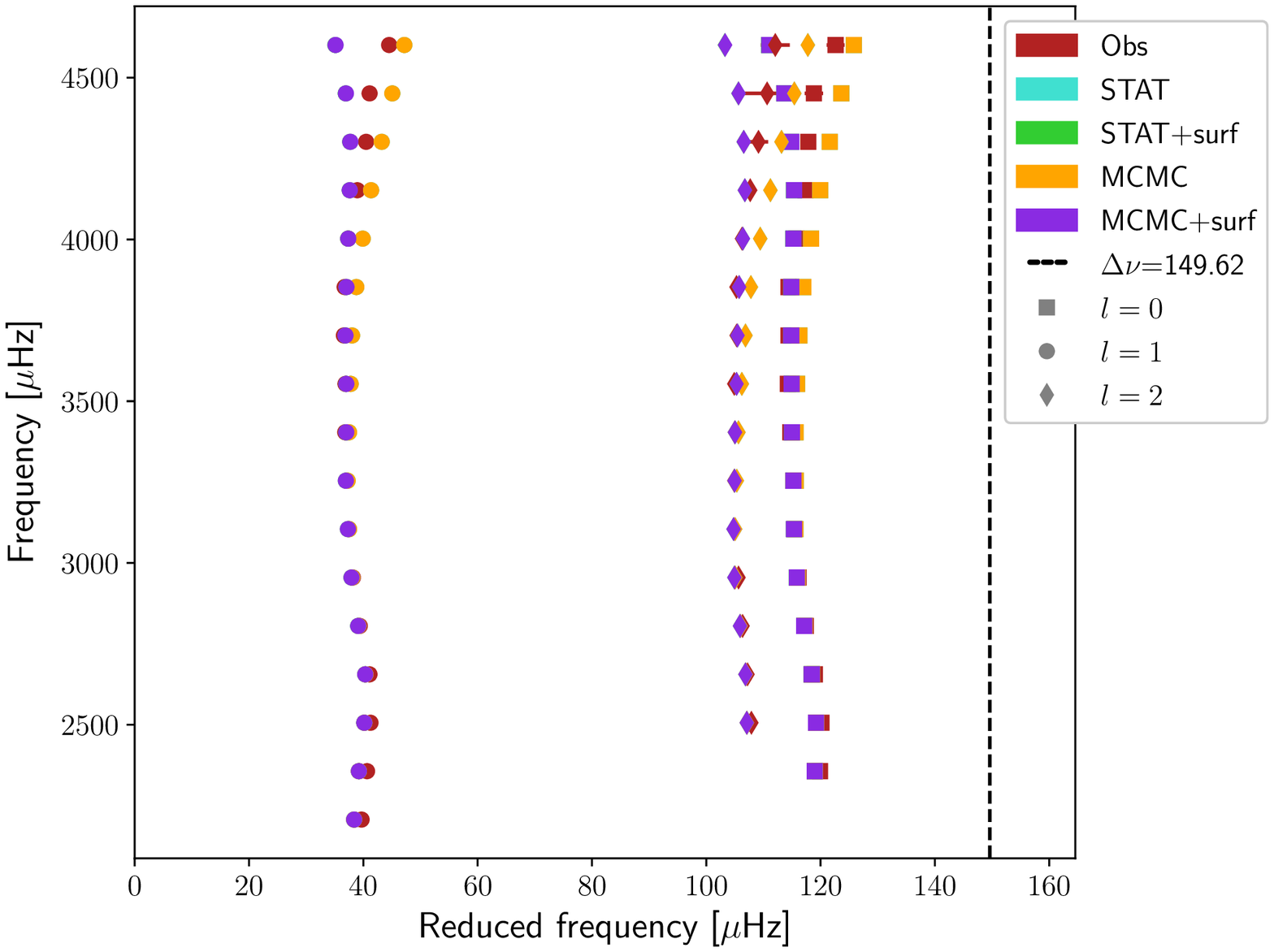}  
  \caption{\centering Echelle diagram}
  \label{fig_appendix_hit_boundaries_echelle}
\end{subfigure}
\begin{subfigure}[b]{.47\textwidth}
  \includegraphics[width=.99\linewidth]{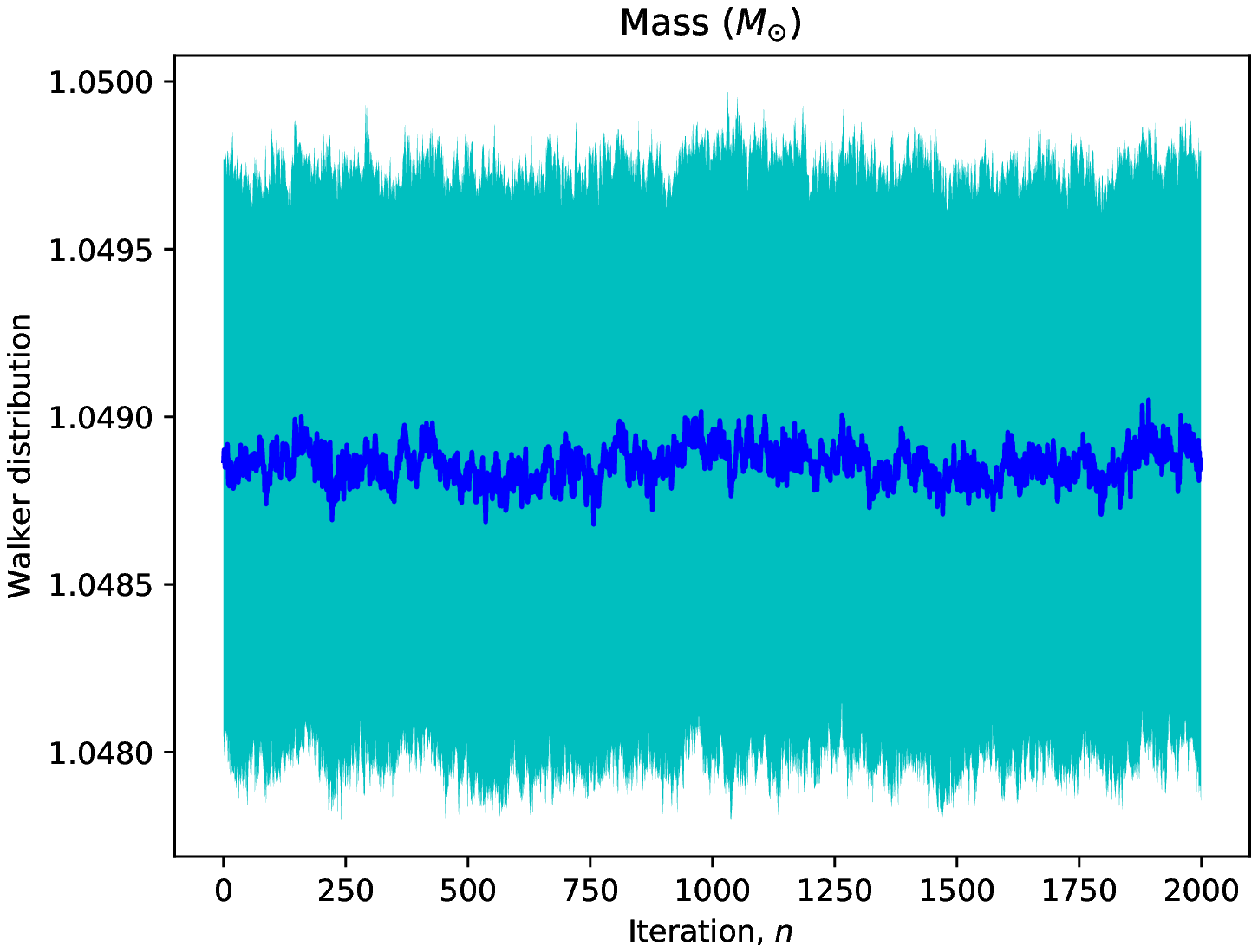} 
  \caption{\centering Distribution of the walkers as a function of the iteration} 
  \label{fig_appendix_hit_boundaries_iter}
\end{subfigure}
\begin{subfigure}[b]{.82\textwidth}
  \includegraphics[width=.99\linewidth]{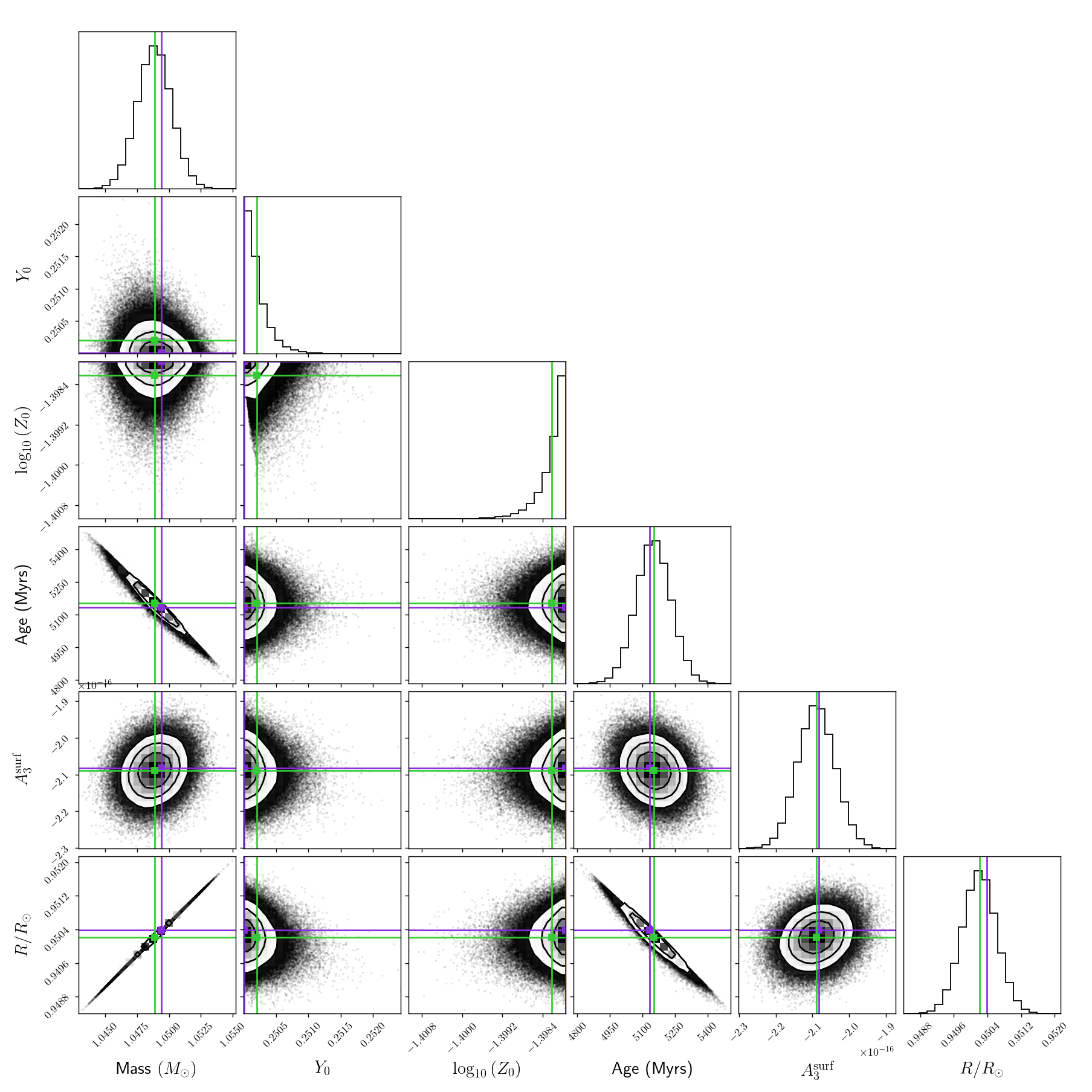} 
  \caption{\centering Triangle plot} 
  \label{fig_appendix_hit_boundaries_triangle}
\end{subfigure}
\caption{Illustration of a run that hit the grid boundaries. The median parameters are denoted in green, and the best MCMC model, for which the $\chi^2$ is minimal, is denoted in purple.}
\label{fig_appendix_hit_boundaries}
\end{figure*}

In Fig. \ref{fig_appendix_hit_boundaries}, we show an illustration of an unsuccessful convergence due to walkers hitting the grid boundaries during the minimisation. This is the main issue that was encountered in our study. Histograms with sharp features are not necessarily the sign of a grid that is too small. It can indicate that there are significant physical differences between the grid models and the observed target, and that the MCMC is trying to compensate with the free variables at its disposal. A typical indicator of this issue is an excessively high metallicity. 

\begin{figure*}[h!]
  \centering
\begin{subfigure}[b]{.45\textwidth}
  \includegraphics[width=.99\linewidth]{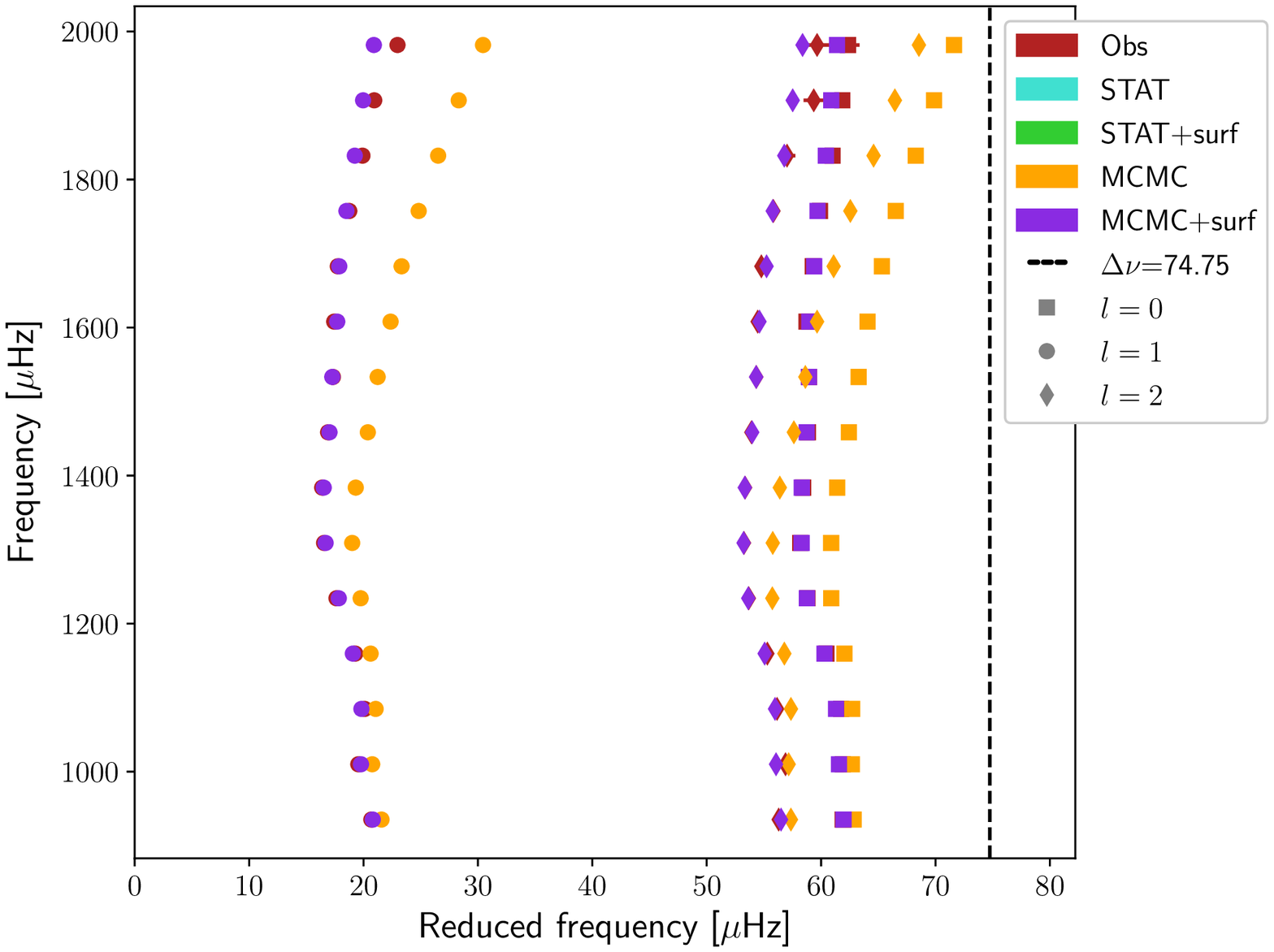}  
  \caption{\centering Echelle diagram}
  \label{fig_appendix_peaked_distributions_echelle}
\end{subfigure}
\begin{subfigure}[b]{.47\textwidth}
  \includegraphics[width=.99\linewidth]{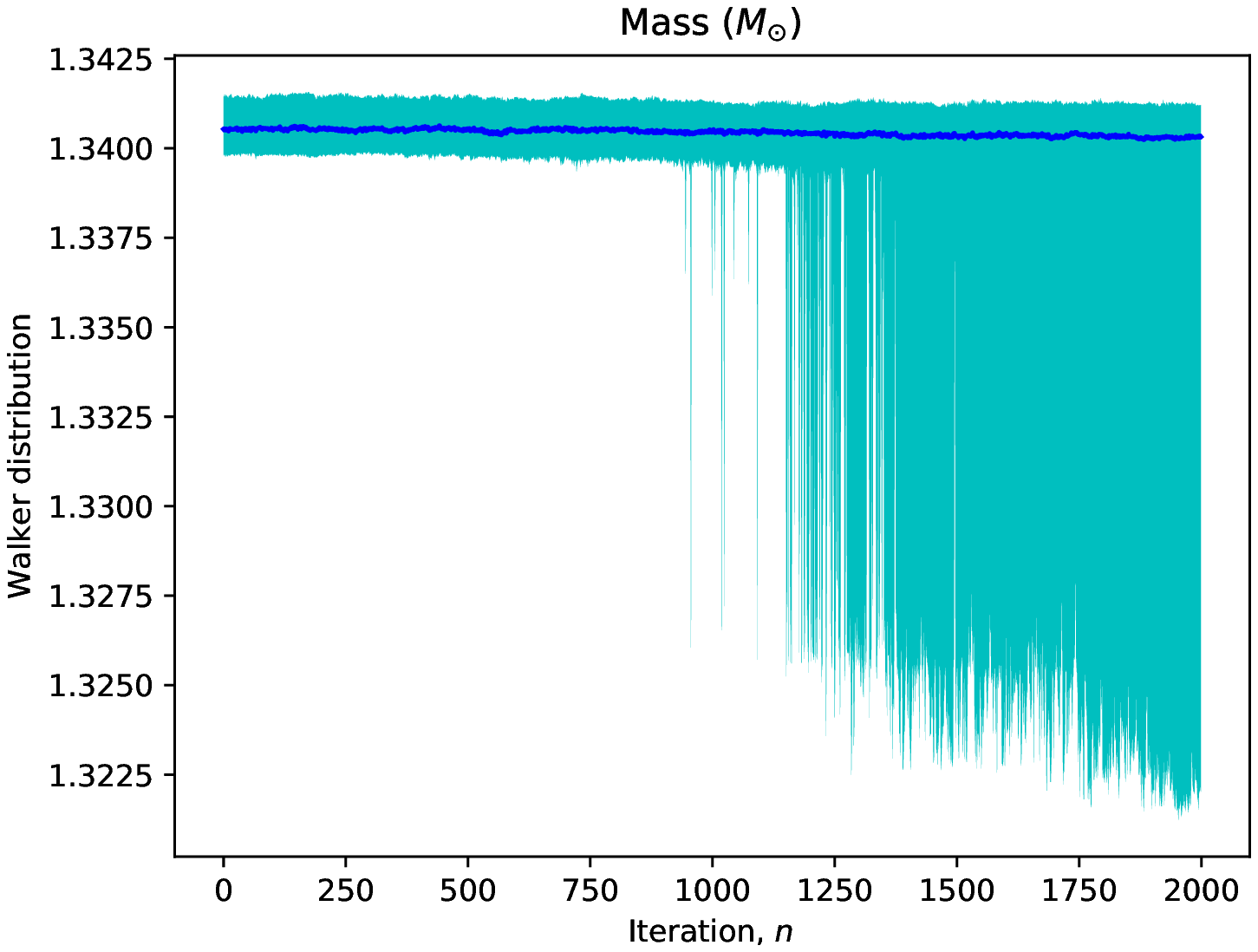} 
  \caption{\centering Distribution of the walkers as a function of the iteration} 
  \label{fig_appendix_peaked_distributions_iter}
\end{subfigure}
\begin{subfigure}[b]{.82\textwidth}
  \includegraphics[width=.99\linewidth]{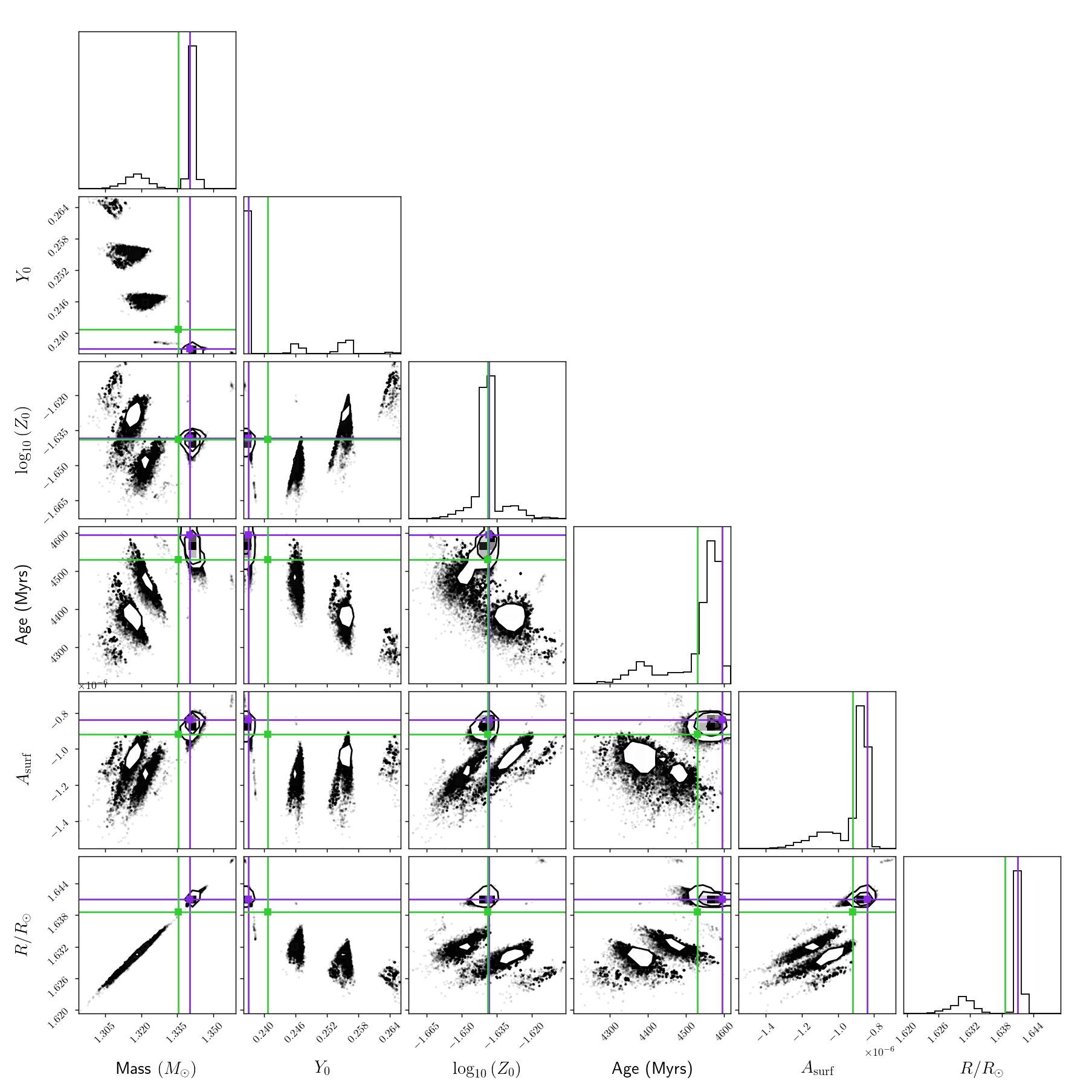} 
  \caption{\centering Triangle plot} 
  \label{fig_appendix_peaked_distributions_triangle}
\end{subfigure}
\caption{Illustration of a run with an excessively peaked posterior distribution. The median parameters are denoted in green, and the best MCMC model, for which the $\chi^2$ is minimal, is denoted in purple.}
\label{fig_appendix_peaked_distributions}
\end{figure*}

In Fig. \ref{fig_appendix_peaked_distributions}, we show an illustration of an unsuccessful convergence due to excessively peaked posterior distributions. This issue is a bit tricky because having excessively peaked posterior distributions does not necessarily imply that the minimisation went wrong. However, it calls into question whether the interpolation was successful. In this illustration, we see that the posterior distributions are multi-modal, indicating that the walkers stuck on grid points and that the interpolation was unsuccessful.

\begin{figure*}[h!]
  \centering
\begin{subfigure}[b]{.45\textwidth}
  \includegraphics[width=.99\linewidth]{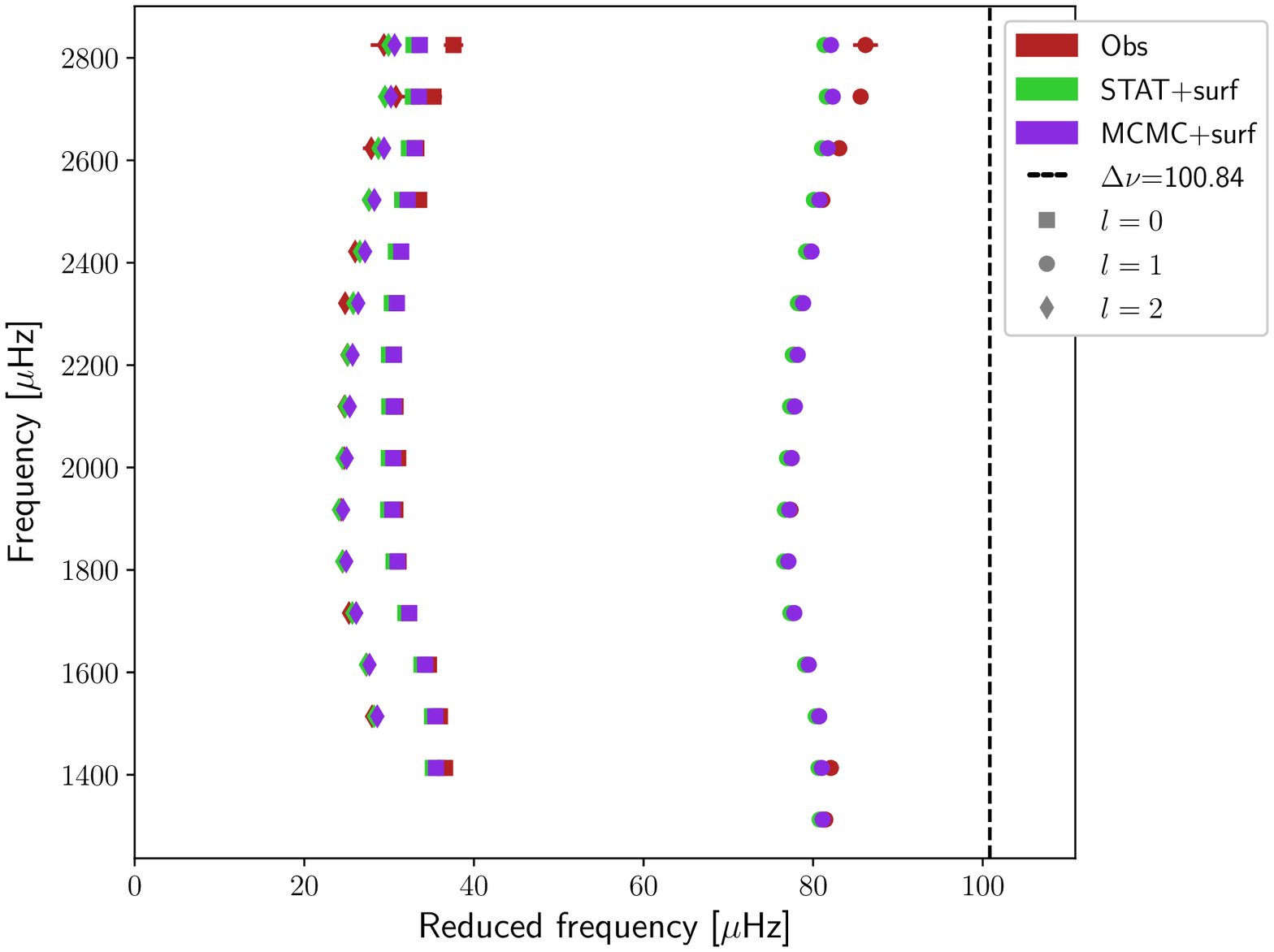}  
  \caption{\centering Echelle diagram}
  \label{fig_appendix_issue_BG1_echelle}
\end{subfigure}
\begin{subfigure}[b]{.47\textwidth}
  \includegraphics[width=.99\linewidth]{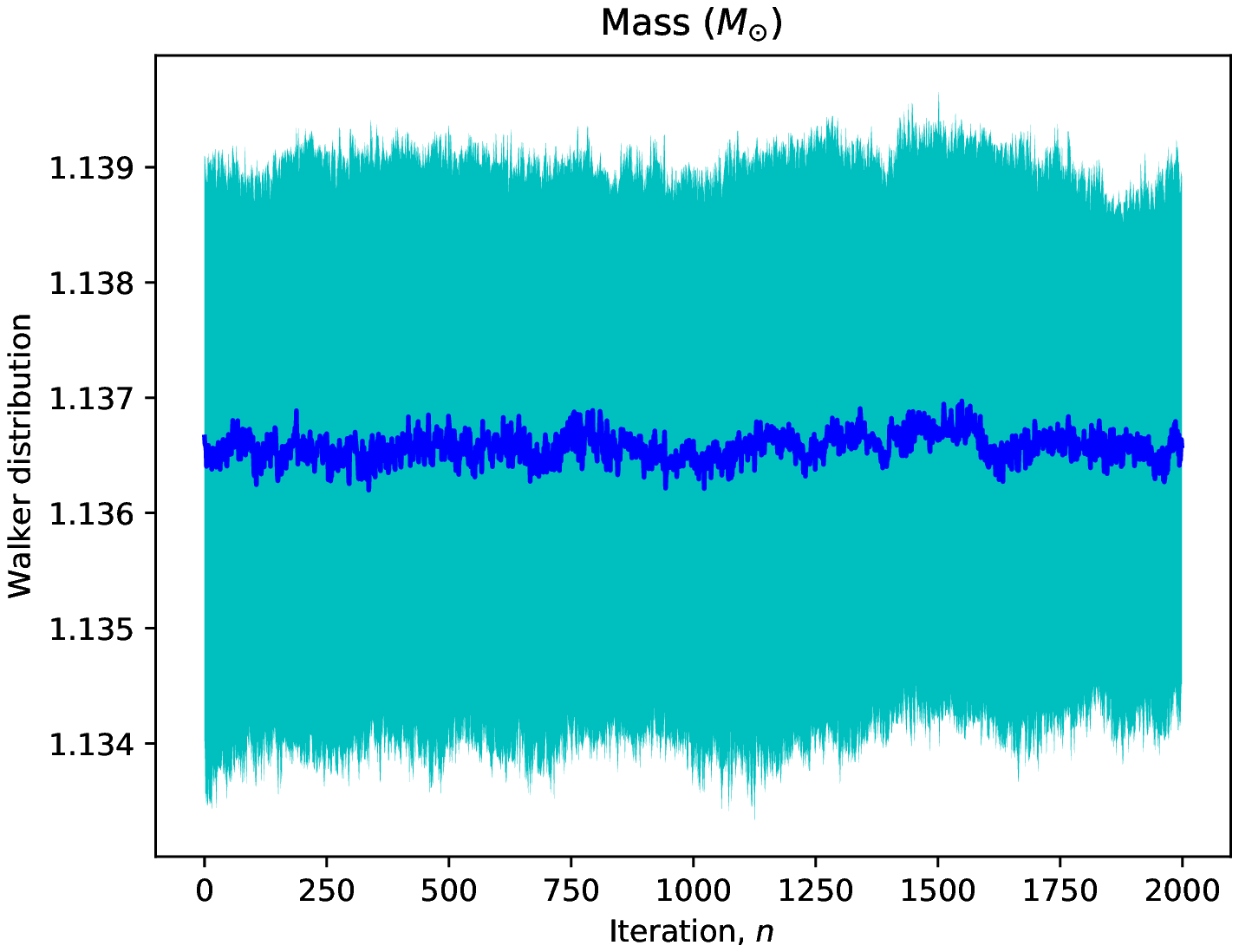} 
  \caption{\centering Distribution of the walkers as a function of the iteration} 
  \label{fig_appendix_issue_BG1_iter}
\end{subfigure}
\begin{subfigure}[b]{.8\textwidth}
  \includegraphics[width=.99\linewidth]{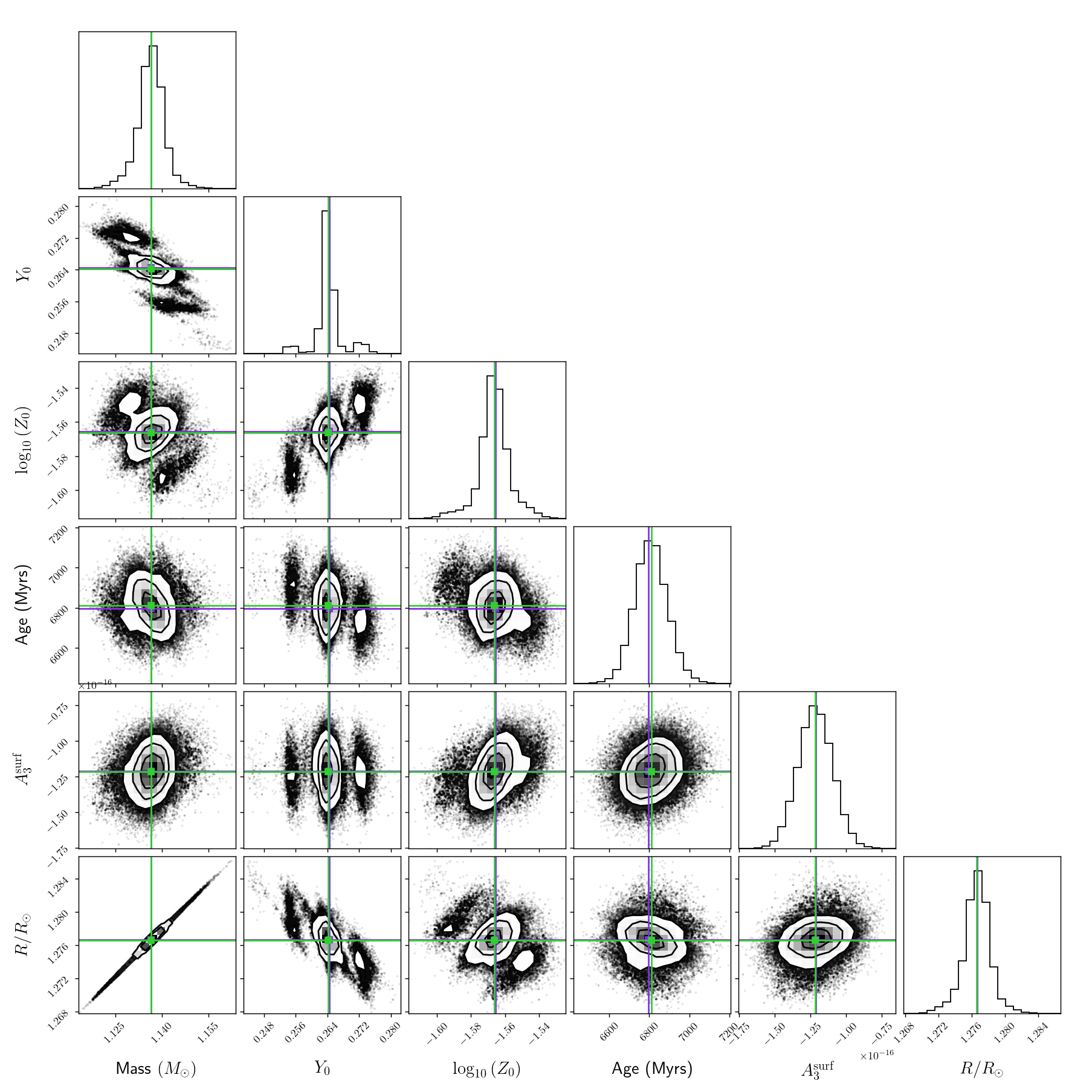} 
  \caption{\centering Triangle plot} 
  \label{fig_appendix_issue_BG1_triangle}
\end{subfigure}
\caption{Illustration of the difficulty with which the BG1 surface prescription reproduces the high frequencies. The median parameters are denoted in green, and the best MCMC model, for which the $\chi^2$ is minimal, is denoted in purple.}
\label{fig_appendix_issue_BG1}
\end{figure*}

In Fig. \ref{fig_appendix_issue_BG1}, we show an illustration of an unsuccessful convergence due
to the surface prescription. In that case, the \textit{BG1} prescription was used. This prescription is known to have difficulties in reproducing the high frequencies, which is what we observe in the illustration. Although the other diagnostic plots do not show irregularities, the fact that the prescription fails to reproduce the high frequencies can significantly bias the stellar parameters.

\end{appendix}
\end{document}